\documentclass[aps]{revtex4}
\usepackage{amsmath,amssymb}
\usepackage{color,graphicx}
\usepackage{multirow}
\usepackage{rotating}
\newcommand{\be}{\begin{equation}}
\newcommand{\ee}{\end{equation}}
\begin{document}
\title{Nonlocally-induced (fractional)  bound states: Shape analysis in the infinite  Cauchy well}
\author{Mariusz  \.{Z}aba  and Piotr Garbaczewski}
\affiliation{Institute of Physics, University of Opole, 45-052
Opole, Poland}
\date{\today }
\begin{abstract}
Fractional (L\'{e}vy-type) operators  are known to be  spatially nonlocal.
This  becomes an issue if  confronted with  a priori imposed exterior Dirichlet boundary data.
We  address spectral properties of the   prototype example of the  Cauchy operator  $(-\Delta )^{1/2}$
 in  the interval $D=(-1,1) \subset R$, with a focus  on  functional shapes of  lowest eigenfunctions
 and their fall-off at the boundaries of $D$.  New  high accuracy
   formulas are deduced  for approximate eigenfunctions.  We  analyze  how their  shape reproduction
   fidelity  is  correlated   with the evaluation finesse
 of  the  corresponding   eigenvalues.
\end{abstract}
 \maketitle

\section{Fractional Laplacians: $R$  versus $D\subset R$.}

The Fourier integral
$  {\frac{1}{\sqrt{2\pi}}} \int_R   |k|^\mu \tilde{f}(k)
  e^{-\imath kx}dk  =   - \partial _{\mu }f(x)/\partial |x|^{\mu }  =  |\Delta |^{\mu /2}f(x)$
  is commonly interpreted   as a definition of  a fractional derivative
  of the $\mu $-th order for $\mu \in (0,2)$. The notation  $- (-\Delta )^{\mu /2} = - |\Delta |^{\mu /2}$
    refers to   a fractional Laplacian  of order $\mu /2$ (restricted to dimension one  i.e. to $R$)  and  two versions  of a fractional dynamics (dimensional constants being scaled away):
   semigroup $\exp(- t |\Delta |^{\mu /2}) \,f$ and unitary  $\exp(- it |\Delta |^{\mu /2})\, f$,  \cite{GS}-\cite{getoor}.
    Here $\tilde{f}$ stands for a Fourier transform of $f\in L^2(R)$ and   $g(k)=|k|^{\mu }\tilde{f}(k)$ is presumed to be $L^2(R)$-integrable.

Apart from the unperturbed (free)  case,  the Fourier  (multiplier) representation of the fractional dynamics
 has proved useful if an infinite or periodic support is  admitted for functions in the  domain, \cite{ZRK}.
  For the  simplest quadratic ($\sim x^2$) perturbation of the fractional Laplacian (the  fractional oscillator problem),
  a complete analytic solution  has been found   in the specialized  Cauchy oscillator  case  \cite{gar,lorinczi},
   by resorting to Fourier space methods.

 For more complicated  perturbations, and likewise for  a deceivingly  simple problem of the fractional Laplacian  in a bounded
  (spatial)  domain,  standard  Fourier techniques seem   to be  of a doubtful or   limited use, \cite{ZRK}.
   A fully-fledged  spatially nonlocal    definition of the   fractional  Laplacian   appears  to be  better suited to handle such problems,
     \cite{K}-\cite{ZG}.  See e.g.  also \cite{garolk} for  a  construction   of Cauchy  semigroups  which   arise
     from  various  perturbations  of the Cauchy operator by  bounded  or  locally bounded positive   functions (i.e.
     external potentials).

\subsection{$|\Delta |^{\mu /2}$  on   $R$.}

The  fractional Laplacian $-|\Delta |^{\mu /2}$,  $\mu \in (0,2)$ is
a pseudo-differential  (integral) operator and its  action   on a
function from  the   $L^2(R)$  domain  is defined as follows:
\begin{equation}
-  |\Delta |^{\mu /2} f(x)\, =  \int_R [f(x+y) - f(x) - {{y\, \nabla f(x)}
\over {1+y^2}}]\, \nu _{\mu }(dy),
\end{equation}
where  $\nu _{\mu }(dx)$ stands for the L\'{e}vy measure.
 This definition is commonly simplified   by  employing   the Cauchy
   principal value  of the  involved  integral (evaluated relative to the   singular points of integrands)
   \begin{equation}
 |\Delta |^{\mu /2} f(x)\, =\, - \int_R  [f(x+y) - f(x) ] \nu _{\mu }(dy)=
 -  {\frac{\Gamma (\mu +1) \sin(\pi \mu/2)}{\pi }} \int_R {\frac{f(z)- f(x)}{|z-x|^{1+\mu }}}\, \label{integral}
 dz
\end{equation}
Here, the L\'{e}vy measure  $d\nu_{\mu }$ has been made explicit and we point out a change of the  integration variable
 $y\rightarrow z+x$.
The Fourier representation of the the integral formula (\ref{integral})
takes the form
\begin{eqnarray}
|\Delta|^{\mu/2}f(x)=  - \frac{\Gamma(1+\mu)\sin\frac{\pi \mu}{2}}{\pi \sqrt{2\pi}}
\int_{-\infty}^{\infty} \tilde{f}(k)e^{-\imath kx}dk \int_{-\infty}^{\infty}\frac{(e^{-\imath ky}-1)dy}{|y|^{1+\mu}} .\label{intf1}
\end{eqnarray}
The integral over $dy$, presuming its existence (which is not the case for $\mu =1$)  can be directly evaluated
\begin{equation}\label{intf2}
 \int_{-\infty}^{\infty}\frac{(e^{-\imath ky}-1)dy}{|y|^{1+\mu}}=
 2|k|^\mu\Gamma(-\mu)\cos\frac{\pi \mu}{2}.
\end{equation}
Since Eq. \eqref{intf1} is  undoubtedly valid for
  all $\mu \in (0,1)\cup (1,2)$,  we can substitute  back an outcome of (\ref{intf2})  and  employ  an identity
$ \Gamma(1+\mu)\Gamma(-\mu)=- \pi /\sin (\pi \mu )$  (remember  that the function  $\Gamma (-\mu )$ has simple  poles at $0, -1, -2$).
 Accordingly,  two potentially divergent entries   compensate   each other  and the limit  $\mu \rightarrow 1$ is now legitimate. Thus
 $ |\Delta|^{\mu/2}f(x)=
   \frac{1}{\sqrt{2\pi}}\int_{-\infty}^{\infty} |k|^\mu \tilde{f}(k)
  e^{-\imath kx}dk$  for all $\mu \in (0,2)$,  and  $|k|^{\mu }$ is a Fourier multiplier  of $|\Delta|^{\mu/2}$  on $R$,  as anticipated,
   see \cite{getoor} and  \cite{stein}.

The fractional Laplacian  $|\Delta |^{\mu /2}$ extends to a self-adjoint operator in $L^2(R)$  and  induces a
strongly continuous  contraction  semigroup $\exp(- t|\Delta |^{\mu /2})$ whose Fourier multiplier  equals  $\exp(-
t|k|^{\mu })$.

\subsection{Interlude:   $- \Delta $ in $D \subset R$.}

The  Hamiltonian-type expression $H=-\Delta + V$,  with $V(x)=0$ for $x\in D=(-1,1)\subset R$,
   is an  encoding    of the Laplacian  with the the  Dirichlet  boundary conditions  (so-called  zero exterior condition on $R\setminus D$) imposed on  $L^2(R)$
  functions $f(x)$  in  the domain of $H$:   $f(x) =0$ for $|x| \geq 1$. The problem is that
   so defined  operator $H$,  if   restricted  to a  domain containing solely  functions $f\in L^2(R)$  with a support in D, is
    not   a self-adjoint operator in $L^2(R)$   (to this end  we need to admit   $C_0^{\infty }(R)$ as a proper domain).

 We recall that $C_0^{\infty }(R)$  comprises  infinitely differentiable functions  that are compactly supported in $R$.
The notation  $C_0^{\infty }(D)$  refers to a definite choice of the support to be $D\subset R$.
The differential operator  $-\Delta $ when acting in $C_0^{\infty }(D)$   (we keep  $D=(-1,1)$) defines a  symmetric operator
   in $L^2(D)$.  The problem of self-adjoint extensions in this case is a classic, c.f. \cite{GK}.
    One of them,  with a domain $D(H)=\{ AC^2[-1,1], f(-1)=0=f(1 )\}$  corresponds  to a standard (quantum mechanical)
    infinite well problem; the  $AC^2$ notation  refers to an absolute
     continuity of $f$ which gives meaning  to the second derivative of $f$  (of relevance at the boundary points of $D$).

     The spectral solution  gives rise to  the    $L^2(D)$ orthonormal
eigenbasis,  composed of real  functions $f_n(x)$, $n=1,2,...$ such
that $f(x)=0$ for $|x|\geq 1$.  More explicitly: $f_n(x)= cos(n\pi x/2)$ for $n$
odd and $sin(n\pi x/2)$ for $n$ even, while respective eigenvalues read $E_n = (n\pi /2)^2$.
It is clear that any $f \in L^2(D)$, in the domain of the
infinite well Hamiltonian,   may be represented as $f(x) =
\sum_{n=1}^{\infty } c_n f_n(x)$, e.g. in the form of   (trigonometric)  Fourier series.

At this point it is useful to mention that, quite independently of the self-adjointness issue,
 in $L^2(D)$ we have two inequivalent ways of making  the Fourier analysis.
If  $R\backslash D$  is neglected and   $L^2(D)$   is considered  on its own  (without any reference to  $L^2(R)$),
 then we can employ the previously mentioned  Fourier series (e.g. the infinite well trigonometric  eigenbasis).

We shall pay more attention to the alternative approach.
Namely, if $L^2(D)$ is considered as a subspace of $L^2(R)$, then  for any  $0\neq f\in L^2(D)$
 we know that $ \tilde{f} \in  L^2(R)$ but $\tilde{f}$ no longer belongs to $L^2(D)$.

In fact, for  any  $f\in C_0^{\infty }(R)$ its Fourier transform $\tilde{f}$  is an entire function
(analytic on the whole complex plane). It does not vanish anywhere in $R$, except at infinity. If $\tilde{f}$ would vanish
on $R\setminus D$, then necessarily it would  be vanishing on  $D$ as well.

Since $f$ is infinitely differentiable it follows that $\tilde{f}$ goes  to zero along the real
axis faster than  the inverse of any polynomial.
These properties hold true irrespective of the choice of the  compact  supporting interval
 $D\subset R$.  Accordingly,  ${\cal{F}}L^2(D)  \subset L^2(R)$,  but ${\cal{F}}L^2(D)\bigcap L^2(D) = \emptyset $.

 Under the infinite well  conditions,   $|k|^2$  still retains  some of the   Fourier multiplier (of $-\Delta $) features.
 This view is supported by an approximation of the infinite well problem by a sequence  of  deepening finite wells, \cite{GK}.
  The convergence can be quantified in the $L^2(R)$ norm.

Useful   examples  worked out  in  \cite{cohen,robinett}, at the first glance,   indicate an undoubtful relevance
  of  $k$ and $k^2$ multipliers. However,   an  the emergence  of
 technical  problems becomes conspicuous,  if the multiplier property is to be elevated to
eigenfunctions which are not $C_0^{\infty}(D)$.

 In fact, for the   infinite well    eigenfunctions $f_n(x)$  ($L^2(D)$-normalized), their
 $L^2(R)$ Fourier transforms $\tilde{f}_n(k)=(2\pi )^{-1/2} \int_D f_n(x)\exp(-ikx)\, dx$  can be directly  evaluated.
 We have $(f,f)_{L^2(D)}= (\tilde{f},\tilde{f})_{L^2(R)}$ and there holds  $\int_R |\tilde{f}_n(k)|^2\,  k^2\, dk =
 (n\pi /2)^2 = (f_n,(-\Delta f_n))$. Likewise  $(f_n,(-i\nabla f_n))=  \int_R |\tilde{f}_n(k)|^2  \, k \, dk=0 $
   for all $n\geq 1$.

One should be aware that the existence of  mean values  of Fourier multipliers  does not mean that we
  can  execute  an inverse Fourier transform  of  e.g.  the function $g(k)= |k|^2 \tilde{f}$  and  retrieve
    $-\Delta f(x)$ as an image-function in $L^2(D)$.
   Actually, the inverse Fourier  transform does not exist in this case, unless we adopt a weaker definition \cite{stein}.
     We note that  $-\Delta _Df(x)$,   at the  boundaries
   $\pm 1$ of D,  needs to  be interpreted as a generalized function (distribution), and has a meaning only if smoothed out
    by a suitable  test function.
     Incidentally, the eigenfunctions   $f_n$ appear   to play such a  smoothing  role,  while evaluating mean values.

\subsection{Cauchy operator  in $D\subset R $:  from $L^2(R)$ to $L^2(D)$.}

The previously indicated   jeopardies related to the Fourier multiplier definition (in case of  $\mu =2$) surely
   extend to fractional Laplacians.   Therefore, in the presence of  spatial  restrictions upon domains
  of  nonlocal operators,  we choose to investigate their  properties  directly  in configuration space   with no recourse to Fourier
   transforms (and multipliers).

The  action of the Cauchy operator  on $C_0^{\infty }(R)$ functions (differentiable, with continuous derivatives of all
 orders and compactly supported) is given by a specialized version of  Eq. (\ref{integral}):
\be
 |\Delta |^{1/2} f(x) = \frac{1}{\pi}\,   \int_R \frac{f(x)- f(x+z)}{z^2}dz =  \frac{1}{\pi}\,
  \int_R  \frac{f(x)-f(z)}{|z-x|^2}dz,\quad x \in R, \label{cauchy}
\ee
and   clearly  has a  Fourier representation with the multiplier  $ |k|\, \tilde{f}(k)$.
 We note in passing that  so defined   $(-\Delta )^{1/2}$
 extends to an unbounded self-adjoint operator in $L^2(R)$.

The Cauchy operator  $|\Delta |^{1/2}$ if restricted to a domain comprising solely  $L^2(R)$ functions
with a support in $D$ and vanishing on $R\backslash D$   is not a self-adjoint  operator in $L^2(R)$.
 However, if we  consider  the action of  $|\Delta |^{1/2}$   on test functions    $f\in C_0^{\infty }(D)$,
  then  the restriction $|\Delta |^{1/2}_Df$ of
$|\Delta |^{1/2}$  to  $D$ is interpreted as the Cauchy operator with the  zero (Dirichlet)  exterior condition
on $R\backslash D$   and is known to extend to a self-adjoint operator in $L^2(D)$, \cite{K}.

The passage from $C_0^{\infty }(R)$ to $C_0^{\infty }(D)$  ultimately  amounts to disregarding   any  $R\backslash D$
 contribution implicit in the formal definition
(\ref{cauchy})  and makes the usage  the   Fourier multiplier  representation   either clumsy or  redundant.

  Let us consider the  $D$ versus $R\backslash D$ interplay in more detail, by  considering  the action of $|\Delta |^{1/2}$
  on   these  $C_0^{\infty }(R)$  functions which are actually supported in $D$. The major problem we  wish to   address is an
  explicit spatial form of the eigenvalue problem for  $|\Delta |^{1/2}_D$,  interpreted
as $|\Delta |^{1/2}_D\, f = E\, f$ where $E\in R^+$ is an eigenvalue and $f\in L^2(D)$.  No closed analytic solutions are here available
 and various approximate methods were invented to  optimize approximations of "true"  eigenvalues and
 shapes of respective "true" eigenfunctions, \cite{K}-\cite{ZG}.

  Each known to date  approximate eigenfunction   $\psi (x)$, \cite{K,KKMS,ZG},  by construction
 belongs to the domain of $|\Delta |^{1/2}_D$ and obeys exterior Dirichlet boundary data. However, generically
 $|\Delta |^{1/2}_D\psi   \in L^2(D)$  no longer respects those data, while such a property is definitely expected from an
  acceptable $L^2(D)$   approximation of   the right-hand -side of the eigenvalue formula $|\Delta |^{1/2}_D\, f = E\, f$.

  This problem is typically bypassed in the mathematical literature, where one considers the spectral problem of finding
an eigenfunction (or its optimal approximation)  in a weaker form.
This (somewhat relaxed) approach   to the spectral problem stems
from the   adopted  definition of the action of  $|\Delta |_D$ in
its $C_0^{\infty }(D)$ domain. Namely, one demands that   for $f$ in
the $L^2(D)$ domain of $|\Delta |_D$ there exists  $|\Delta |_D\, f
\in L^2(D)$ such that
 $(g, |\Delta |_D\, f)_{L^2(D)} = (|\Delta |\, g,f)_{L^2(D)}$   holds true for  $g\in C_0^{\infty }(D)$, \cite{Kbis,KKM}.

Let us tentatively consider  the action of  $|\Delta |^{1/2}$  on $C_0^{\infty }(R)$ functions $\psi (x)$,  supported
in $D=(-1,1)$. Accordingly, for all $x\in (-1,1)$ we have:
\be
  A_D \psi (x)= \frac{2}{\pi}\frac{\psi(x)}{1-x^2} + \frac{1}{\pi}\int_{-1-x}^{1-x}\frac{\psi(x)-\psi(x+y)}{y^2}dy. \label{stu2}
\ee
The integral in \eqref{stu2} should be understood as the  Cauchy principal value evaluated with respect to $0$, according to
$\int_{-1-x}^{1-x}=\lim_{\varepsilon \to 0} \left[\int_{-1-x}^{-\varepsilon} +\int_{-\varepsilon}^{1-x}\right]$.

Let us  change the integration variable  $y=t-x$ in Eq.
\eqref{stu2}. We have: \be \label{int}
 A_D \psi (x)
=\frac{2}{\pi}\frac{\psi(x)}{1-x^2}+\frac{1}{\pi}\int_{-1}^{1}\frac{\psi(x)-\psi(t)}{(t-x)^2}dt
\ee where   the $R\setminus D$ and $D$ contributions  are now
clearly isolated.   The Cauchy principal value of the integral in Eq. (\ref{int})  is no longer evaluated with respect to $0$,
 but with respect to $x$.
 The integral expression in Eq. (\ref{int}) which is now restricted
 to $t\in (-1,1)$ while $x\in (-1,1)$ by definition,  is a special (Cauchy) case of  so-called  regional fractional  Laplacian
 for $D$, \cite{dyda,dyda1,guan}.

\section{Trial  analytic forms  of approximate  eigenfunctions.}

\subsection{The ground state function.}

In the present paper, we shall employ  Eq.  (\ref{stu2})  as the definition of    the   Cauchy operator in action on functions  $\psi  \in {\cal{D}}(D)$,  in the spatially restricted domain $D=(-1,1)$.
 We no longer require  $\psi $  to belong to   $C_0^{\infty }(D)$,  we need  however  $\psi (x)$ to be  infinitely differentiable
  in $D$  and identically vanish for $|x|\geq 1$.
 Let us address the eigenvalue problem   $A_D\psi(x)=E\psi(x)$, whose approximate solution will be sought for subsequently.

The term $\frac{2}{\pi}\frac{1}{1-x^2}$  in Eq. (\ref{stu2}) is a sum of the geometric series $\frac{2}{\pi}(1+x^2+x^4+\ldots)$, $x\in(-1,1)$.    Therefore it seems natural to assume that
a solution $\psi (x)$ of $A_D\psi(x)=E\psi(x)$  might be represented in the form of the  power series $\psi(x)=\sum\limits_{n=0}^\infty c_{n}x^n$ as well,   with a proviso that $\psi (x)$ needs to be identically $0$
at the boundaries $\pm 1$ of $D$. Thus  we   allow  $\psi (x)$ to  have a  domain   $\bar{D}= [-1,1]$.

The ground state is an even concave  function, \cite{BK,BKM}, therefore  we actually  expect  that
$\psi(x)=\sum\limits_{n=0}^\infty c_{2n}x^{2n}$.  Since $\psi (\pm 1)=0$,
there holds $c_0+c_2+\ldots = 0$.   Inserting  $\psi$  to the eigenvalue formula and keeping $x\in (-1,1)$
we get
\be \label{series}
\frac{1}{\pi}\int\limits_{-x-1}^{-x+1}\frac{-\psi'(x)z-\psi''(x)\frac{z^2}{2!}-\psi'''(x)\frac{z^3}{3!}-\ldots}{z^2}dz + \frac{2}{\pi}\frac{1}{1-x^2}\psi(x)=E\psi(x).
\ee
presuming that the integral (Cauchy principal value)  and series summation can be interchanged, next setting $x=0$, we arrive at the series expansion
 which defines the ground state eigenvalue $E$, given $\psi (x)$:
\be
-\frac{2}{\pi}\left(\frac{c_2}{1}+\frac{c_4}{3}+\frac{c_6}{5}+\ldots\right)+\frac{2}{\pi}c_0=E c_0.
\ee
With $c_0 \neq 0$, we have
\be \label{value}
E=\frac{2}{\pi}\left[1-\frac{1}{c_0}\left(\frac{c_2}{1}+\frac{c_4}{3}+\frac{c_6}{5}+\ldots\right)\right].
\ee
The  series converge, which follows  (via the D'Alembert criterion)   from  the convergence of
$\sum\limits_{n=0}^\infty c_{2n}x^{2n} = \psi (x)$ for all $x\in [-1,1]$.

\begin{figure}[h]
\begin{center}
\centering
\includegraphics[width=70mm,height=70mm]{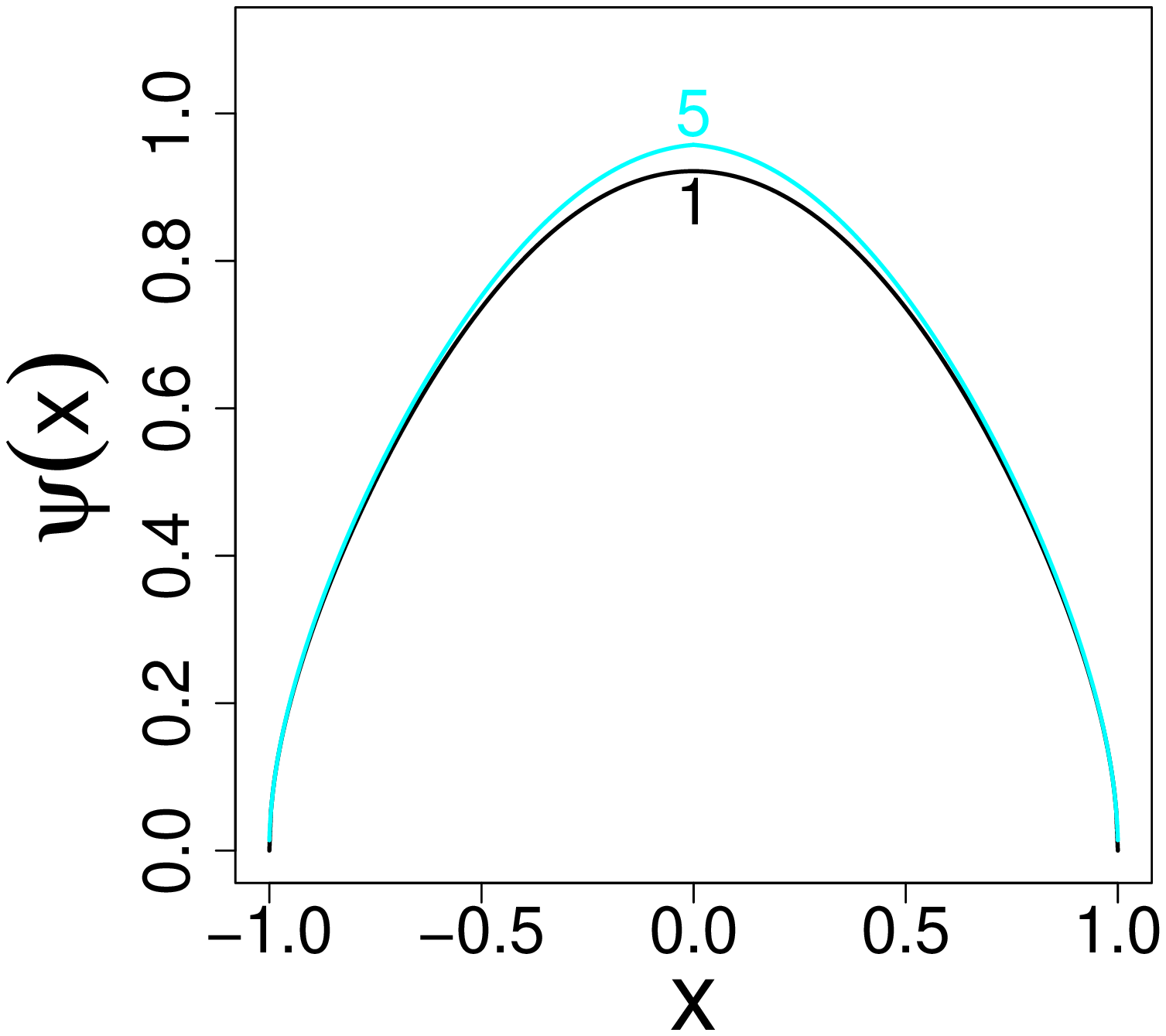}
\includegraphics[width=70mm,height=70mm]{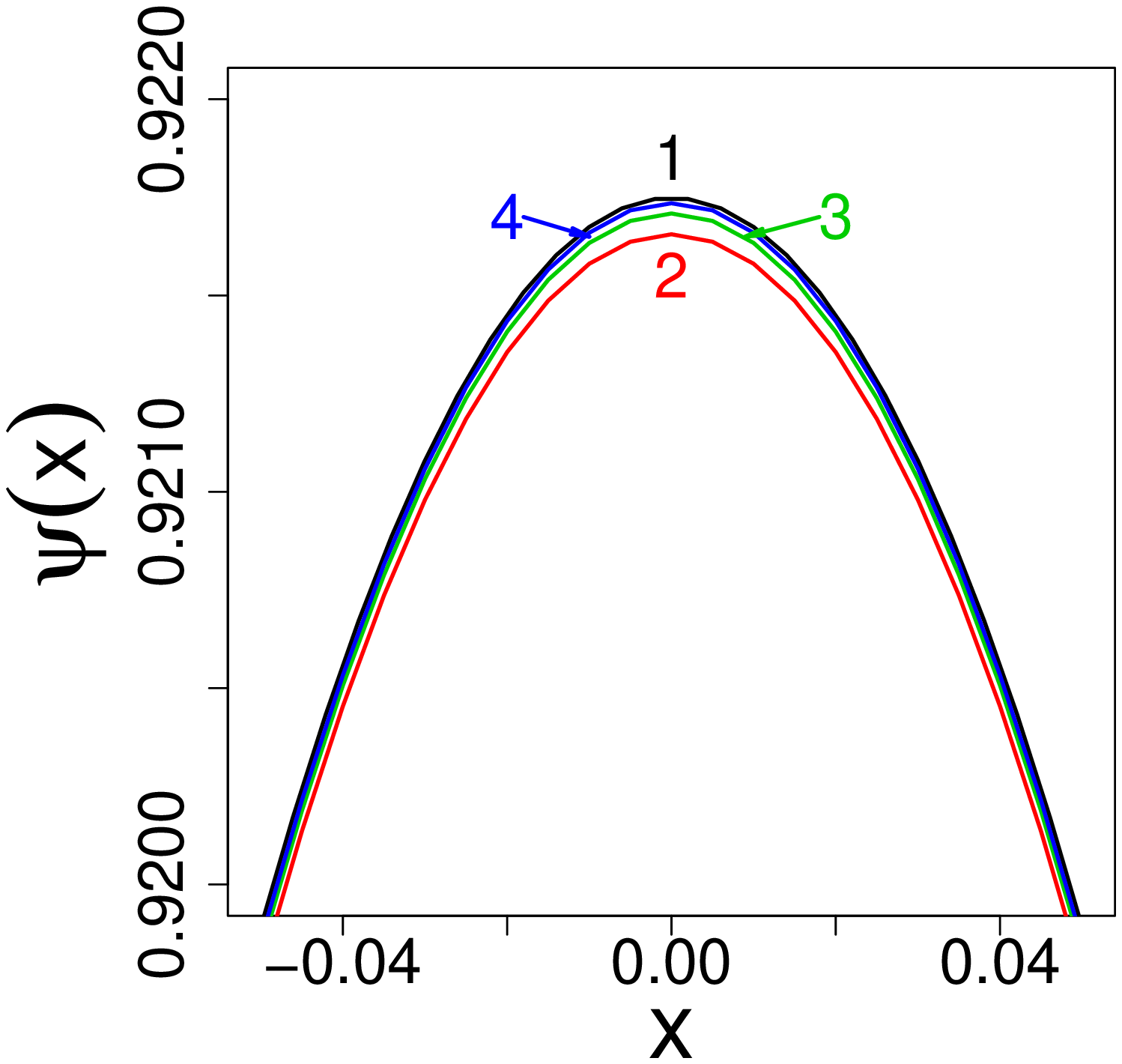}
\caption{Approximate ground states for  Cauchy wells. Numbers refer to:  1- infinite well  $\psi(x)$ of  Eq.(\ref{ground}); 2, 3, 4 - finite wells with  depths  $V_0=5000, 10000, 20000$ respectively, \cite{ZG};
 5 - infinite well proposal of \cite{K}. In the right  panel, the  curve $5$ is out of the  frame.}
\end{center}
\end{figure}

By independent arguments, we know that the   ground state function should  be close  (loosely speaking)  to  the $\cos  (x) $   while away  from the boundaries of $D$
and $(1-x^2)^{1/2}$ in the close  vicinity of the boundaries, \cite{ZRK,K,ZG}.  Let us consider the trial  approximation of  ground state function, given in an analytic form:
\be \label{ground}
\psi(x)=C\sqrt{(1-x^2)\cos(\alpha x)},
\ee
where the coefficient $\alpha $  has been  adjusted to differ slightly from  ${\frac{\pi}{2}}-{\frac{\pi}{8}}$, known  to be the leading term in the asymptotic eigenvalue formula for the infinite Cauchy well, \cite{K}:
\be
\alpha=\frac{1443}{4096}\pi=\left(\frac{\pi}{2}-\frac{\pi}{8}\right)-\frac{\pi}{64}-\frac{\pi}{256}-\frac{\pi}{512}-\frac{\pi}{1024}-\frac{\pi}{4096},
\ee
$C=0.921749$  being the $L^2(D)$ normalization constant.

In the recent paper \cite{ZG} we have  introduced  an algorithm for evaluating approximate eigenfunctions of  {\it finite} Cauchy wells of arbitrary depth.  The idea was to implement as close approximation of the
infinite well  spectral properties in terms of those for very deep finite wells.  In Fig. 1 we have depicted  approximate ground state functions for finite wells of  depths $V_0=5000,10000,20000$,
an approximate ground state for the infinite Cauchy  well proposed in Ref. \cite{K}  and  our approximate formula $\psi (x)$  of  Eq. (\ref{ground})  for the infinite well.

In the left panel, curves for  $V_0=5000,10000,20000$  and that for $\psi (x)$  are  graphically  indistinguishable in the adopted scales, while the proposal of \cite{K}  is conspicuously different.
In the right panel a vicinity of the maximum has been enlarged  and the proposal of \cite{K}, in the adopted scale,  is out of frame.

Let us expand the approximate ground state $\psi$  of Eq. (\ref{ground}) into   power series   $\psi(x)=\sum\limits_{n=0}^{\infty}c_{2n}x^{2n}$
 with $x \in [-1,1]$. The expansion  coefficients can be explicitly identified and
we reproduce  numerical values for first few of them:
\be
\begin{array}{llllll}
&c_0=0.921749,\quad &c_2=-0.743145,\quad &c_4=0.011510,\quad &c_6=-0.020710, \quad &c_8=-0.015567,\\
&c_{10}=-0.012318,\quad &c_{12}=0.009969,\quad &c_{14}=-0.008234,\quad &c_{16}=-0.006922,\quad &c_{18}=-0.005910.
\end{array}
\ee
Although   $\psi(x)$  is not a "true" eigenfunction but an approximation of the ground state, by  employing  merely  first $10$  expansion coefficients
 in  the  series expansion  Eq. (\ref{value}), we obtain  the very rough outcome  $E=1.15318$.
  The ground  state eigenvalues  have been approximated by other (more accurate)
  methods and, up to four decimal digits we have e.g.:  $1.1577$ according to \cite{K}, $1.1573$  according to \cite{ZG}.

We point out that a convergence of the series  (\ref{value})  is  very slow.   To  get more accurate approximation of the eigenvalue
associated with the approximate ground state  $\psi (x)$   (\ref{ground})
we need to account for much larger number of expansion coefficients  $c_{2n}$.

\subsection{Analysis of $A_D\psi (x)$.}

At the moment we are not that much interested in  producing high  accuracy approximate formulas (this issue will be addressed  subsequently
in the present paper).  The analytic expression  Eq. (\ref{ground}) for $\psi (x)$  is extremely useful for another purpose.

Namely, we can make explicit the action od $A_D$ upon   functions with definite geometric shapes  and analyze not only  how much $A_D\psi (x)$ deviates from $\psi (x)$ and ultimately from  $E\, \psi (x)$
(with a properly adjusted value $E$), but also the boundary behavior of those functions. See e.g. \cite{dyda,dyda1} for  some hints in this connection.

Given $\psi (x)=  C\sqrt{(1-x^2)\cos(\alpha x)}$, $|x| \leq  1$,  we would like to know whether  the Dirichlet boundary data (e.g. vanishing of a function for $|x|\geq 1$) are respected by   $A_D\, \psi (x)$.
To this end let us consider
\be
\frac{1}{\pi}  \int\limits_{-x-1}^{-x+1}\frac{\psi(x)-\psi(x+z)}{z^2}dz=  \frac{1}{\pi} \int\limits_{-x-1}^{-x+1}\frac{\psi(x)   - C  \sqrt{1-(x+z)^2} (1-\gamma_2(x+z)^2-\gamma_4(x+z)^4-\ldots)}{z^2}dz,
\ee
where we have expanded  $\sqrt{\cos\alpha(x+z)}$ into power series whose coefficients are denoted $\gamma_{2n}$,  ($\gamma _0=1$).

First few coefficients are given explicitly,
\begin{eqnarray}
\gamma_2=\frac{\alpha^2}{4},\quad \gamma_4=\frac{\alpha^4}{96}, \quad \gamma_6=\frac{19\alpha^6}{5760},\quad
 \gamma_8=\frac{559\alpha^8}{645120}.
\end{eqnarray}
We integrate each term of the series separately.
The  integral corresponding to $\gamma _0=1$   can be rewritten as follows  (see e.g. \cite{GR})
\be
{\frac{\psi (x)}{\pi }} \,  \int\limits_{-x-1}^{-x+1}\frac{1 - \sqrt{p + qz +  rz^2}}{z^2}dz,
\ee
where   $p=1/\cos(\alpha x)$, $q=-2x/(1-x^2)\cos(\alpha x)$, $r=-1/(1-x^2)\cos(\alpha x)$.   We evaluate the
integral in the sense of its  Cauchy principal value  (see e.g. \cite{GR}),  temporarily skipping the factor $\psi (x)/\pi $:
\be
   \int\limits_{-x-1}^{-x+1}\frac{1-\sqrt{p+qz+rz^2}}{z^2}dz=-\frac{2}{(1-x^2)}+\frac{\pi}{\sqrt{(1-x^2)\cos(\alpha x)}}.
\ee
We note that the first term   in the above (after restoring $\psi /\pi $) cancels its negative in the defining expression
(\ref{stu2}) for $A_D\psi (x)$.

Subsequent integrals can be evaluated analogously,  but with $\psi (x)$ fully incorporated  in the integrated expressions.
 We merely disregard (but keep in mind)  an omnipresent  coefficient  $C/\pi $  and denote  $a=1-x^2, b=-2x, c=-1$.
 With this proviso  other integrals follow:

\be
\gamma_2\int\limits_{-x-1}^{-x+1}\frac{\sqrt{a +bz +cz^2}(x+z)^2}{z^2}=\gamma_2 \int\limits_{-1}^1\frac{u^2\sqrt{1-u^2}}{(u-x)^2}du=\gamma_2 \frac{\pi(1-6x^2)}{2},
\ee
\be
\gamma_4\int\limits_{-x-1}^{-x+1}\frac{\sqrt{a +bz +cz^2}(x+z)^4}{z^2}= \gamma_4 \int\limits_{-1}^1\frac{u^4\sqrt{1-u^2}}{(u-x)^2}du= {\frac{\gamma_4}{8}} \pi(1+12x^2-40x^4),
\ee
\be
\gamma_6\int\limits_{-x-1}^{-x+1}\frac{\sqrt{a +bz +cz^2}(x+z)^6}{z^2}=\gamma_6 \int\limits_{-1}^1\frac{u^6\sqrt{1-u^2}}{(u-x)^2}du={\frac{\gamma_6}{16}}  \pi(1+6x^2+40x^4-112x^6).
\ee
Accordingly, after reintroducing the factor $C/\pi $ we arrive at the polynomial expansion of $A_D\psi $
\be
A_D\psi(x)=C\sum\limits_{n=0}^{\infty}\gamma_{2n}w_{2n}(x),
\ee
where coefficients  $\gamma_{2n}$ for $0\leq n\leq 4$  have been  explicitly identified  before  while $w_{2n}(x)$  are polynomials of degree $2n$, like e.g.
\be
w_0(x)=1,\quad w_2(x)=\frac{1-6x^2}{2},\quad w_4(x)=\frac{1+12x^2-40x^4}{8},\quad w_6(x)=\frac{1+6x^2+40x^4-112x^6}{16}.
\ee

\begin{figure}[h]
\begin{center}
\centering
\includegraphics[width=55mm,height=55mm]{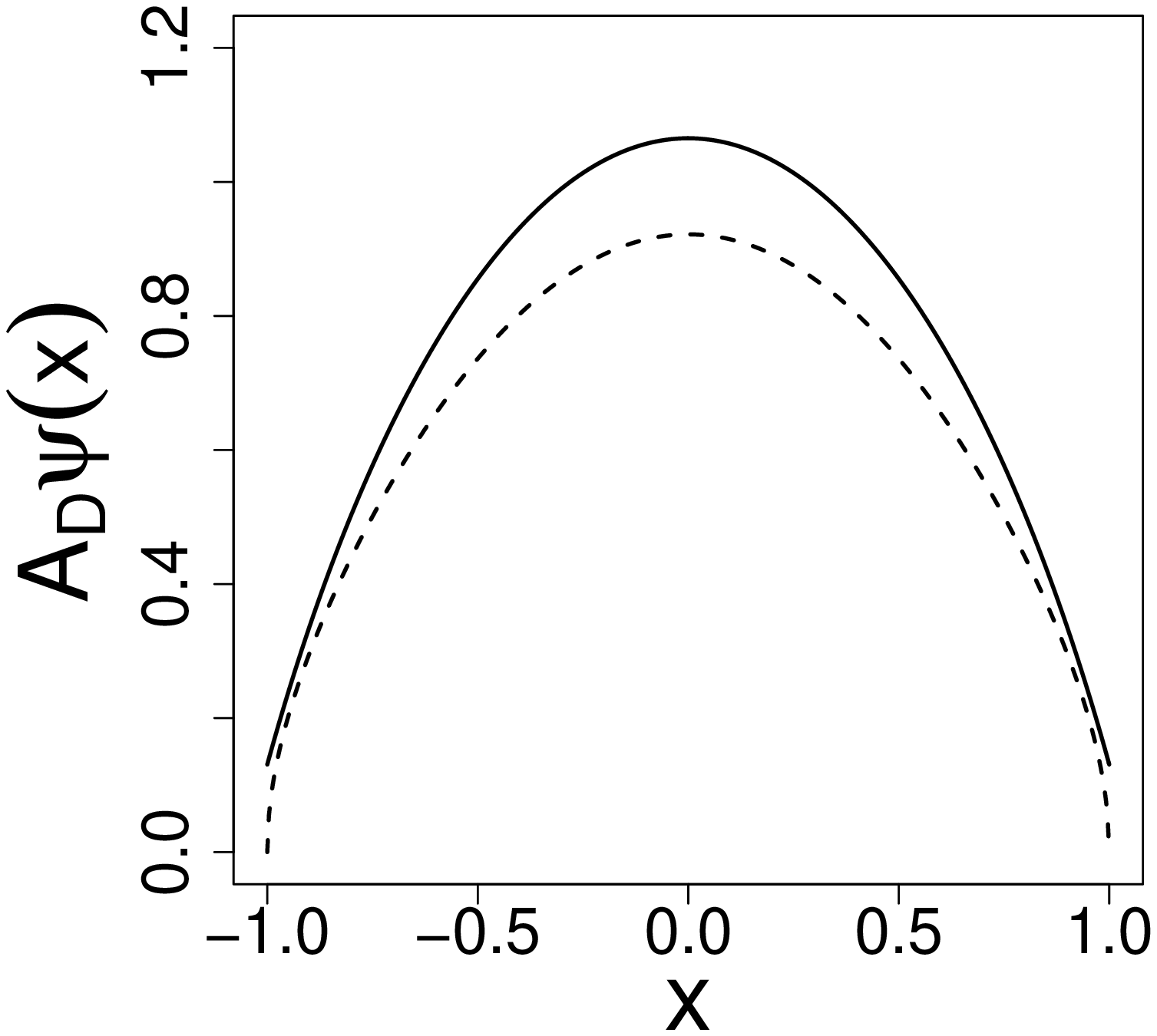}
\includegraphics[width=55mm,height=55mm]{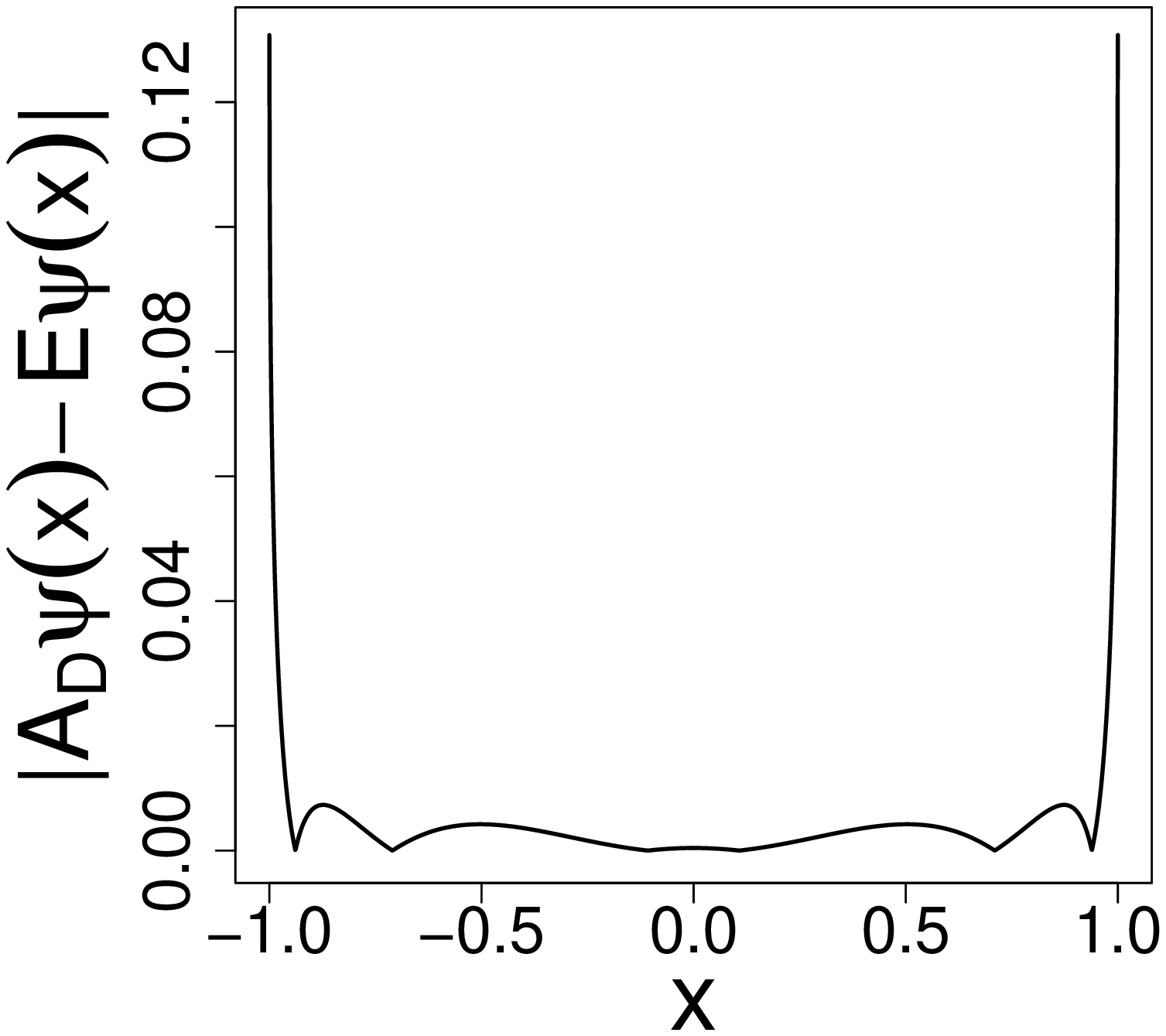}
\includegraphics[width=55mm,height=55mm]{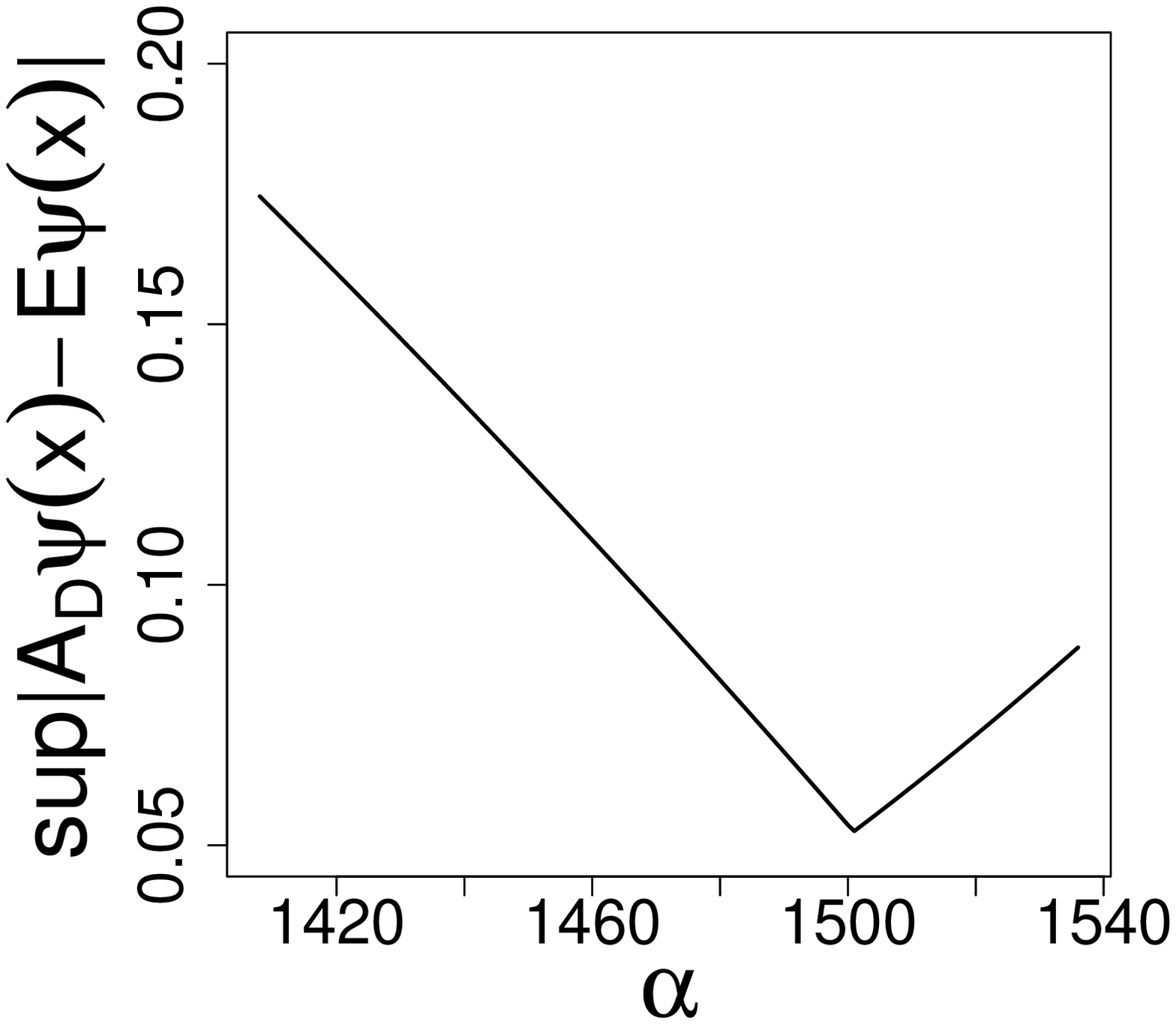}
\caption{Left panel: a comparison of   $\psi(x)=C\sqrt{(1-x^2)cos(\alpha x)}$ (dotted line) and $A_D\psi $ (solid line).
 Middle  panel: $|A_D \psi (x) - E\psi (x)|$ with   $E=1.156$. Right panel: supremum of $|A_D \, \psi  - E\, \psi(x)|(\alpha )$  for $E=1.156$.
The $\alpha $-axis is scaled in units $\pi/4096$. }
\end{center}
\end{figure}

The series  expansion of  $\sqrt{\cos(\alpha x)}$  converges very  fast for  $x\in[-1,1]$.  Accordingly,
by accounting for  only   first few expansion terms we get  quite  good approximation of $A_D\, \psi $. In Fig. 2, we have depicted both
  $\psi(x)=C\sqrt{(1-x^2)cos(\alpha x)}$
 (a dotted line) and the resultant $A_D\, \psi $ (solid line).

Since in Ref. \cite{ZG} we have numerically identified an approximate ground state eigenvalue to be close to
 $E=1.156$ (eventually correctable to $E=1.1573$, \cite{ZG}), let us employ the latter value instead of less
  accurate  $E=1.15318$ obtained in a rough reasoning presented  before.
    Then, we can   point-wise compare $\psi (x)$ against
   $A_D\psi (x)$ by  depicting a curve   $|A_D \psi(x)-E\psi(x)|, x\in D$, see e.g. Fig. 3.

The deviation of $A_D\psi (x)$ form $ E\, \psi (x)$ appears to be  rather small and effectively
 concentrates   in the vicinity of the boundaries   $x=\pm 1$.
Since we have $ \lim\limits_{x\to\pm 1} A_D\psi(x)=0.130753,\quad \lim\limits_{x\to\pm 1} E\psi(x)=0$,
there  holds $|A_D \psi(x)-E\psi(x)|\leqslant 0.130753,\, x\in D$ which is the best  point-wise  estimate ever obtained in the literature on the
(shape) subject, compare e.g. \cite{KKMS} (Lemma 1, formulas  8.9 and  8.10) and \cite{K}. We emphasize that the behavior of $|A_D \psi (x)  -  E \psi |$  is fairly robust with respect to the specific choice of $E$.
The dominant contribution comes to the upper bound  comes form the behavior of  $A_D \, \psi (x)$ at the boundaries od $D$.

The shape of the approximate ground state, as proposed in  \cite{KKMS,K},  while  away from the boundaries (i.e. around $x=0$)
 is significantly different from our  present finding and  from the  numerically-deduced behavior of eigenfunctions in finite but
  deep Cauchy wells, \cite{ZG}.

\begin{figure}[h]
\begin{center}
\centering
\includegraphics[width=70mm,height=70mm]{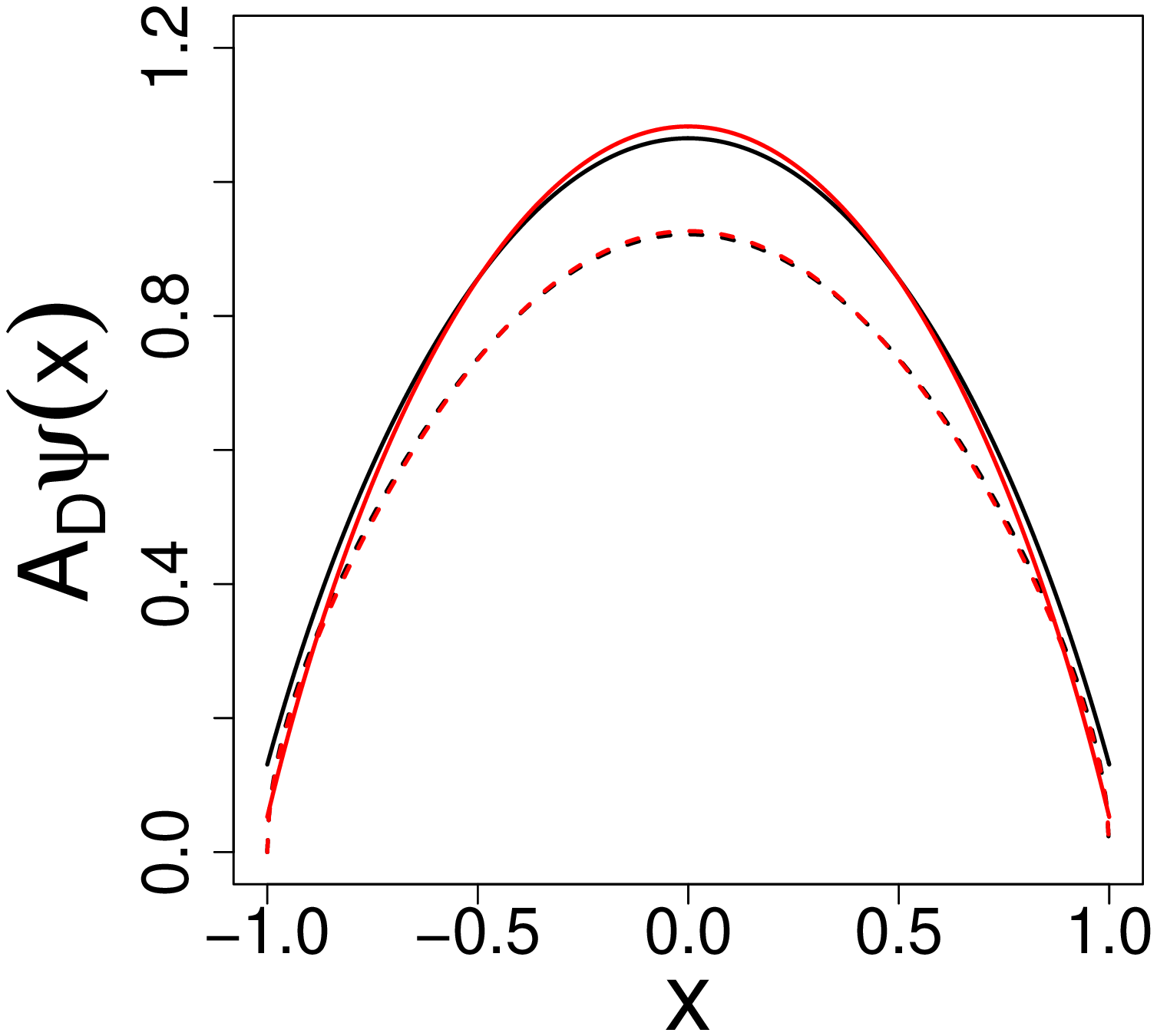}
\includegraphics[width=70mm,height=70mm]{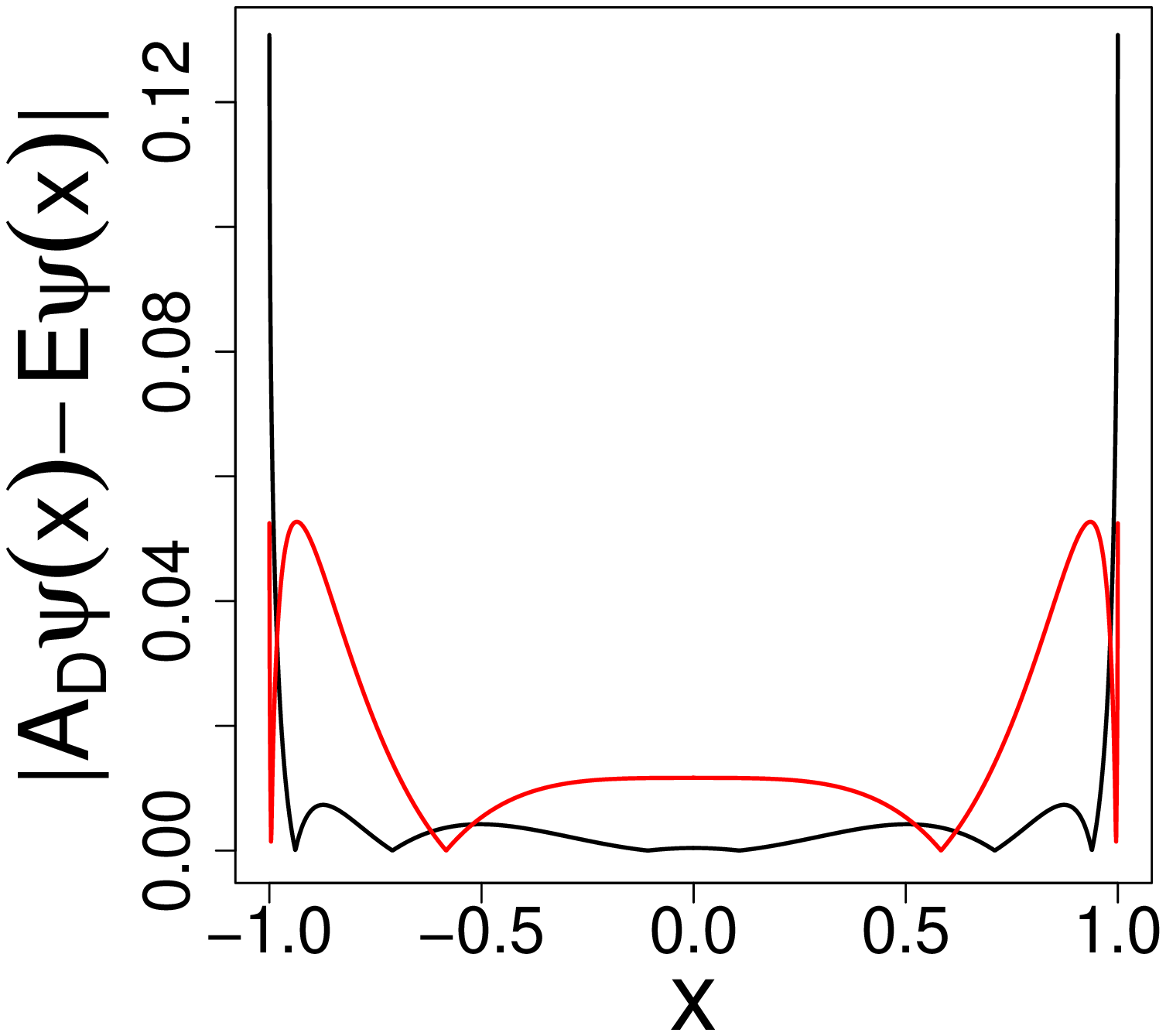}
\caption{Left panel: $\psi (x)$ is represented by dotted lines (black and red variants are practically indistinguishable),
$A_D\psi (x)$ for $\alpha=1443\pi/4096$  (solid black)  and  $\alpha=1501\pi/4096$ (solid red). Right panel:
$|A_D\psi (x) - E \psi (x)|$ for $E=1.156$  and previous $\alpha $s, respectively in black and red.}
\end{center}
\end{figure}

We note that in the  definition  (\ref{ground}) we can in principle vary  $\alpha $. Taking
a supremum of $|A_D\, \psi (x) - E\psi (x) |(\alpha )$ over $x\in D$  as a  criterion   for how close $A_D\psi $ is to $E\psi $,
we   realize that the optimal  $\alpha $ choice would be $\alpha=1501\pi/4096$, see e.g. at the location of the minimum in  Fig. 4.

To see better how the  choice of  $\alpha $ may affect the shape of $A_D\psi(x)$ and  $|A_D \psi(x)-E\psi(x)|$,
we display  the behavior of these functions comparatively  for  $\alpha=1443\pi/4096$ and $\alpha=1501\pi/4096$.
We note that for  $\alpha=1501\pi/4096$  we have  $|A_D\psi (x) - E \psi (x)|< 0.06$ which is much better  point-wise estimate
than  previously obtained  $0.13$ (for $\alpha=1443\pi/4096$). The price paid is slightly worse fitting away form the $\pm 1$ boundaries.

\subsection{First excited state.}

The ground state function, previously denoted $\psi (x)$, in fact should be labeled by $n=1$, hence denoted $\psi _1(x)$.
To avoid  notational  confusion, the first excited state ($n=2$) will be denoted $\psi _2(x)$.

\begin{figure}[h]
\begin{center}
\centering
\includegraphics[width=55mm,height=55mm]{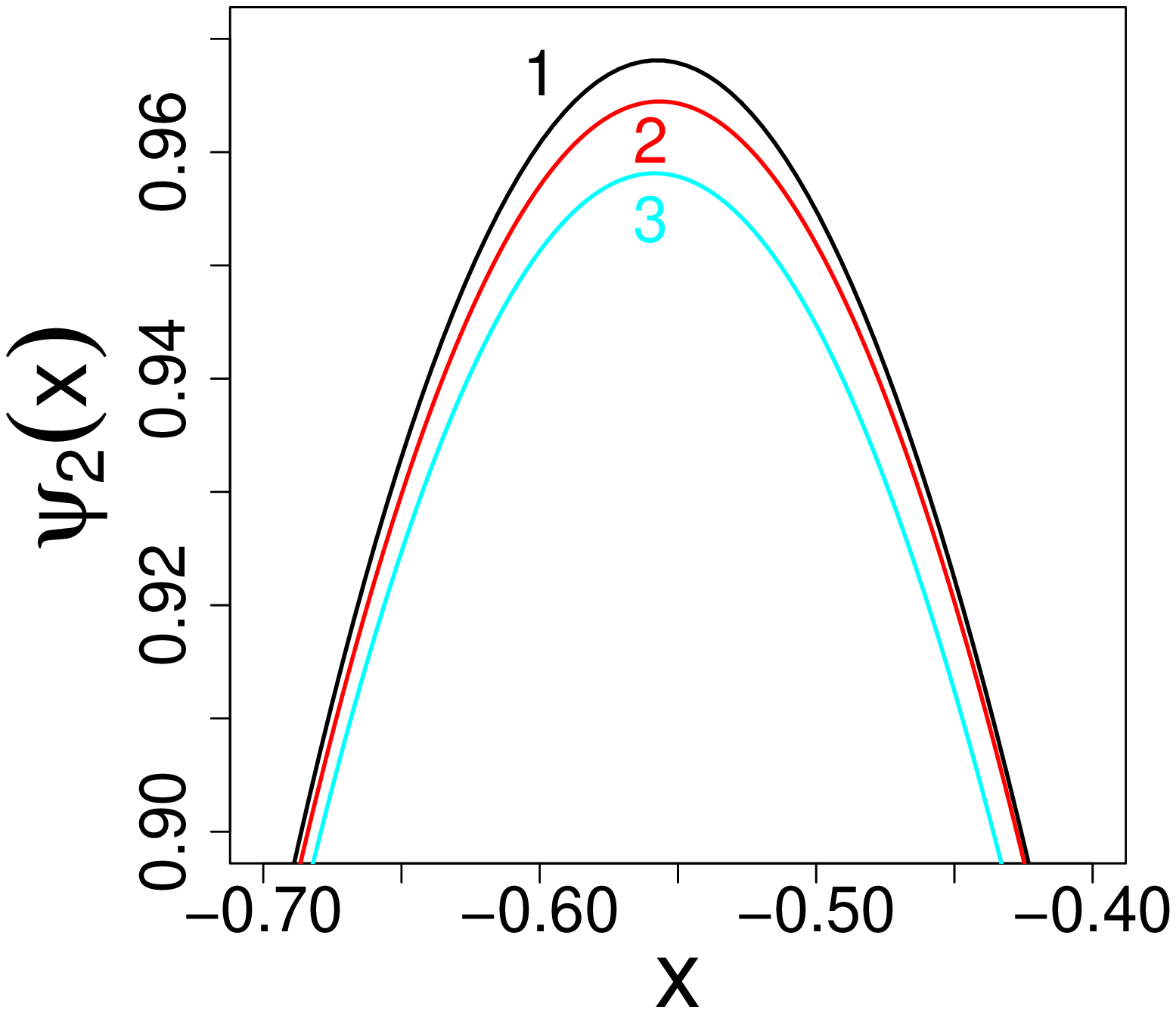}
\includegraphics[width=55mm,height=55mm]{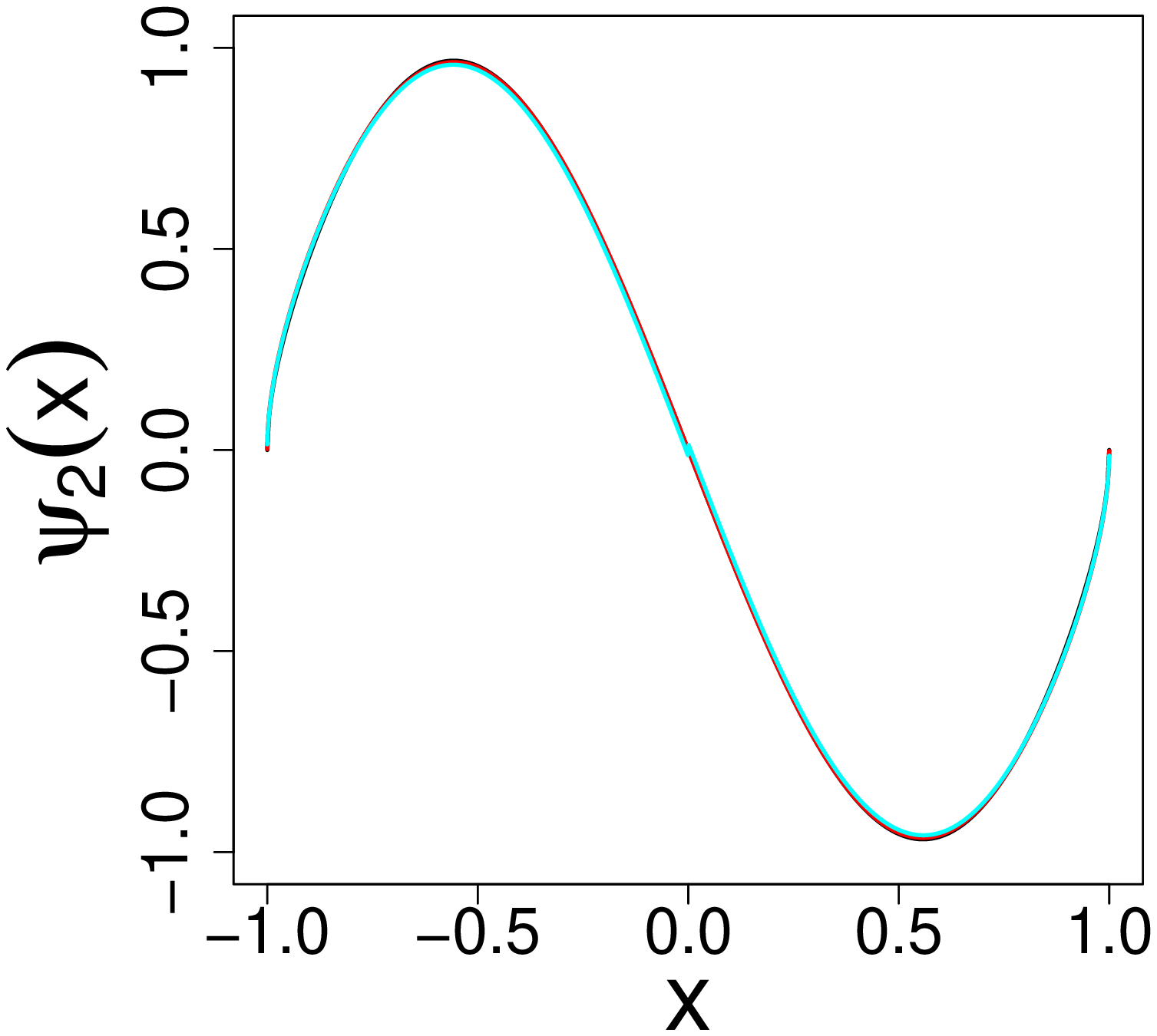}
\includegraphics[width=55mm,height=55mm]{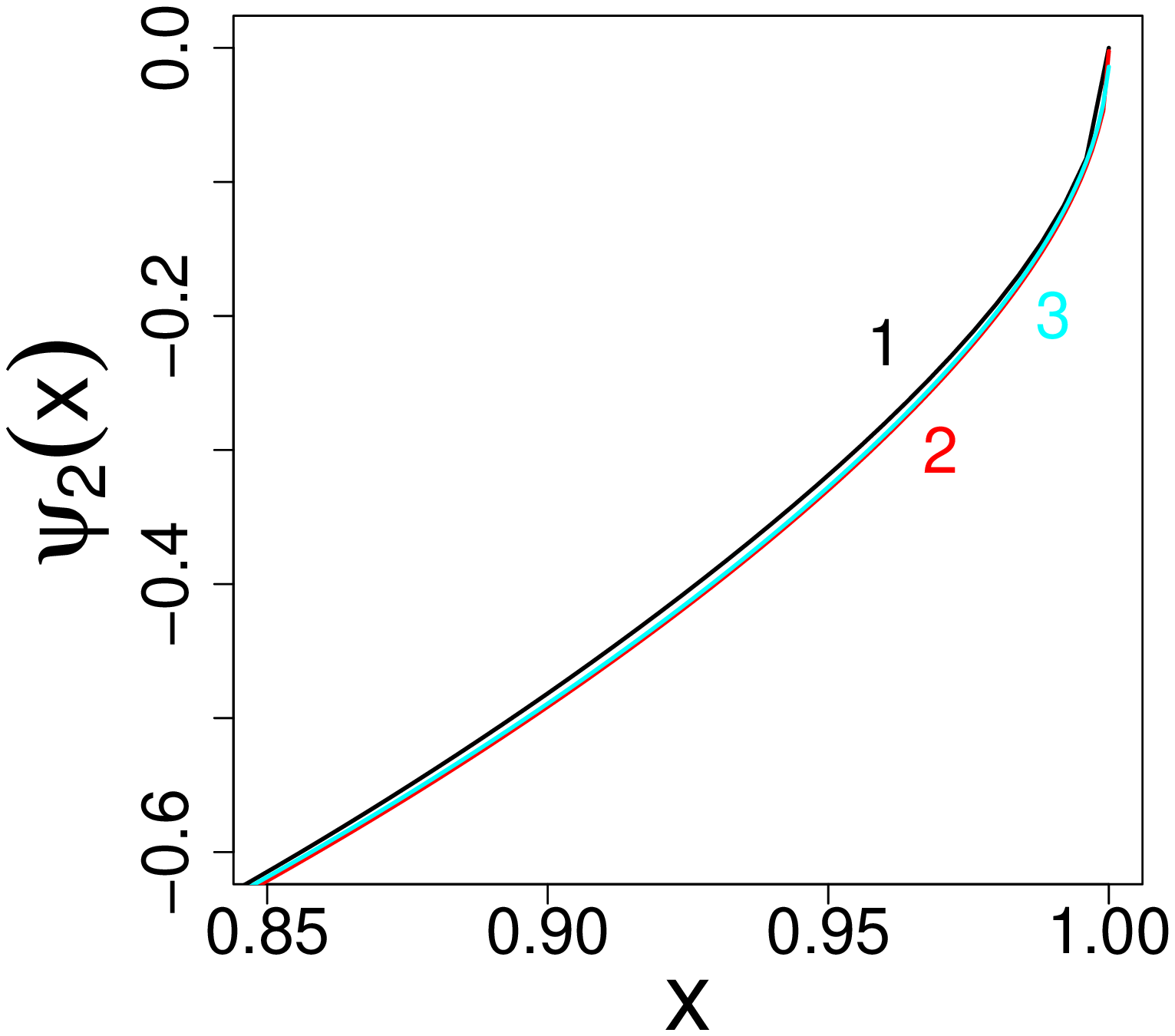}
\caption{Approximate expressions for the first excited state function. Numbers refer to:  1 -  $\psi_2(x)$ (\ref{ex}),
 2 - finite Cauchy well outcome, depth $V_0=5000$, \cite{ZG}, 3 - an approximation proposed  in  \cite{K}.
 Left panel: magnified vicinity of a maxium.  Right panel:  magnified vicinity of the boundary $x=1$.}
\end{center}
\end{figure}

We introduce a trial  analytic  approximation of a "true" first excited state in the form:
\be \label{ex}
\psi_2(x)=-C \sin(\beta x)\sqrt{(1-x^2)\cos(\beta x)},
\ee
where
\be
\beta=\frac{1760}{4096}\pi=\frac{\pi}{2}-\frac{\pi}{16}-\frac{\pi}{128},
\ee
and  $C=1.99693$ is a normalization constant.
 The minus   sign  is basically irrelevant, but is introduced   for graphical   comparison purposes.
 The parameter $\beta$  fitting comes from our data for deep but finite Cauchy wells, \cite{ZG}.
In Fig. 6 we display various approximate formulas for the "true" excited eigenfunction, comparing our analytic guess with  approximate
curves proposed in \cite{K} and \cite{ZG}.

\subsection{Analysis of  $A_D \, \psi _2(x)$.}

Main arguments follow these of Section II.A and II.B.  The first excited  function $\psi _2(x)$ is odd, hence its power series
expansion $\psi_2(x)=\sum\limits_{n=0}^{\infty}c_{2n+1}x^{2n+1}$ contains only odd-labeled coefficients $c_n$.
Like in section IIA, we deduce  the eigenvalue $E_2$
\be
E_2=\frac{4}{\pi}\left[1-\frac{1}{c_1}\left(\frac{c_3}{1}+\frac{c_5}{3}+\frac{c_7}{5}+\ldots\right)\right].
\ee
Since we know explicitly the numerical values of coefficients $c_2$, let us list few of them
\be
\begin{array}{llllll} \label{E}
&c_1=-2.695662,\quad &c_3=3.394555,\quad &c_5=-1.040718,\quad &c_7=0.152499, \quad &c_9=0.000755,\\
&c_{11}=0.010042,\quad &c_{13}=0.008479,\quad &c_{15}=0.007567 ,\quad &c_{17}=0.006774,\quad &c_{19}=0.006095,
\end{array}
\ee
and next insert them directly to the expansion (\ref{E}), while disregarding the remainder of the series.
 An approximate eigenvalue reads  $E_2=2.72874$, to be compared (even though the result is  very rough)
 with  the deep finite  Cauchy well prediction $E_2= 2.7534$ of Ref. \cite{ZG} (eventually correctable by $0.0013$  to
 $E_2= 2.7547$)  and $E_2= 2.7547$ of Ref. \cite{K}.
The series (\ref{E}) converge slowly, therefore much larger number of coefficients $c_n$ need to be accounted for,
to make reliable  the numerical value for $E_2$.

To deduce  $A_D\psi _2(x)$, let us  first analyze the  integral
\be
\int\limits_{-x-1}^{-x+1}\frac{\psi_2(x)-\psi_2(x+z)}{z^2}dz
\ee
 alone, see e.g. (\ref{stu2}) for comparison. For clarity, in the power series expansion
\be
\sin(\beta x)\sqrt{\cos(\beta x)}=\sum\limits_{n=0}^{\infty}\gamma_{2n+1}x^{2n+1},\quad x\in D,
\ee
we enlist few  $\gamma_{2n+1}$ in their explicit numerical form:
\be
\gamma_1=\beta,\quad \gamma_3=-\frac{5\beta^3}{12}, \quad \gamma_5=\frac{19\beta^5}{480},\quad \gamma_7=-\frac{31\beta^7}{8064}.
\ee
The Cauchy principal value can be evaluated for each expansion term of  $\sin(\beta x)\sqrt{\cos(\beta x)}$  separately.
 Accordingly, for the first term we have
 \be
-C\gamma_1\lim\limits_{\varepsilon\to 0}\left(\,\int\limits_{-x-1}^{-\varepsilon}+\int\limits_{\varepsilon}^{-x+1}\,\right)\frac{x\sqrt{1-x^2}-(x+z)\sqrt{1-(x+z)^2}}{z^2}dz.
\ee
Since
\be \label{31}
\left(\,\int\limits_{-x-1}^{-\varepsilon}+\int\limits_{\varepsilon}^{-x+1}\,\right)\frac{x\sqrt{1-x^2}}{z^2}dz=x\sqrt{1-x^2}\left(-\frac{2}{1-x^2}+\frac{2}{\varepsilon}\right),
\ee
and
\be
\left(\,\int\limits_{-x-1}^{-\varepsilon}+\int\limits_{\varepsilon}^{-x+1}\,\right)\frac{-(x+z)\sqrt{1-(x+z)^2}}{z^2}dz=-x\left(\frac{\sqrt{1-(x-\varepsilon)^2}}{\varepsilon}+\frac{\sqrt{1-(x+\varepsilon)^2}}{\varepsilon}\right)+2\pi x+u(x,\varepsilon),
\ee
where (here unspecified) $u(x,\epsilon )$  approaches $0$ if  $\varepsilon \rightarrow 0$,  for all $x\in D$.

The first term in (\ref{31}), if multiplied by $-C\gamma _1/\pi $, cancels its negative in the expansion of $2\psi _2(x)/\pi (1-x^2)$.
Because of
\be \label{cancel}
\lim\limits_{\varepsilon\to 0}\left(\frac{2\sqrt{1-x^2}-\sqrt{1-(x-\varepsilon)^2}-\sqrt{1-(x+\varepsilon)^2}}{\varepsilon}\right)=0,
\ee
the first expansion term    has  an ultimate  form  $-C\gamma_1w_1(x)$, where  $w_1(x)=2x$.

In connection with (\ref{cancel}) we point out that    troublesome (divergent)  $2/\varepsilon$  entries (related to the Cauchy principal value evaluation)
  are cancelled by  their negatives coming from  the principal value procedure  of the form (\ref{cauchy}) while adopted to  $ - \int\limits_{-x-1}^{-x+1}\frac{\psi_2(x+z)}{z^2}dz$.
 The remaining   expansion terms of $ \int\limits_{-x-1}^{-x+1}\frac{\psi_2(x)}{z^2}dz$  are cancelled by their  negatives  that originate from $\psi_2(x)/(1-x^2)$.

\begin{figure}[h]
\begin{center}
\centering
\includegraphics[width=55mm,height=55mm]{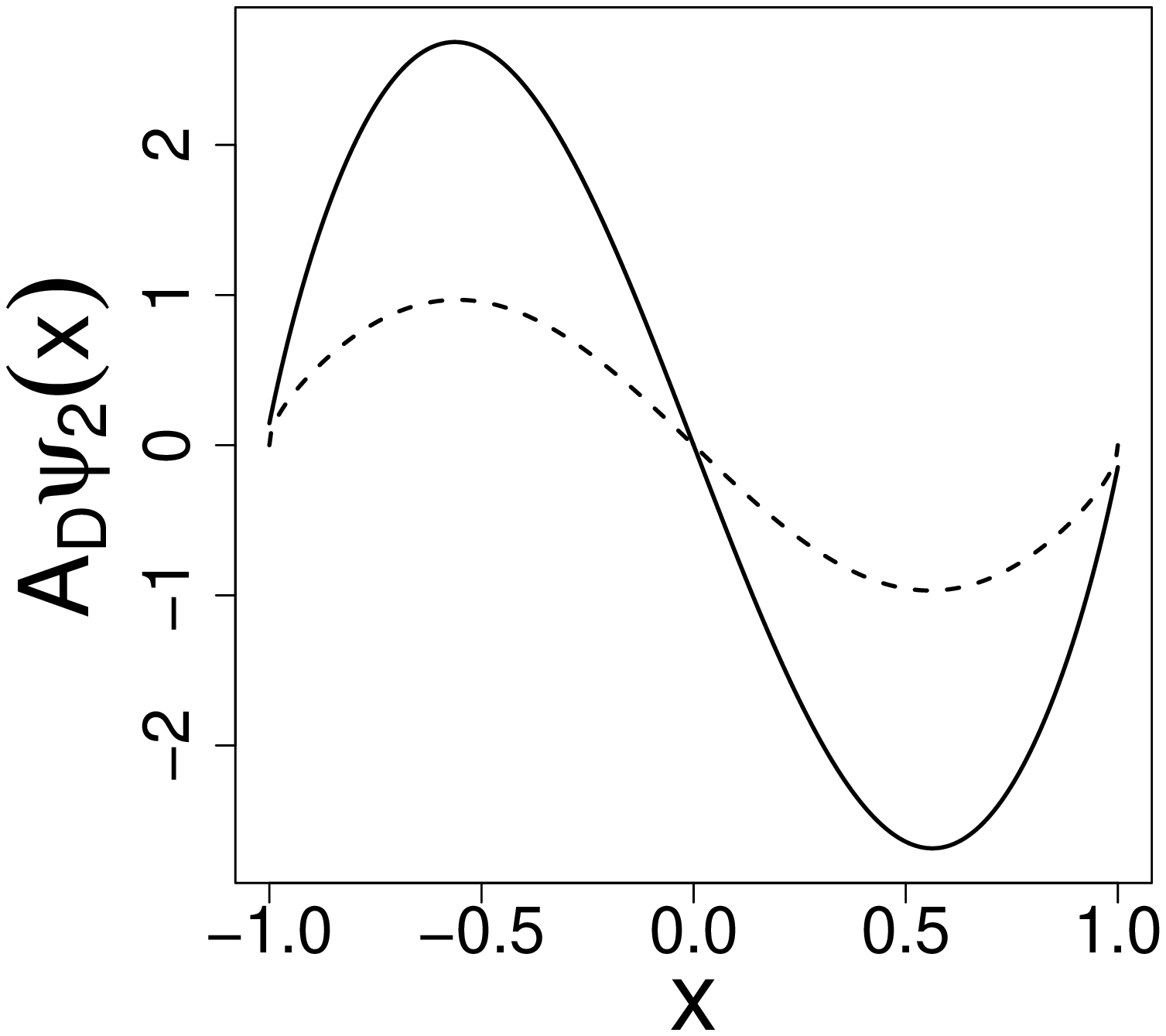}
\includegraphics[width=55mm,height=55mm]{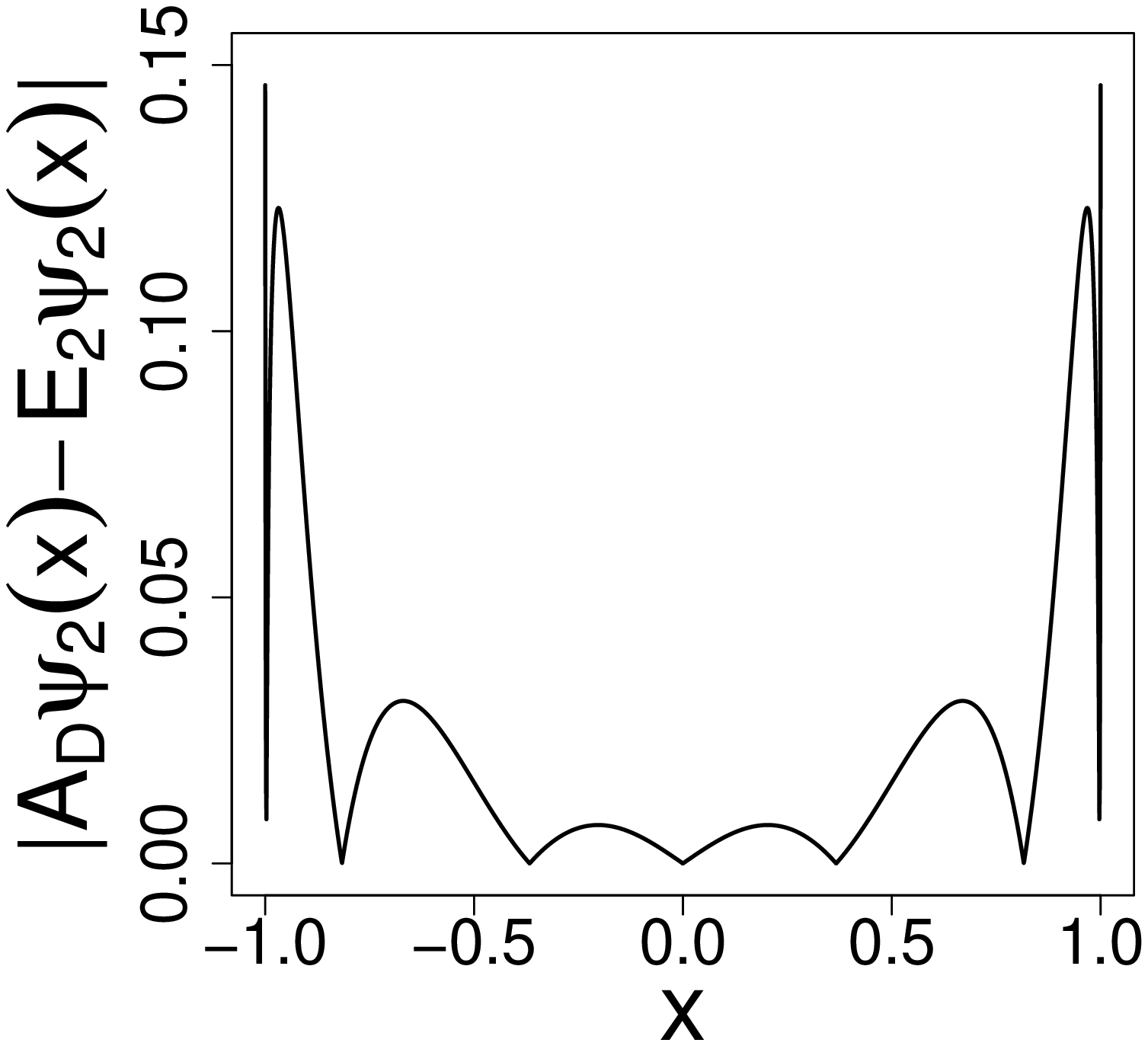}
\includegraphics[width=55mm,height=55mm]{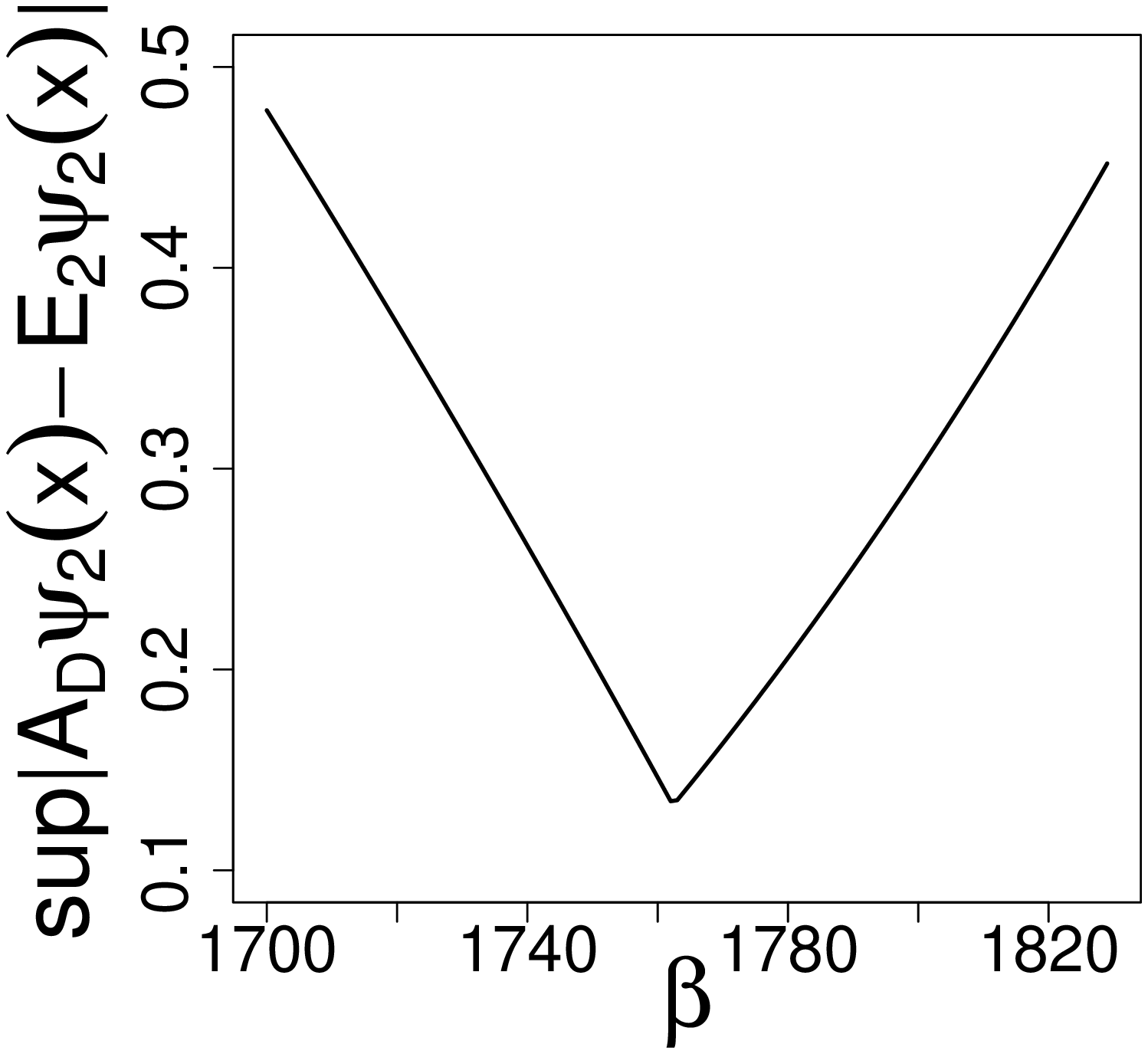}
\caption{Left panel: a comparison of the  approximate function  $\psi _2(x)$
(dotted line) with $A_D\psi _2(x)$  (solid line).  Middle panel: we depict  $|A_D \psi_2   -  E_2\psi_2(x)|$,  where $E=2.75$.
Right apnel: supremum of  $|A_D\psi_2(x)- E_2\psi_2(x)|$, $E=2.75$ as a function of $\beta$. The horizontal axis is scaled in units  $\pi/4096$.}
\end{center}
\end{figure}

An ultimate form of $A_D\psi _2(x)$ is
\be \label{conv}
A_D\psi_2(x)=\sum\limits_{n=0}^{\infty}\gamma_{2n+1}w_{2n+1}(x),\quad x\in D,
\ee
where the coefficients  $\gamma_{2n+1}$, $n=0,1,\ldots$ have been  introduced before (while expanding $\sin(\beta x)\sqrt{\cos(\beta x)}$)
and first few polynomials  $w_{2n+1}$ have  the form:
\be
w_1(x)=2x,\quad w_3(x)=-x(1-4x^2),\quad w_5(x)=\frac{x(-1-8x^2+24x^4)}{4},\quad w_7(x)=\frac{x(-1-4x^2-24x^4+64x^6)}{8}.
\ee

The convergence of (\ref{conv}) is much worse than that encountered in connection with the ground state function.
Therefore a number of polynomials employed in the approximation of $\psi _2$ must be relatively large to make  that approximation reliable.
In Fig. 7, we compare  $A_D\psi_2 (x)$ with $\psi _2(x)$, for an approximation restricted to first $15$  series expansion terms only.

Like in case of the ground state function, we ask for an affinity of  $A_D\psi _2(x)$ with $E_2\psi _2(x)$, where we adopt the value $E_2=2.75$.
In Fig. 8   the  affinity  function  $|A_D\psi_2(x)-E_2\psi_2(x)|$ is depicted and found to be bounded point-wise by $0.1462$ which is much better estimate
than any  ever obtained, \cite{K,KKMS}.

In the definition of an approximate function $\psi _2(x)$ we have still some flexibility allowed with respect to the choice of the
parameter $\beta $.  In Fig. 9, $\sup|A_D\psi_2(x)-E_2\psi_2(x)|$  is depicted as a function of $\beta $ in the vicinity of
 $\beta=1760\pi/4096$. A minimum is achieved for  $\beta=1762\pi/4096$ and sets the upper bound  $0.1344$.

In principle, we can provide analytic approximations for (consecutive) higher excited state functions.
However, our  discussion of Section II  should be considered merely as a warm-up,  a preparatory step to address more serious
goals.

 Let us  note that  all ultimate formulas
 have involved  the polynomial expansions. Interestingly, we could not associate them with any members of a hypergeometric family
 or  other  orthogonal polynomials in $D=(-1,1)$).  We shall  follow  this polynomial expansion strategy in the next section
  to get  most accurate to date approximations of eigenvalues and eigenfunctions in the infinite Cauchy well problem.

We shall impose one more demand, deliberately  absent in the existing literature on  approximate  Cauchy well eigenfunctions. We need
that actually not only $\psi (x)$ is  to vanish identically for $|x|\geq 1$, but $A_D\psi (x)$ as well, to become in all respects
 as close as possible to $E\,  \psi (x)$, where $E$ stands for an approximate eigenvalue.

\section{Polynomial expansions of eigenfunctions in the infinite Cauchy  well: pushing ahead approximation  finesse.}

\subsection{Ground state function addressed anew.}

We wish to  solve the eigenvalue problem
 $A_D\psi(x)=E\psi(x)$,  with the Cauchy operator $A_D$ in $D$,   defined  in Section II.
  This   means  that  in the serch for approximate solutions,  an approximation accuracy  can  be  made  arbitrarily fine,
with the growth of the   degree  of  the  truncated  polynomial expansion of  the sought for eigenfunction.
 Functions in the  domain of $A_D$  are restricted by the exterior Dirichlet condition $\psi(x)=0,\>|x|\geqslant 1$,
  but we impose the very same  restriction  upon the resultant $A_D \, \psi (x)$, given  $\psi (x)\in D$.

We take Eq. (\ref{stu2}) as a working   definition of $A_D$  and proceed with its integral part, here denoted
\be
B_D\psi(x)=\frac{1}{\pi}\int\limits_{-x-1}^{-x+1}\frac{\psi(x)-\psi(x+z)}{z^2}dz.
\ee
Let us consider  the action of $B_D$ upon   functions $\psi (x) = x^{2n} \sqrt{1-x^2}$. We get:
\be
B_D\sqrt{1-x^2}=-\frac{2}{\pi}\frac{\sqrt{1-x^2}}{1-x^2}+1,
\ee
\be
B_Dx^2\sqrt{1-x^2}=-\frac{2}{\pi}\frac{x^2\sqrt{1-x^2}}{1-x^2}-\frac{1-6x^2}{2},
\ee
\be
B_Dx^4\sqrt{1-x^2}=-\frac{2}{\pi}\frac{x^4\sqrt{1-x^2}}{1-x^2}-\frac{1+12x^2-40x^4}{8}.
\ee
\be
B_Dx^6\sqrt{1-x^2}=-\frac{2}{\pi}\frac{x^6\sqrt{1-x^2}}{1-x^2}-\frac{1+6x^2+40x^4-112x^6}{16}.
\ee
Accordingly we have
\be  \label{trial}
B_Dx^{2n}\sqrt{1-x^2}=-\frac{2}{\pi}\frac{x^{2n}\sqrt{1-x^2}}{1-x^2}+(c_{2n}+3c_{2n-2}x^2+5c_{2n-4}+\ldots+(2n+1)c_0x^{2n}),
\ee
where  $c_{2n}$  are  expansion coefficients of the Taylor series for $\sqrt{1-x^2}$. Namely, we have
\be
\sqrt{1-x^2}=\sum_{n=0}^{\infty}c_{2n}x^{2n}=\sum_{n=0}^{\infty}\frac{(2n)!}{(1-2n)(n!)^24^n}x^{2n},
\ee
which allows to rewrite   $B_Dx^{2n}\sqrt{1-x^2}$  as follows
\be
B_Dx^{2n}\sqrt{1-x^2}=-\frac{2}{\pi}\frac{x^{2n}\sqrt{1-x^2}}{1-x^2}+\sum_{k=0}^n\frac{(2k)!(2n+1-2k)}{(1-2k)(k!)^24^k}x^{2n-2k}.
\ee

Then ground state function is even, hence we can expect its  power series expansion in the form:
\be  \label{g}
\psi(x)=C\sqrt{1-x^2}\sum_{n=0}^\infty\alpha_{2n}x^{2n},\qquad \alpha_0=1.
\ee
where our major task is to   deduce the expansion   coefficients $\alpha_{2n}$.

Coming back to the definition (\ref{stu2}),  of $A_D$, we realize that $A_D\psi (x)= B_D\psi (x) + 2\psi (x)/\pi (1-x^2)$.
  The second term of this expression,  upon employing  the  trial   $\psi (x)$, as in  Eq. (\ref{trial}) or (42),  clearly cancels  the
 first term of  $B_D\psi (x)$, compare e.g. (36)-(42).

  In view of this, the action of $A_D$ upon  the ground state candidate-function  $\psi (x)$
of  (\ref{g})  greatly simplifies.  Ultimately, the eigenvalue problem $A_D\psi (x)= E \psi (x)$ takes the form:
\be
\sum_{n=0}^\infty\alpha_{2n}\sum_{k=0}^n\frac{(2k)!(2n+1-2k)}{(1-2k)(k!)^24^k}x^{2n-2k}=
E\sum_{k=0}^\infty\frac{(2k)!}{(1-2k)(k!)^24^k}x^{2k}\sum_{n=0}^\infty\alpha_{2n}x^{2n},
\ee
which can be re-ordered as follows
\be \label{system}
\sum_{k=0}^\infty\sum_{n=k}^\infty\alpha_{2n}\frac{(2k)!(2n+1-2k)}{(1-2k)(k!)^24^k}x^{2n-2k}=\sum_{k=0}^\infty\sum_{n=0}^\infty E\alpha_{2n}\frac{(2k)!}{(1-2k)(k!)^24^k}x^{2k+2n}.
\ee
This system of  equations  (from which the coefficients $\alpha _{2n}$ are to follow for all $n$)
 looks hopelessly discouraging, if we are interested in a fully-fledged solution of the eigenvalue problem.
 However, things simplify if we  look for approximate solutions, readily  accessible  upon a truncation of
 otherwise infinite series.

By definition we  know that any solution  $\psi (x)$ is defined in the domain $\bar{D}= [-1,1] $ and obeys the boundary condition
  $\psi (\pm 1)=0$.   We extend this restriction to $A_D\psi (x)$
and demand
\be \label{c}
\lim\limits_{x\to\pm 1}A_D\psi(x)=0.
\ee
In what follows we shall assume that $\alpha _0=1$.  This choice is possible, in view of the presumed normalization
(we have $C$ involved).
 Each considered  truncated series approximating $\psi (x)$ will be $l^2(D)$ normalized, this
operation being  encoded in a multiplicative constant $C$.\\

{\bf Example 1:} Let us exemplify our procedure by truncating the series after the polynomial of degree $2$.  We have
$ A_D(1+\alpha_2 x^2)\sqrt{1-x^2}=1-\alpha_2\left(\frac{1}{2}-3x^2\right)$ and   the condition (\ref{c}) implies
$ 1+\frac{5}{2}\alpha_2=0$. Accordingly we end up with an approximate eigenstate  $\psi(x)=C(1-2/5x^2)\sqrt{1-x^2}$
where  $C=\sqrt{875/996}\approx 0.937291$. The approximate eigenvalue reads  $E=1.2$.\\
{\bf Example 2:} An analogous procedure for series terminating at the polynomial of the $4$-th degree  gives rise to
$A_D(1+\alpha_2 x^2+\alpha_4 x^4)\sqrt{1-x^2}$ and (\ref{c}) implies  $1+\frac{5}{2}\alpha_2+\frac{27}{8}\alpha_4=0$.
Moreover, we have $ 1-\frac{1}{2}\alpha_2-\frac{1}{8}\alpha_4=E$ and  $3\alpha_2-\frac{3}{2}\alpha_4=E\left(-\frac{1}{2}+\alpha_2\right)$.
The coefficients readily follow with values $\alpha_2\approx -0.353189$  and  $\alpha_4\approx -0.0346746$.  The approximate
eigenvalue reads $E\approx 1.18093$. The normalized approximate eigenfunction takes the form
 $\psi(x)=C(1-0.353189x^2-0.0346746x^4)\sqrt{1-x^2}$,  where  $C=0.931331$.\\

It is clear, that the procedure can be continued indefinitely by increasing the polynomial degree at the series  truncation "point".
An approximation accuracy grows with the polynomial degree. The polynomial degree growth increases the  number of linear equations
to solve.

 Let $(a_{k,n})$  denote a  matrix  with  elements
\be
a_{k,n}=(2n+1-2k)c_k=\frac{(2k)!(2n+1-2k)}{(1-2k)(k!)^2 4^k},\qquad n\geqslant k.
\ee
We  recall that
\be
c_k=\frac{(2k)!}{(1-2k)(k!)^2 4^k}.
\ee

If we consider an approximation of $\psi (x)$ by series  terminating at the polynomial of degree $2n$, the eigenvalue problem we address takes the form
of a   linear  system of equations with  unknown $E$ and $\alpha _{2n}$ (we recall our choice of  $\alpha _0=1$):
\begin{eqnarray} \label{eigen}
\sum\limits_{k=i}^n \alpha_{2k}a_{k-i,k}=E\sum\limits_{k=0}^i \alpha_{2k}c_{i-k},\qquad i=0,1,\ldots,n-1,\nonumber\\
\sum\limits_{m=0}^n \left(\alpha_{2m}\sum\limits_{k=0}^m a_{k,m}\right)=0.
\end{eqnarray}
The last identity in (\ref{eigen})  comes from our demand (\ref{c}), here adopted to  $A_D w_{2n}\sqrt{1-x^2}=0 $  at  $x=\pm 1$.\\

\begin{table}[h]
\begin{center}
\begin{tabular}{|c||c|c|c|c|c|c|c|c|c|c|}
  \hline
  - & C & E & $\alpha_2$ & $\alpha_4$ & $\alpha_6$ & $\alpha_8$ & $\alpha_{10}$ & $\alpha_{12}$ & $\alpha_{14}$ & $\alpha_{16}$\\
  \hline
  $w_2$ & 0.937291 & 1.200000 & -0.400000 & - & - & - & - & - & - & - \\
  $w_4$ & 0.931331 & 1.180929 & -0.353189 & -0.03467461 & - & - & - & - & - & -\\
  $w_6$ & 0.927253 & 1.170127 & -0.333863 & -0.00891937 & -0.0332900 & - & - & - & - & -\\
  $w_8$ & 0.925363 & 1.165443 & -0.326159 & -0.00332500 & -0.0173088 & -0.0221718 & - & - & - & -\\
  $w_{10}$ & 0.924339 & 1.162981 & -0.322268 & -0.00097523 & -0.0134732 & -0.0111668 & -0.0163303 & - & - & -\\
  $w_{12}$ & 0.923728 & 1.161534 & -0.320035 & 0.00025555 & -0.0117497 & -0.0084661 & -0.0081667 & -0.0126748 & - & -\\
  $w_{14}$ & 0.923337 & 1.160614 & -0.318637 & 0.00098488 & -0.0107978 & -0.0072137 & -0.0061348 & -0.0063114 & -0.0102120 & -\\
  $w_{16}$ & 0.923071 & 1.159993 & -0.317704 & 0.00145367 & -0.0102098 & -0.0065016 & -0.0051721 & -0.0047139 & -0.0050726 & -0.0084590\\
  $w_{18}$ & 0.922884 & 1.159555 & -0.317051 & 0.00177313 & -0.0098192 & -0.0060507 & -0.0046131 & -0.0039456 & -0.0037753 & -0.0041958\\
  $w_{20}$ & 0.922746 & 1.159234 & -0.316576 & 0.00200068 & -0.0095458 & -0.0057448 & -0.0042525 & -0.0034927 & -0.0031448 & -0.0031160\\
  $w_{30}$ & 0.922409 & 1.158447 & -0.315422 & 0.00253637 & -0.0089180 & -0.0050710 & -0.0035043 & -0.0026342 & -0.0021154 & -0.0017921\\
  $w_{40}$ & 0.922868 & 1.158159 & -0.315006 & 0.00272257 & -0.0087053 & -0.0048519 & -0.0032737 & -0.0023882 & -0.0018495 & -0.0015006\\
  $w_{50}$ & 0.922230 & 1.158022 & -0.314810 & 0.00280842 & -0.0086084 & -0.0047537 & -0.0031724 & -0.0022828 & -0.0017390 & -0.0013839\\
  $w_{60}$ & 0.922198 & 1.157948 & -0.314703 & 0.00285494 & -0.0085562 & -0.0047014 & -0.0031190 & -0.0022279 & -0.0016822 & -0.0013250\\
  $w_{70}$ & 0.922179 & 1.157902 & -0.314638 & 0.00288292 & -0.0085249 & -0.0046702 & -0.0030874 & -0.0021957 & -0.0016492 & -0.0012910\\
  $w_{80}$ & 0.922166 & 1.157872 & -0.314595 & 0.00290106 & -0.0085047 & -0.0046501 & -0.0030671 & -0.0021751 & -0.0016283 & -0.0012696\\
  $w_{90}$ & 0.922158 & 1.157852 & -0.314566 & 0.00291348 & -0.0084909 & -0.0046364 & -0.0030534 & -0.0021612 & -0.0016142 & -0.0012552\\
  $w_{100}$& 0.922152 & 1.157837 & -0.314545 & 0.00292235 & -0.0084810 & -0.0046267 & -0.0030437 & -0.0021514 & -0.0016042 & -0.0012451\\
  $w_{150}$& 0.922137 & 1.157802 & -0.314496 & 0.00294331 & -0.0084578 & -0.0046039 & -0.0030208 & -0.0021285 & -0.0015811 & -0.0012218\\
  $w_{200}$& 0.922132 & 1.157789 & -0.314478 & 0.00295063 & -0.0084497 & -0.0045959 & -0.0030130 & -0.0021206 & -0.0015732 & -0.0012139\\
  $w_{300}$& 0.922129 & 1.157781 & -0.314466 & 0.00295585 & -0.0084440 & -0.0045903 & -0.0030074 & -0.0021151 & -0.0015677 & -0.0012083\\
  $w_{400}$& 0.922127 & 1.157778 & -0.314461 & 0.00295767 & -0.0084419 & -0.0045884 & -0.0030055 & -0.0021132 & -0.0015658 & -0.0012064\\
  $w_{500}$& 0.922127 & 1.157776 & -0.314459 & 0.00295851 & -0.0084410 & -0.0045875 & -0.0030046 & -0.0021123 & -0.0015649 & -0.0012056\\
  \hline
\end{tabular}
\end{center}
\caption{Approximate solutions of the eigenvalue equation $A_D\psi (x) = E \psi (x)$. The approximating polynomial of degree $2n$ is indicated by $w_{2n}(x)$.
We report first few values of coefficients $\alpha _{2k}$ for each $2n$-th case, together with an approximate eigenvalue $E$ and the normalization coefficient $C$.
For comparison we report the ground state eigenvalue reported in Ref. \cite{KKMS}, E= 1.157773883697 (based on a diagonalization of the $900 \times 900 $-matrix).   Our
ultimate result actually is $E=1.1577764$.}
\end{table}

\begin{sidewaystable*}
\begin{center}
\begin{tabular}{|c||c|c|c|c|c|c|c|c|c|c|}
  \hline
  $n$ & 1 & 2 & 3 & 4 & 5 & 6 & 7 & 8 & 9 & 10\\
  \hline
$\alpha_{2n}$    & $-0.3144595$ & $0.00295851$ & $-0.0084410$ & $-0.0045875$ & $-0.0030046$ & $-0.0021123$ & $-0.0015649$ & $-0.0012056$ & $-0.0009571$ & $-0.0007784$ \\
$\alpha_{20+2n}$ & $-0.0006454$ & $-0.0005439$ & $-0.0004647$ & $-0.0004016$ & $-0.0003506$ & $-0.0003088$ & $-0.0002740$ & $-0.0002449$ & $-0.0002201$ & $-0.0001990$ \\
$\alpha_{40+2n}$ & $-0.0001808$ & $-0.0001650$ & $-0.0001512$ & $-0.0001391$ & $-0.0001284$ & $-0.0001189$ & $-0.0001104$ & $-0.0001029$ & $-0.0000960$ & $-0.0000899$ \\
$\alpha_{60+2n}$ & $-0.0000843$ & $-0.0000793$ & $-0.0000747$ & $-0.0000705$ & $-0.0000666$ & $-0.0000631$ & $-0.0000598$ & $-0.0000568$ & $-0.0000540$ & $-0.0000515$ \\
$\alpha_{80+2n}$ & $-0.0000491$ & $-0.0000469$ & $-0.0000448$ & $-0.0000429$ & $-0.0000411$ & $-0.0000394$ & $-0.0000378$ & $-0.0000363$ & $-0.0000349$ & $-0.0000336$ \\
$\alpha_{100+2n}$& $-0.0000324$ & $-0.0000312$ & $-0.0000301$ & $-0.0000291$ & $-0.0000281$ & $-0.0000272$ & $-0.0000263$ & $-0.0000255$ & $-0.0000247$ & $-0.0000239$ \\
$\alpha_{120+2n}$& $-0.0000232$ & $-0.0000225$ & $-0.0000219$ & $-0.0000213$ & $-0.0000207$ & $-0.0000201$ & $-0.0000196$ & $-0.0000191$ & $-0.0000186$ & $-0.0000181$ \\
$\alpha_{140+2n}$& $-0.0000177$ & $-0.0000172$ & $-0.0000168$ & $-0.0000164$ & $-0.0000160$ & $-0.0000157$ & $-0.0000153$ & $-0.0000150$ & $-0.0000147$ & $-0.0000143$ \\
$\alpha_{160+2n}$& $-0.0000140$ & $-0.0000137$ & $-0.0000135$ & $-0.0000132$ & $-0.0000129$ & $-0.0000127$ & $-0.0000125$ & $-0.0000122$ & $-0.0000120$ & $-0.0000118$ \\
$\alpha_{180+2n}$& $-0.0000116$ & $-0.0000114$ & $-0.0000112$ & $-0.0000110$ & $-0.0000108$ & $-0.0000106$ & $-0.0000104$ & $-0.0000103$ & $-0.0000101$ & $-9.97*10^{-6}$ \\
$\alpha_{200+2n}$& $-9.81*10^{-6}$ & $-9.67*10^{-6}$ & $-9.53*10^{-6}$ & $-9.39*10^{-6}$ & $-9.26*10^{-6}$ & $-9.13*10^{-6}$ & $-9.00*10^{-6}$ & $-8.88*10^{-6}$ & $-8.76*10^{-6}$ & $-8.65*10^{-6}$ \\
$\alpha_{220+2n}$& $-8.54*10^{-6}$ & $-8.43*10^{-6}$ & $-8.33*10^{-6}$ & $-8.23*10^{-6}$ & $-8.13*10^{-6}$ & $-8.03*10^{-6}$ & $-7.94*10^{-6}$ & $-7.85*10^{-6}$ & $-7.76*10^{-6}$ & $-7.68*10^{-6}$ \\
$\alpha_{240+2n}$& $-7.59*10^{-6}$ & $-7.51*10^{-6}$ & $-7.43*10^{-6}$ & $-7.36*10^{-6}$ & $-7.28*10^{-6}$ & $-7.21*10^{-6}$ & $-7.14*10^{-6}$ & $-7.08*10^{-6}$ & $-7.01*10^{-6}$ & ${-6}.95*10^{-6}$\\
$\alpha_{260+2n}$& ${-6}.88*10^{-6}$ & ${-6}.82*10^{-6}$ & ${-6}.77*10^{-6}$ & ${-6}.71*10^{-6}$ & ${-6}.65*10^{-6}$ & ${-6}.60*10^{-6}$ & ${-6}.55*10^{-6}$ & ${-6}.50*10^{-6}$ & ${-6}.45*10^{-6}$ & ${-6}.40*10^{-6}$ \\
$\alpha_{280+2n}$& ${-6}.35*10^{-6}$ & ${-6}.31*10^{-6}$ & ${-6}.27*10^{-6}$ & ${-6}.22*10^{-6}$ & ${-6}.18*10^{-6}$ & ${-6}.14*10^{-6}$ & ${-6}.10*10^{-6}$ & ${-6}.07*10^{-6}$ & ${-6}.03*10^{-6}$ & ${-6}.00*10^{-6}$ \\
$\alpha_{300+2n}$& $-5.96*10^{-6}$ & $-5.93*10^{-6}$ & $-5.90*10^{-6}$ & $-5.87*10^{-6}$ & $-5.84*10^{-6}$ & $-5.81*10^{-6}$ & $-5.79*10^{-6}$ & $-5.76*10^{-6}$ & $-5.73*10^{-6}$ & $-5.71*10^{-6}$ \\
$\alpha_{320+2n}$& $-5.69*10^{-6}$ & $-5.67*10^{-6}$ & $-5.65*10^{-6}$ & $-5.63*10^{-6}$ & $-5.61*10^{-6}$ & $-5.59*10^{-6}$ & $-5.57*10^{-6}$ & $-5.56*10^{-6}$ & $-5.54*10^{-6}$ & $-5.53*10^{-6}$ \\
$\alpha_{340+2n}$& $-5.51*10^{-6}$ & $-5.50*10^{-6}$ & $-5.49*10^{-6}$ & $-5.48*10^{-6}$ & $-5.47*10^{-6}$ & $-5.46*10^{-6}$ & $-5.45*10^{-6}$ & $-5.45*10^{-6}$ & $-5.44*10^{-6}$ & $-5.44*10^{-6}$ \\
$\alpha_{360+2n}$& $-5.43*10^{-6}$ & $-5.43*10^{-6}$ & $-5.43*10^{-6}$ & $-5.43*10^{-6}$ & $-5.43*10^{-6}$ & $-5.43*10^{-6}$ & $-5.43*10^{-6}$ & $-5.44*10^{-6}$ & $-5.44*10^{-6}$ & $-5.45*10^{-6}$\\
$\alpha_{380+2n}$& $-5.46*10^{-6}$ & $-5.46*10^{-6}$ & $-5.47*10^{-6}$ & $-5.48*10^{-6}$ & $-5.49*10^{-6}$ & $-5.51*10^{-6}$ & $-5.52*10^{-6}$ & $-5.54*10^{-6}$ & $-5.56*10^{-6}$ & $-5.57*10^{-6}$ \\
$\alpha_{400+2n}$& $-5.59*10^{-6}$ & $-5.62*10^{-6}$ & $-5.64*10^{-6}$ & $-5.66*10^{-6}$ & $-5.69*10^{-6}$ & $-5.72*10^{-6}$ & $-5.75*10^{-6}$ & $-5.78*10^{-6}$ & $-5.82*10^{-6}$ & $-5.86*10^{-6}$ \\
$\alpha_{420+2n}$& $-5.90*10^{-6}$ & $-5.94*10^{-6}$ & $-5.98*10^{-6}$ & ${-6}.03*10^{-6}$ & ${-6}.08*10^{-6}$ & ${-6}.13*10^{-6}$ & ${-6}.19*10^{-6}$ & ${-6}.25*10^{-6}$ & ${-6}.32*10^{-6}$ & ${-6}.39*10^{-6}$ \\
$\alpha_{440+2n}$& ${-6}.46*10^{-6}$ & ${-6}.54*10^{-6}$ & ${-6}.62*10^{-6}$ & ${-6}.71*10^{-6}$ & ${-6}.81*10^{-6}$ & ${-6}.91*10^{-6}$ & $-7.03*10^{-6}$ & $-7.15*10^{-6}$ & $-7.28*10^{-6}$ & $-7.42*10^{-6}$ \\
$\alpha_{460+2n}$& $-7.57*10^{-6}$ & $-7.74*10^{-6}$ & $-7.92*10^{-6}$ & $-8.12*10^{-6}$ & $-8.35*10^{-6}$ & $-8.59*10^{-6}$ & $-8.87*10^{-6}$ & $-9.18*10^{-6}$ & $-9.54*10^{-6}$ & $-9.95*10^{-6}$ \\
$\alpha_{480+2n}$& $-0.0000104$ & $-0.0000110$ & $-0.0000117$ & $-0.0000125$ & $-0.0000136$ & $-0.0000150$ & $-0.0000171$ & $-0.0000204$ & $-0.0000271$ & $-0.0000540$ \\
  \hline
\end{tabular}
\end{center}
\caption{For the approximate ground state function, the corresponding   polynomial  $w_{500}$  is displayed in detail, in terms of its
expansion coefficients  $\alpha_{2n}$.  Note a numbering convention: in the first row  we have displayed consecutively
$\alpha _2, \alpha _4...$ up to $\alpha _{20}$.}
\end{sidewaystable*}

\begin{figure}[h]
\begin{center}
\centering
\includegraphics[width=55mm,height=55mm]{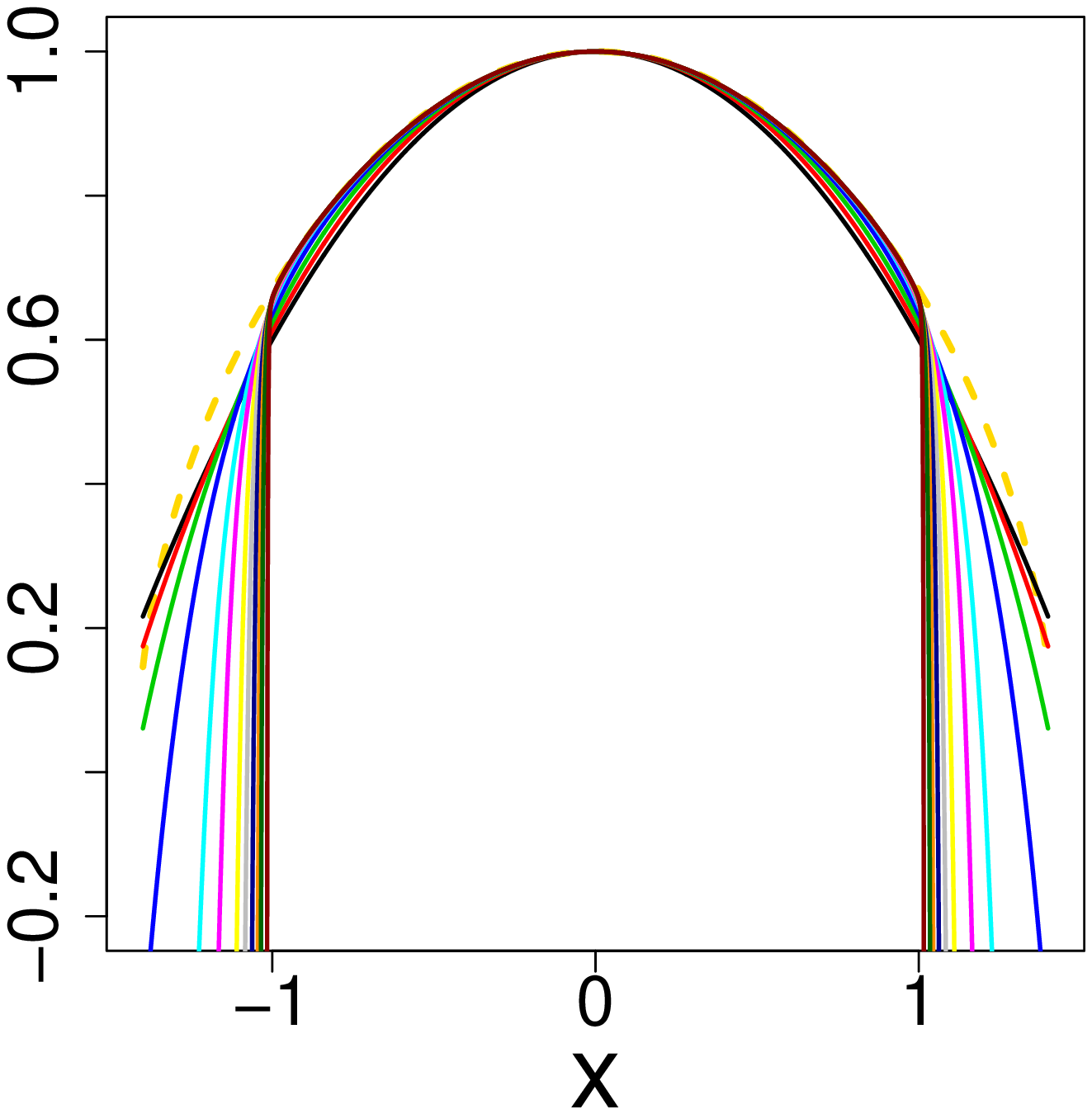}
\includegraphics[width=55mm,height=55mm]{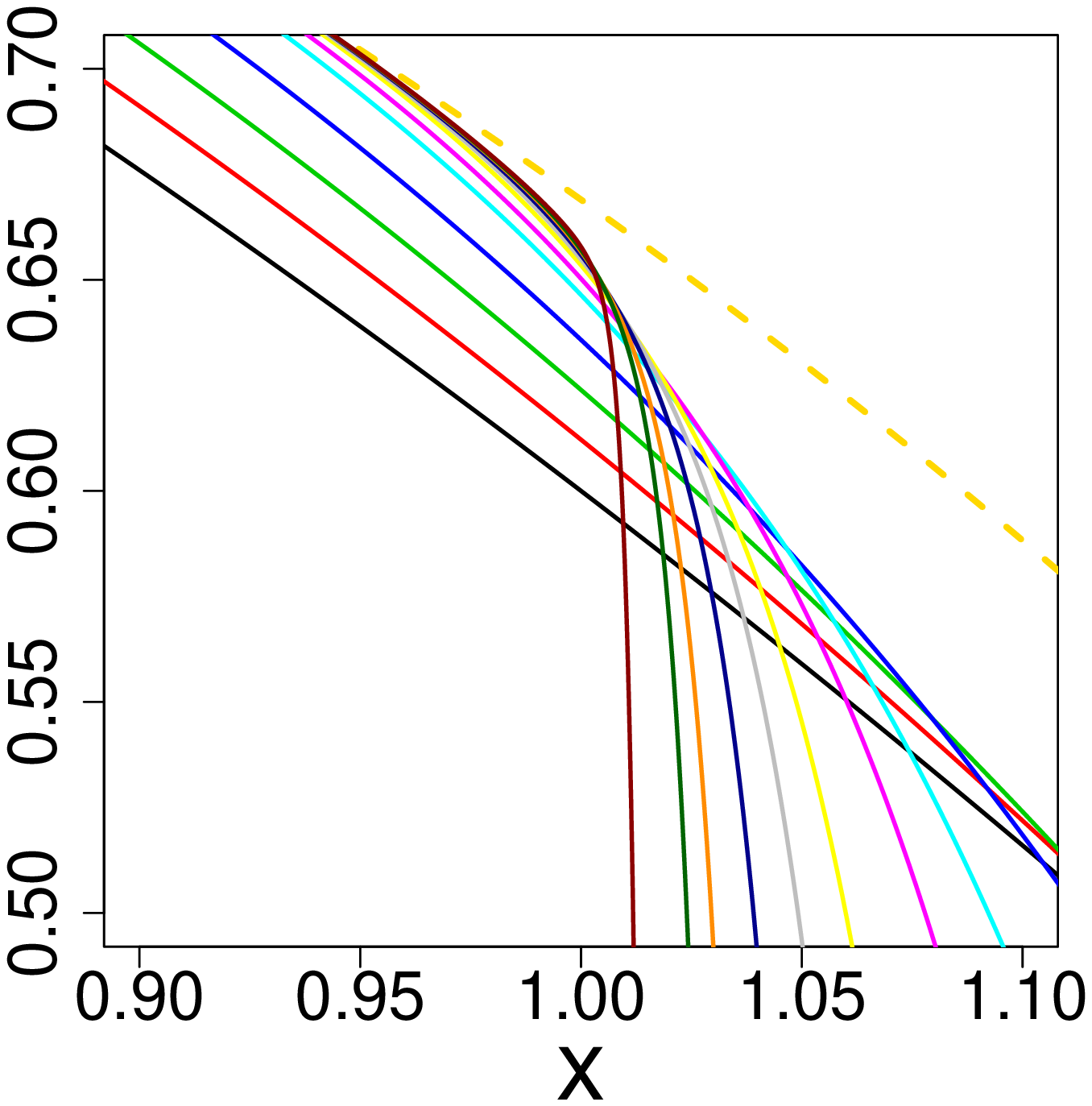}
\includegraphics[width=55mm,height=55mm]{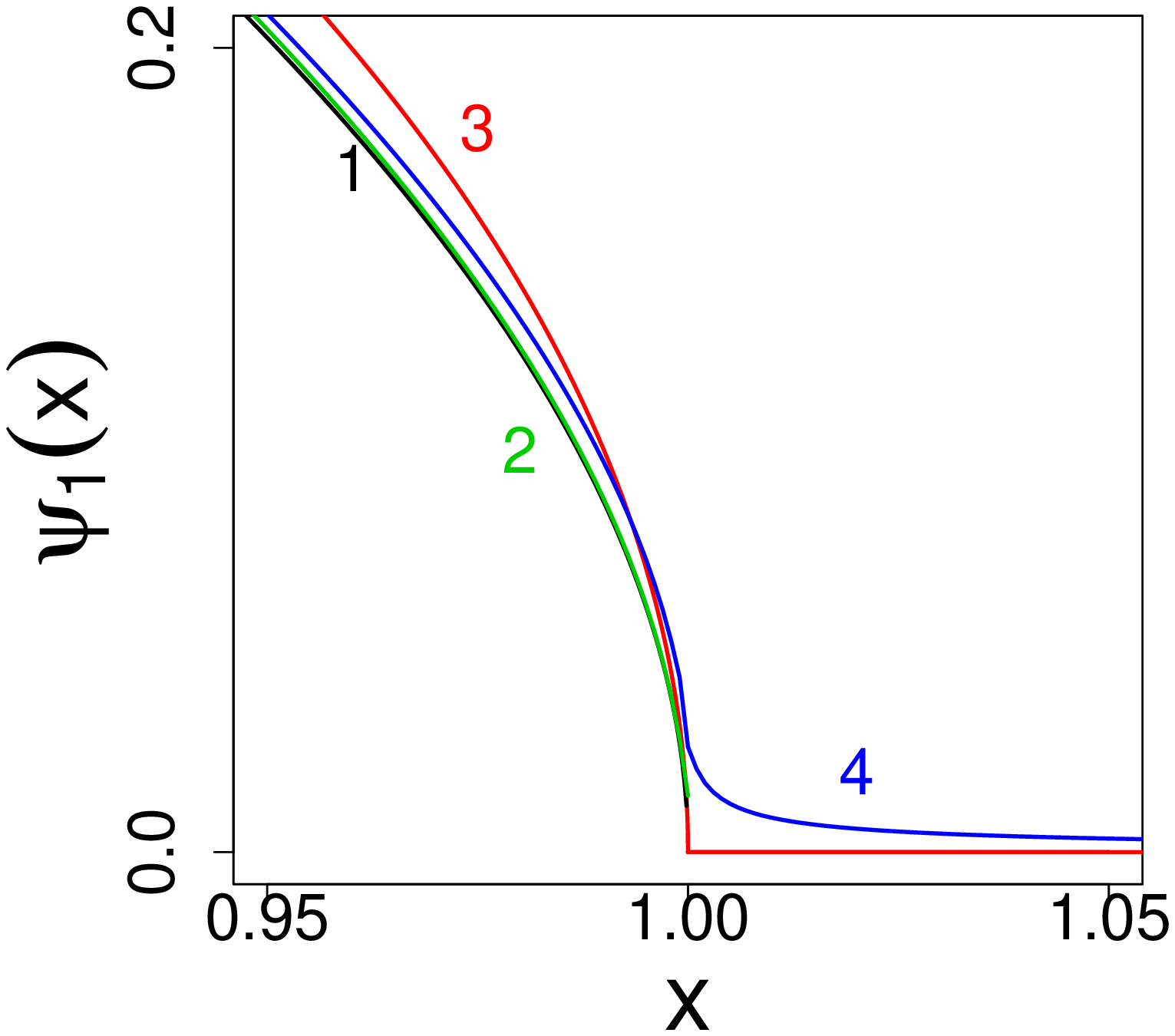}
\caption{Left panel: a comparative display of polynomials $w_{2n}(x)$ of degrees  $2,4,6,10,20,30,50,70,100,150,200,500$  and
the curve  $\sqrt{\cos{(1443\pi x/4096)}}$ (gold) which has been a building block in  the formula (\ref{ground}).
Middle  panel provides an enlargement in  the vicinity of the right  boundary. Right panel depicts various approximations of the ground state function at the right boundary $x=1$:
 1 - curve  $Cw_{500}(x)\sqrt{1-x^2}$, 2 - curve of  \cite{K}, 3 - $\psi_1(x)\sim (1-|x|)^{1/2}$ of  \cite{ZRK}, 4  - $V_0=500$
finite well  ground state of \cite{ZG}.}
\end{center}
\end{figure}

The   system   (\ref{eigen})  can  be solved  for  various  series  truncation choices,  up to the  $2n=500$  polynomial degree.
All computations have been carried out  by employing the routines of Wolfram Mathematica, which appear to be dedicated to solving even very large
linear systems of equations.

{\bf Remark:} One needs to be aware that (\ref{eigen}),  as a system of $n+1$ equations, has more than one solution. We select an optimal approximation of the ground
 state function   by selecting a solution with then least value of $E$.
The same system of equations produces   solutions that approximate higher (excited) even eigenfunctions. There appear also complex solutions which we
discard as physically irrelevant.\\

Our findings are gathered in Table I, where we report explicit values for first few expansion coefficients
 $\alpha_{2n}$ of approximating polynomials, the  approximate eigenvalue $E$  and related normalization constant  $C$. Symbols $w_{2n}$  refer
 to approximating polynomials of degree $2n$.
The computed eigenvalues definitely drop down with the growth of the polynomial  degree $2n$, with a visible stabilization tendency.
In our opinion, our result is much sharper than that reported in Ref. \cite{KKMS}, see also \cite{K}.   Numerical values of the coefficients $\alpha_{2n}$
grow as well with $2n$ growth, with a clear stabilization tendency.

In Table II we make explicit the functional form of the polynomial $w_{500}$. All  coefficients  $\alpha_{2n}$ are reproduced as well. At the moment that provides the best
available approximation of the ground state function in the infinite Cauchy well problem.

Since we are interested in  fine details of the eigenfunction shape, it is instructive to display the behavior of
 the major eigenfunction building blocks, i.e.
the  polynomials $w_{2n}(x)$,  in the vicinity of the boundaries of $D$. In Fig. 6, we   admit   $x$  from the exterior of $\bar{D}$,
 i.e. $|x|>1$.
We compare the  near-the-boundaries  behavior of polynomials of degrees  $2,4,6,10,20,30,50,70,100,150,200,500$,   with a function
 $\sqrt{\cos{(1443\pi x/4096)}}$ (colored gold) appearing in the definition of  the trial ground state function   (\ref{ground})
 (c.f. Section II). That clearly explains deficiencies of the cosine factor  in the adopted definition  and   obvious
 virtues of the present polynomial expansion method. We note that the polynomial degree growth,  is accompanied by a   steeper   decent
of the representative  curves  at  the  boundaries.

At this point it is also  instructive  to have a comparison of the  boundary behavior of the approximate ground state function proposed in the literature so far.
 We note that $\psi (x)$   of  Ref.  \cite{K}, at the boundaries,  is fapp (for all practical purposes) identical
with our  $ \psi (x) \sim Cw_{500}(x)\sqrt{1-x^2}$.

As before, we can analyze a deviation of  $A_D\psi (x)$  from $E\, \psi (x)$ in  terms of $|A_D\psi(x)-E\psi(x)|,\quad x\in  \bar{D}$.
Results are depicted in Fig. 7.  Note that for the polynomial of degree $2n=500$ we get an upper bound $|A_D\psi(x)-E\psi(x)|<0.01$.
We note that with the growth of the polynomial degree, the near-the-boundary maximum of $|A_D\psi (x) - E\psi (x)|$ decreases.
\begin{figure}[h]
\begin{center}
\centering
\includegraphics[width=70mm,height=70mm]{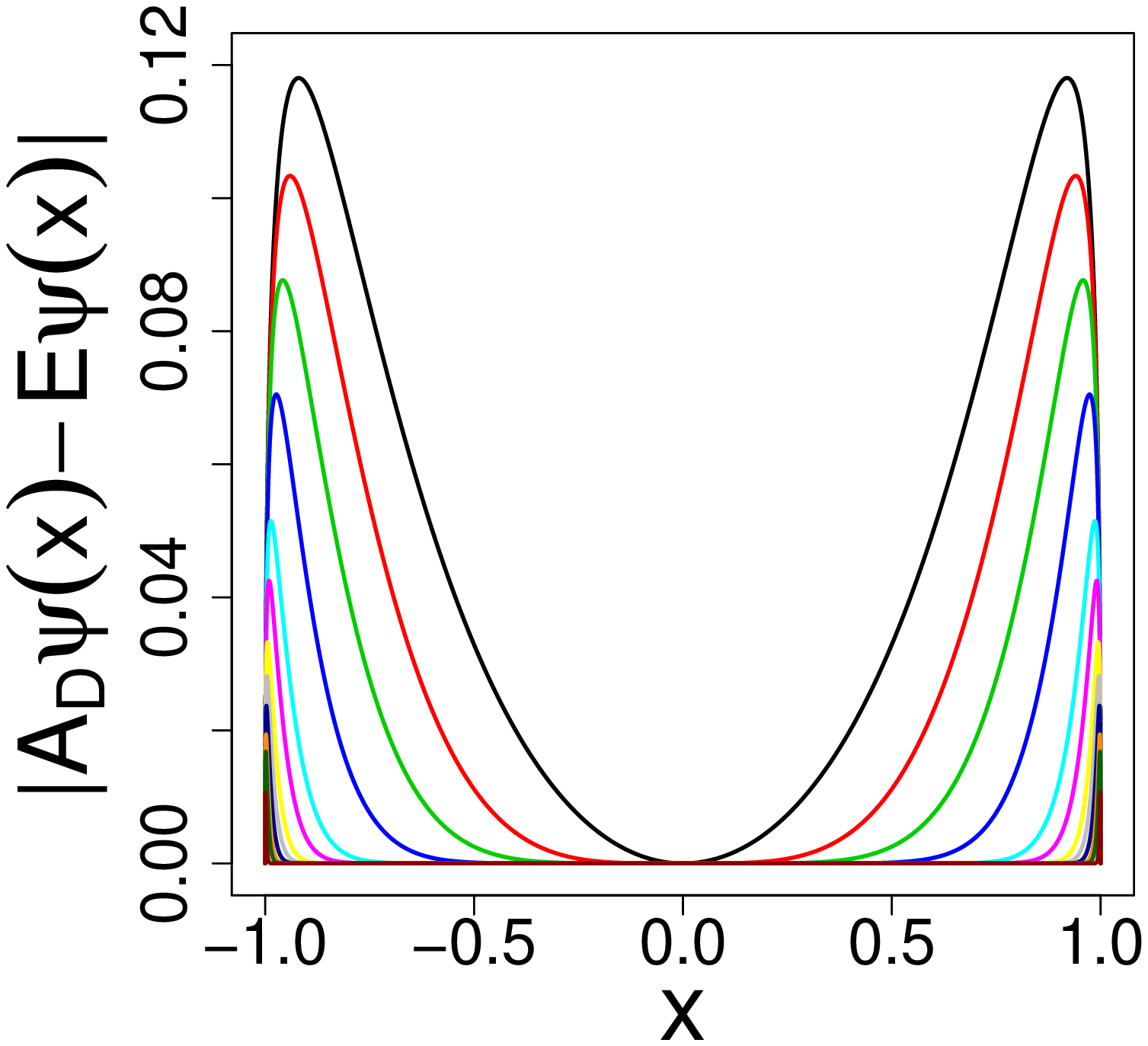}
\includegraphics[width=70mm,height=70mm]{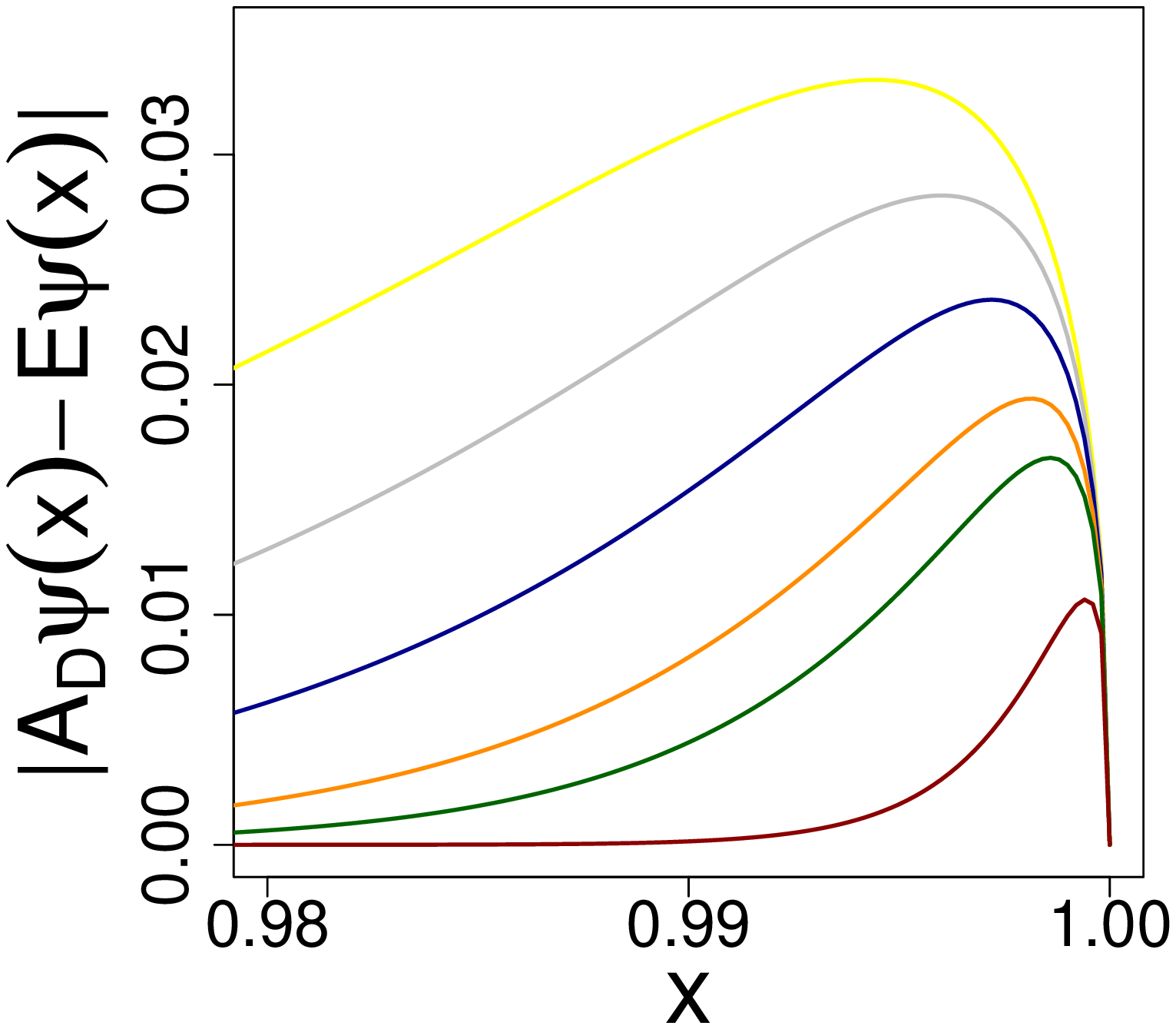}
\caption{Left panel: $|A_D\psi(x)-E\psi(x)|$ where  $\psi = C\sqrt{1-x^2} w_{2n}(x)$, with $2n=  2,4,6,10,20,30,50,70,100,150,200,500$.
Right panel: polynomial  degrees  $2n= 50,70,100,150,200,500$, $\psi (x)$  in the vicinity of the right boundary $x=1$.}
\end{center}
\end{figure}

\subsection{Other  even eigenfunctions $\psi _{2k+1}(x), \quad k>1$.}

The system (50) of equations  has been  dedicated to obtain even eigenfunctions. As mentioned before it  has  infinitely many solutions,  both real and complex-valued.
 Each real solution is interpreted as an approximation of a certain  eigenfunction.  Since for each resolved   polynomial  of  degree $2n$,   we  can   jointly compute
approximate eigenfunctions and the corresponding eigenvalues, there appears a natural ordering with respect to increasing   $E$-values which we enumerate
by  consecutive odd numbers $2k$,  with  $k=1$ corresponding to the ground state. That allows   for a  systematic selection of higher rank even  eigenfunctions.
We keep intact the notation $w_{2n}(x)$ for an approximating polynomial of degree $2n$, although  one should keep in mind that for each consecutive $E_{2k+1}$, we deal
we the  corresponding $2k+1$-th  polynomial  (and  appropriate   $2k+1$-th set of expansion coefficients).
The previous Table II data refer to the ground state solution $\psi _1(x)$  only.  The coefficients tables for other polynomials area available upon request.

\begin{figure}[h]
\begin{center}
\centering
\includegraphics[width=55mm,height=55mm]{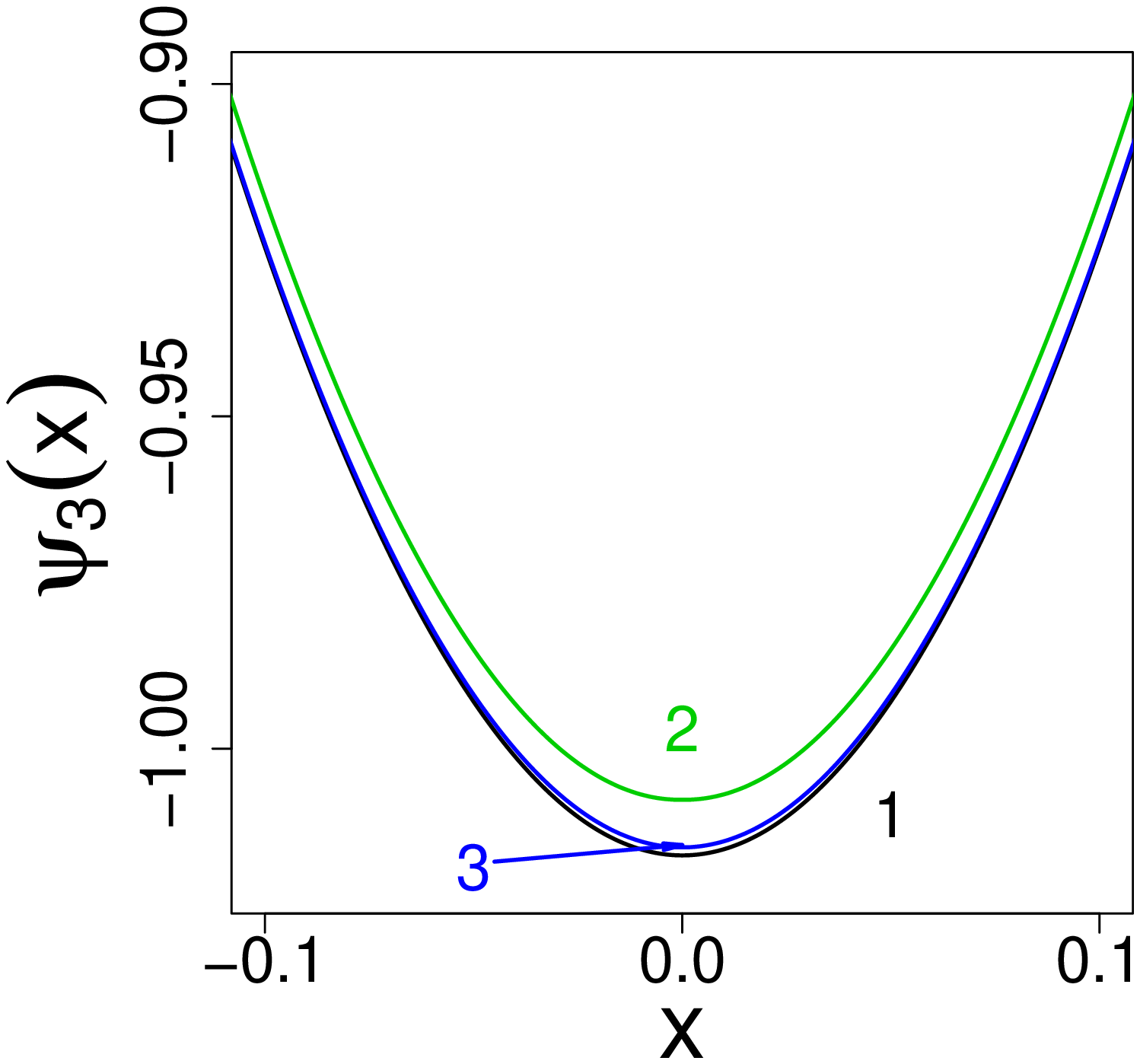}
\includegraphics[width=55mm,height=55mm]{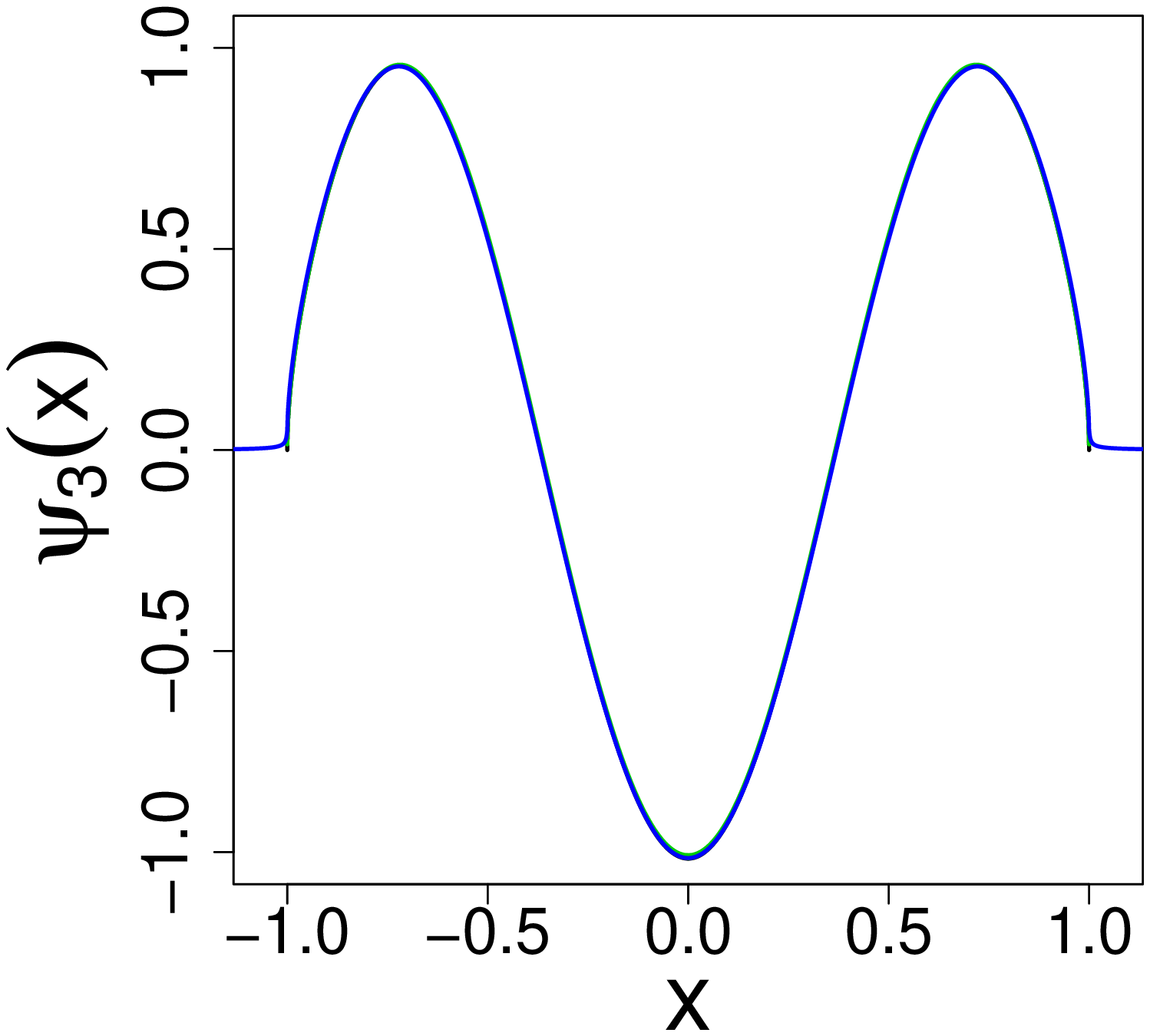}
\includegraphics[width=55mm,height=55mm]{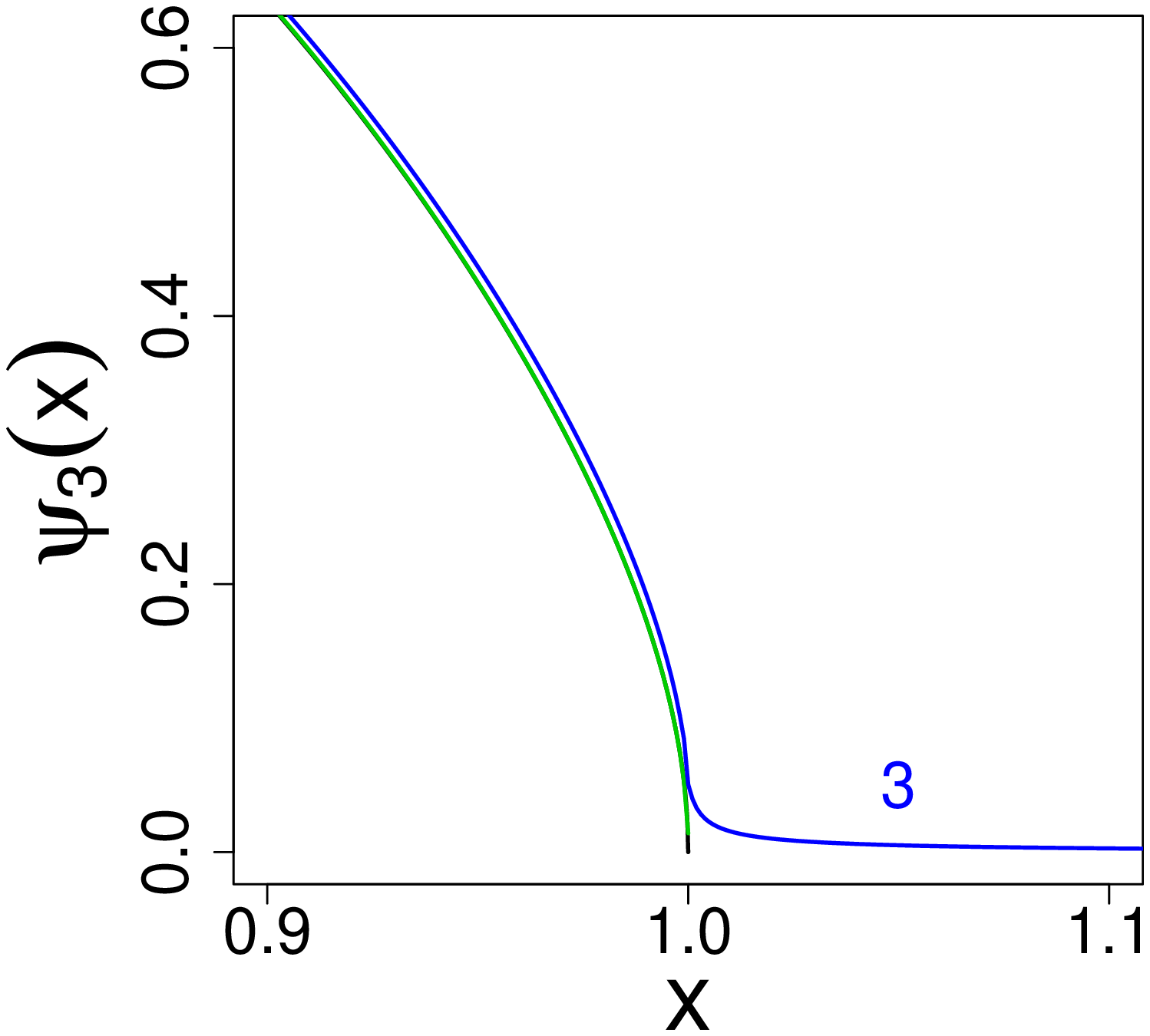}
\caption{An approximation of the third eigenfunction $\psi _3(x)$. Numbers refer to: $1$ - $\psi _3\sim C w_{500}(x)\sqrt{1-x^2}$, $2$  -  $\psi _3(x)$\
 according to  \cite{K}, $3$ - finite  $V_0=500$  Cauchy well ground state,  \cite{ZG}. Left panel: enlargement of the vicinity
of the minimum. Right panel - enlargement of the vicinity of $x=1$.}
\end{center}
\end{figure}

\begin{figure}[h]
\begin{center}
\centering
\includegraphics[width=70mm,height=70mm]{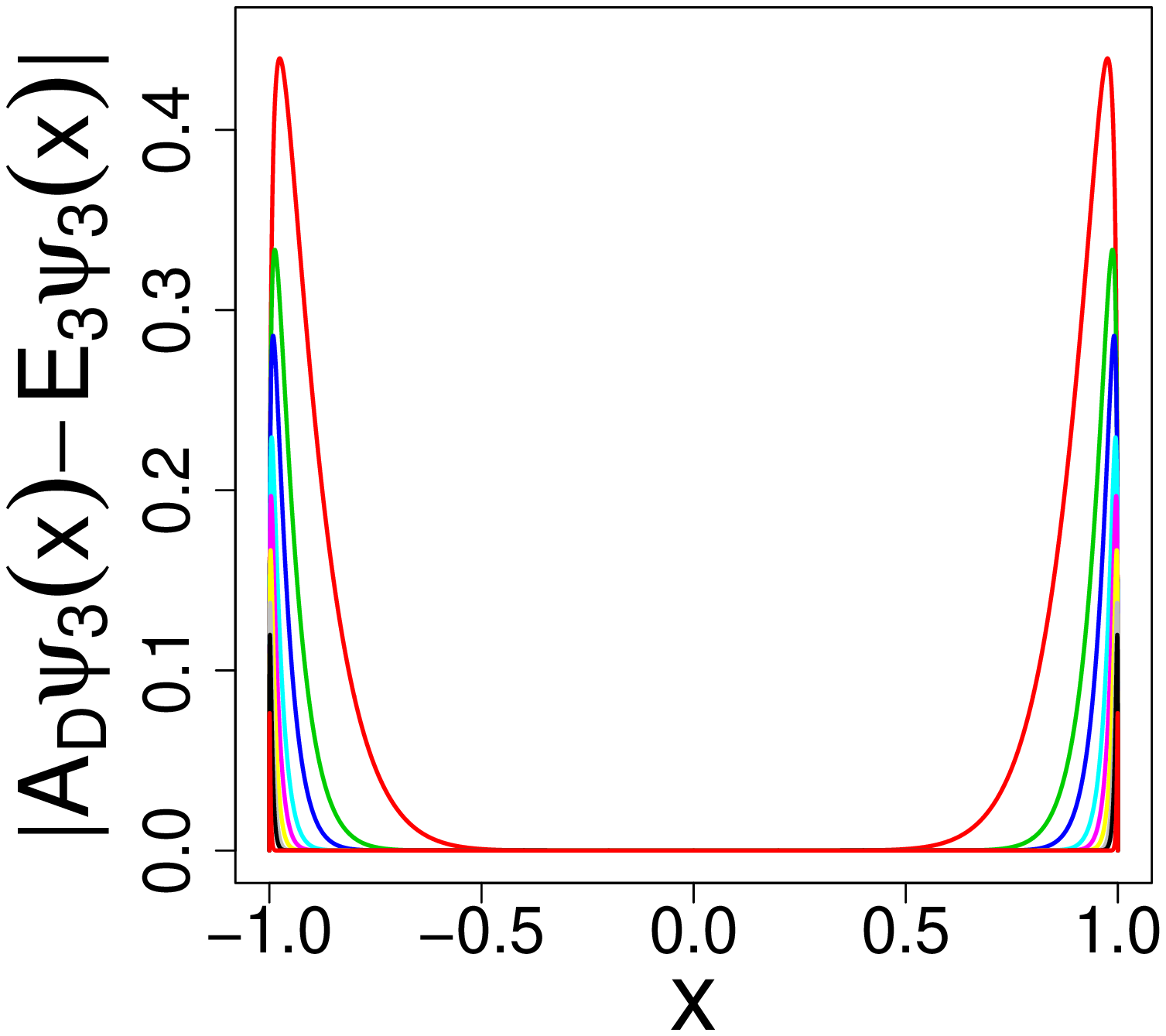}
\includegraphics[width=70mm,height=70mm]{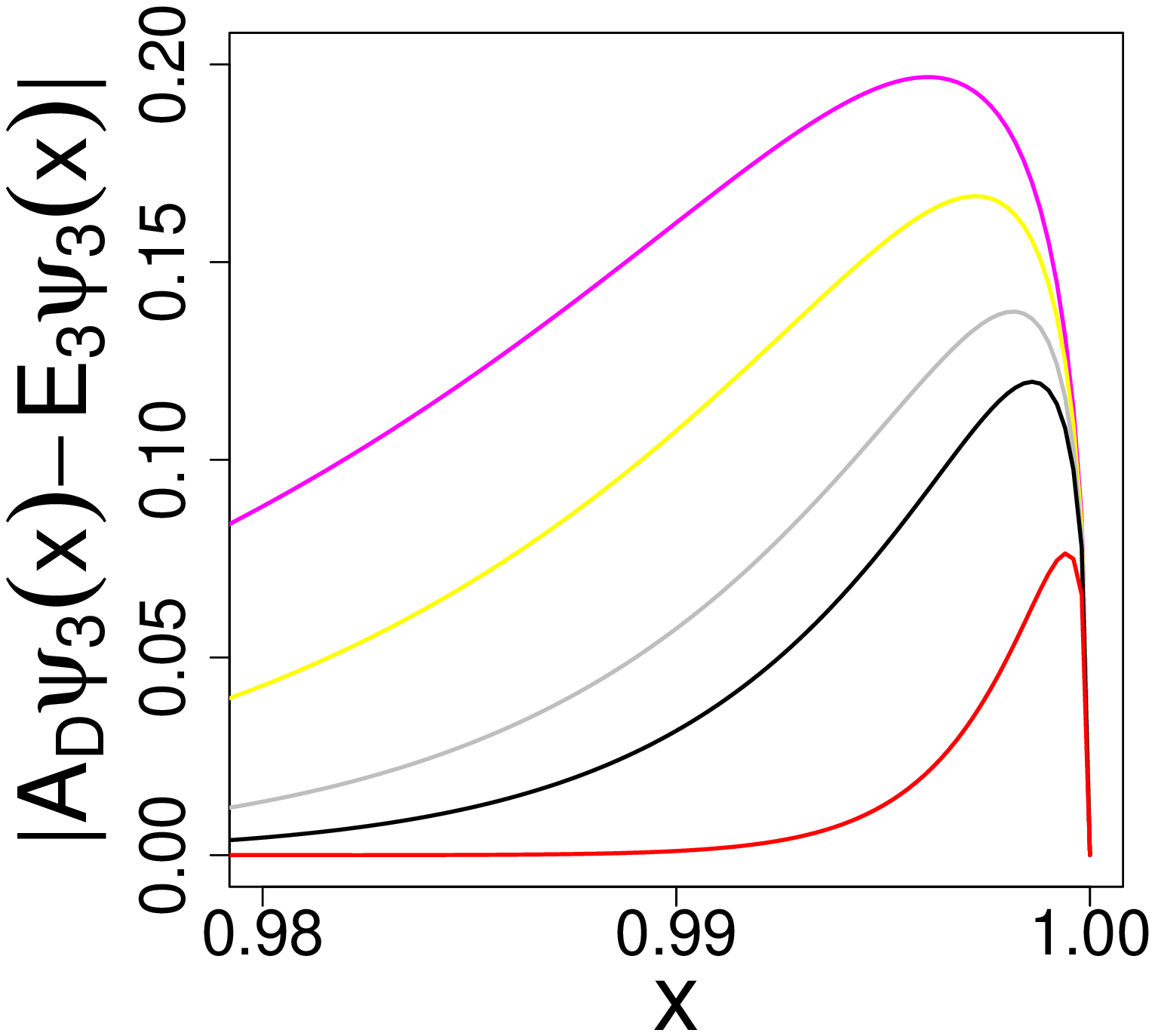}
\caption{$|A_D\psi_3(x)-E_3\psi_3(x)|$ where  $\psi_3$  is a product  of   $C \sqrt{1-x^2}$  and a polynomial of degree $2n$, we depict  $2n= 10,20,30,50,70,100,150,200,500$.  Right panel
refers to  $2n= 70,100,150,200,500$. Note that for $2n=500$, we have $|A_D\psi _3(x) - E_3\psi _3(x)|<0.07$.}
\end{center}
\end{figure}

\begin{figure}[h]
\begin{center}
\centering
\includegraphics[width=55mm,height=55mm]{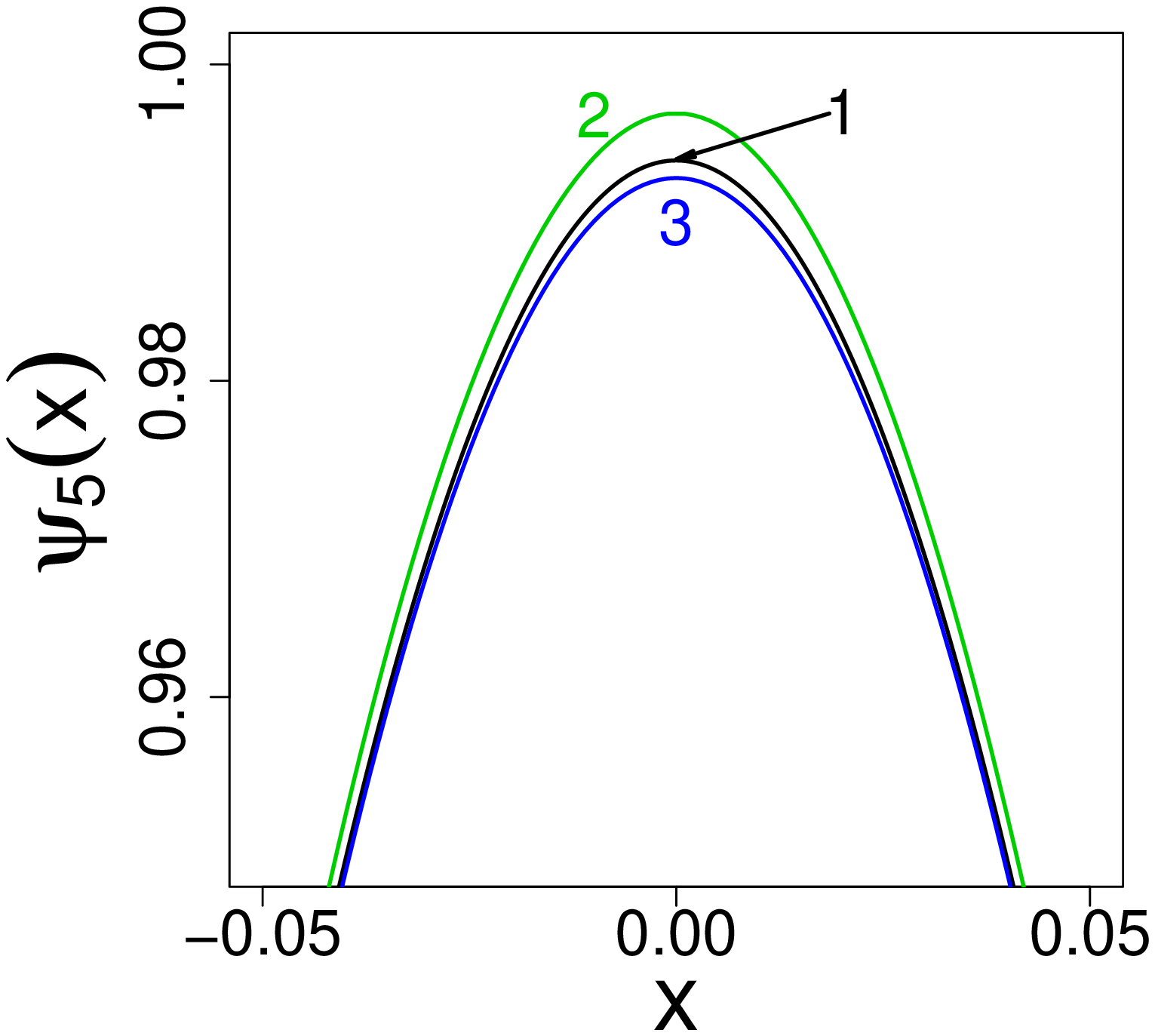}
\includegraphics[width=55mm,height=55mm]{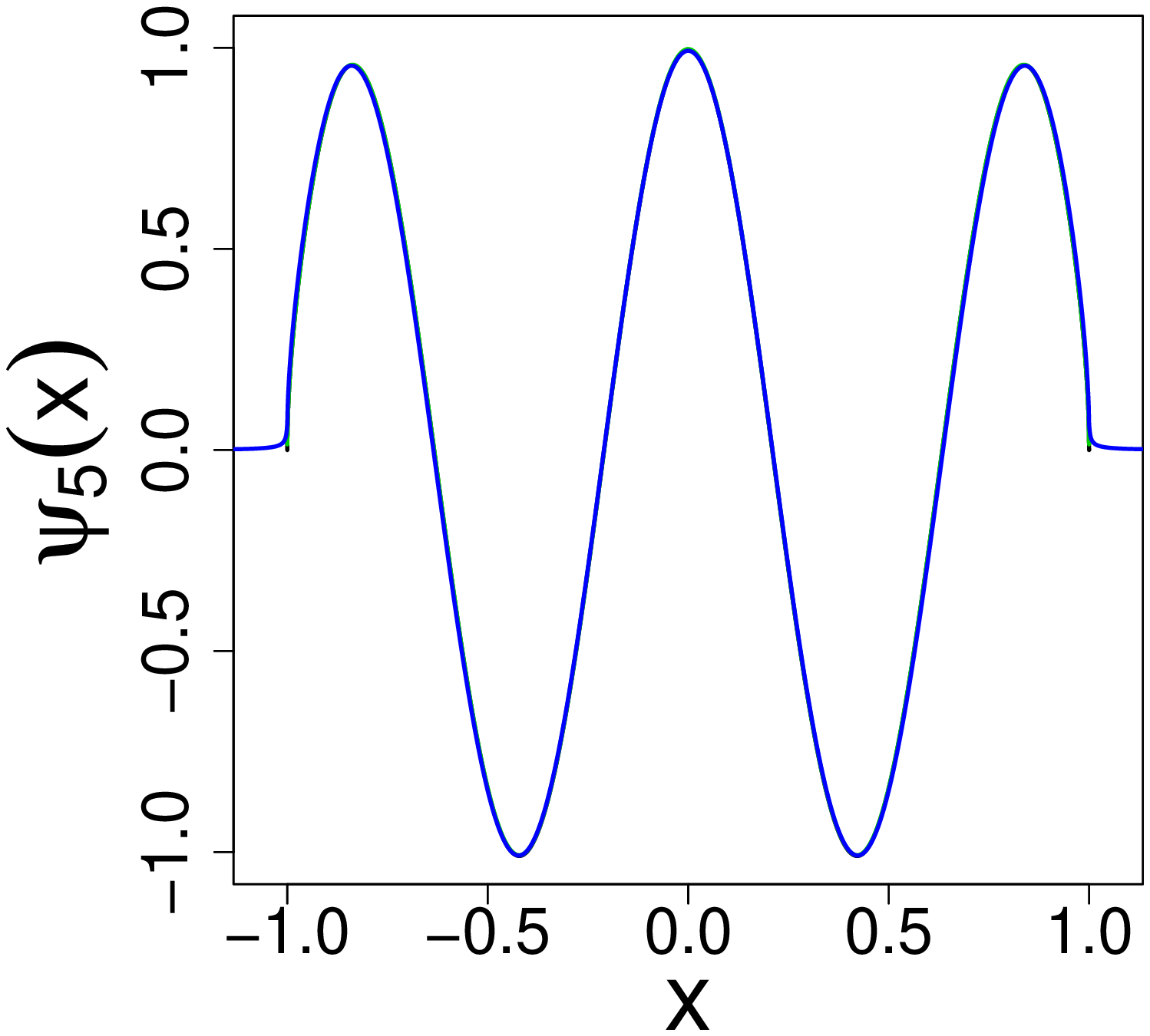}
\includegraphics[width=55mm,height=55mm]{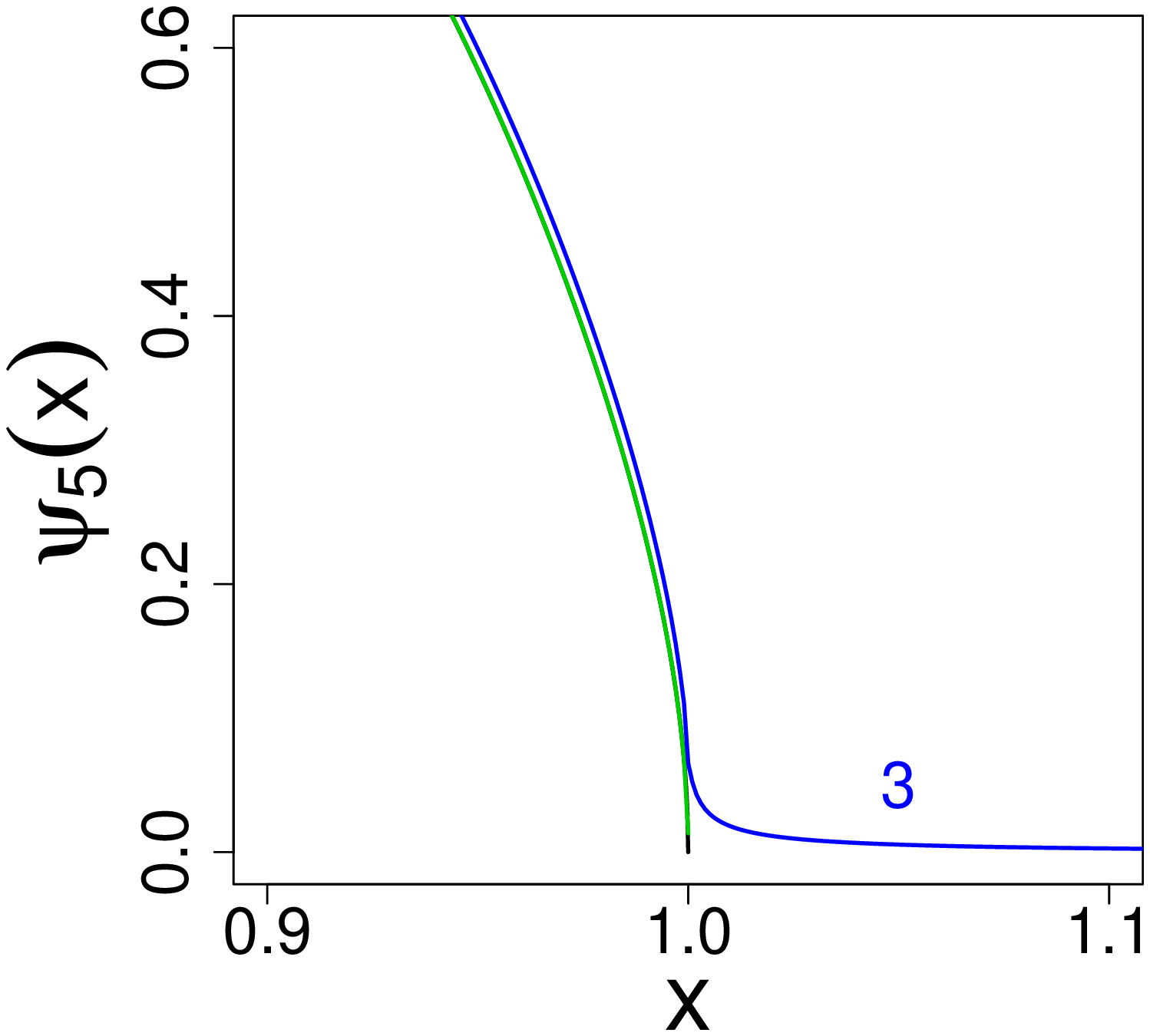}
\caption{For an approximate eigenfunction  $\psi _5(x)$  we display:  $1$  -  $2n=500$,  $2$ - $\psi _5(x)$ of Ref. \cite{K}, $3$ - the  fifth  finite  $V_0=500$ Cauchy well eigenfucntion (computed, but
not reproduced in \cite{ZG}). Left panel - minimum vicinity enlargement. Right  panel -  $x=1$ vicinity enlargement.}
\end{center}
\end{figure}

\begin{figure}[h]
\begin{center}
\centering
\includegraphics[width=70mm,height=70mm]{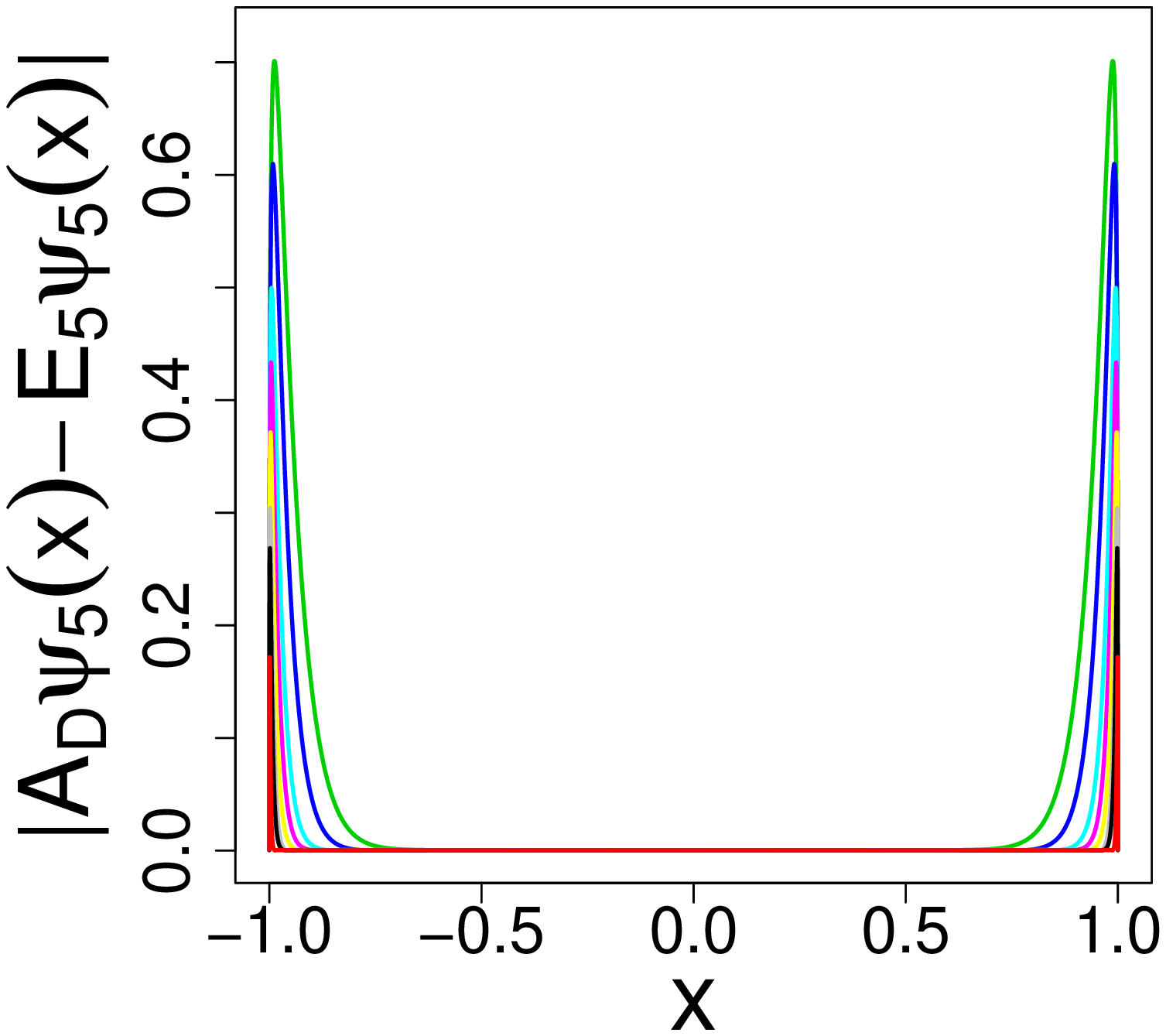}
\includegraphics[width=70mm,height=70mm]{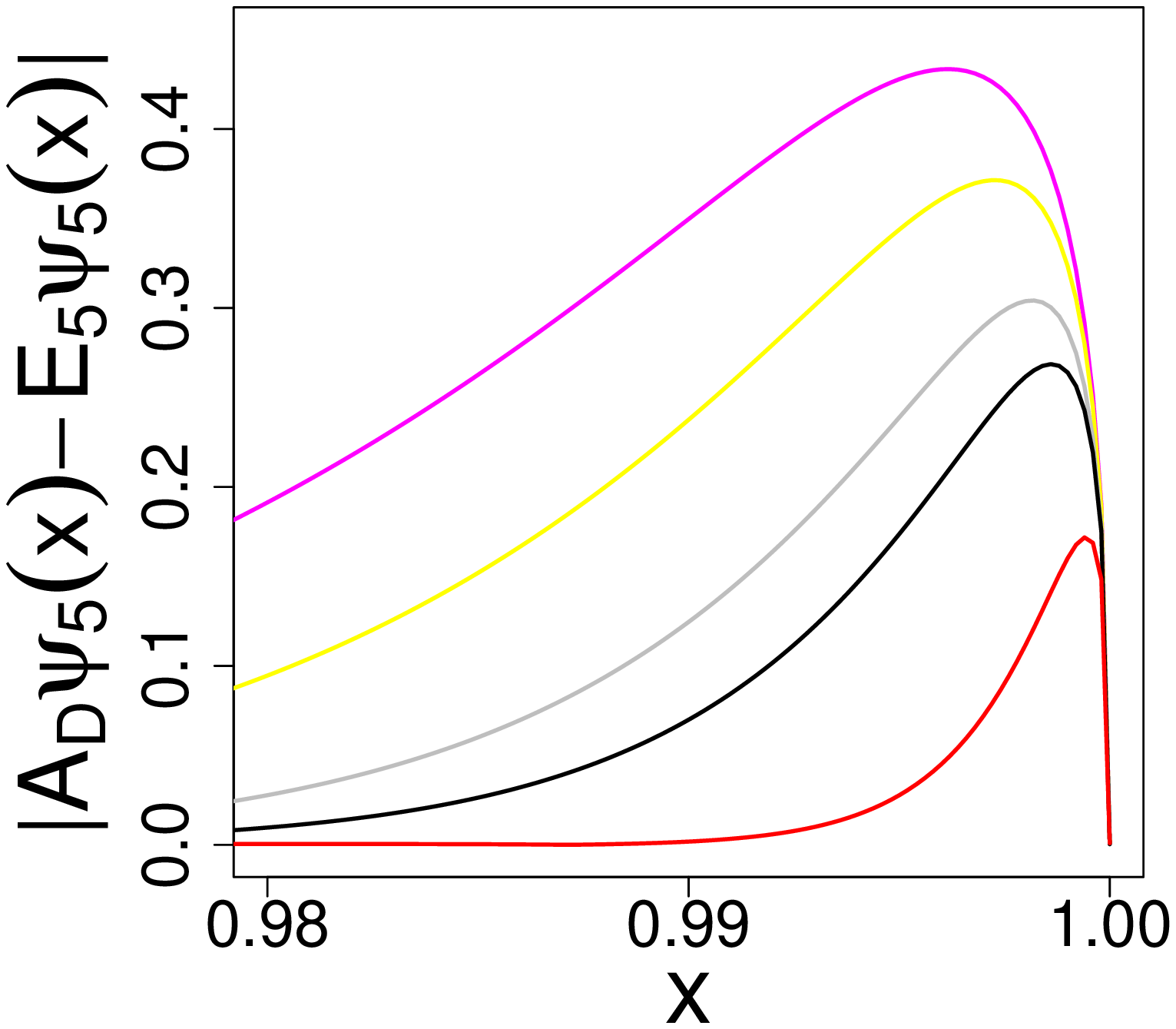}
\caption{$|A_D\psi_5(x)-E_5\psi_5(x)|$   where  $\psi_5(x)$  is inferred for  $2n=20,30,50,70,100,150,200,500$. Right  panel - $2n=70,100,150,200,500$.}
\end{center}
\end{figure}

\subsection{Odd eigenfunctions, $\psi _{2k}(x), \quad k\geq 1$.}

\begin{figure}[h]
\begin{center}
\centering
\includegraphics[width=55mm,height=55mm]{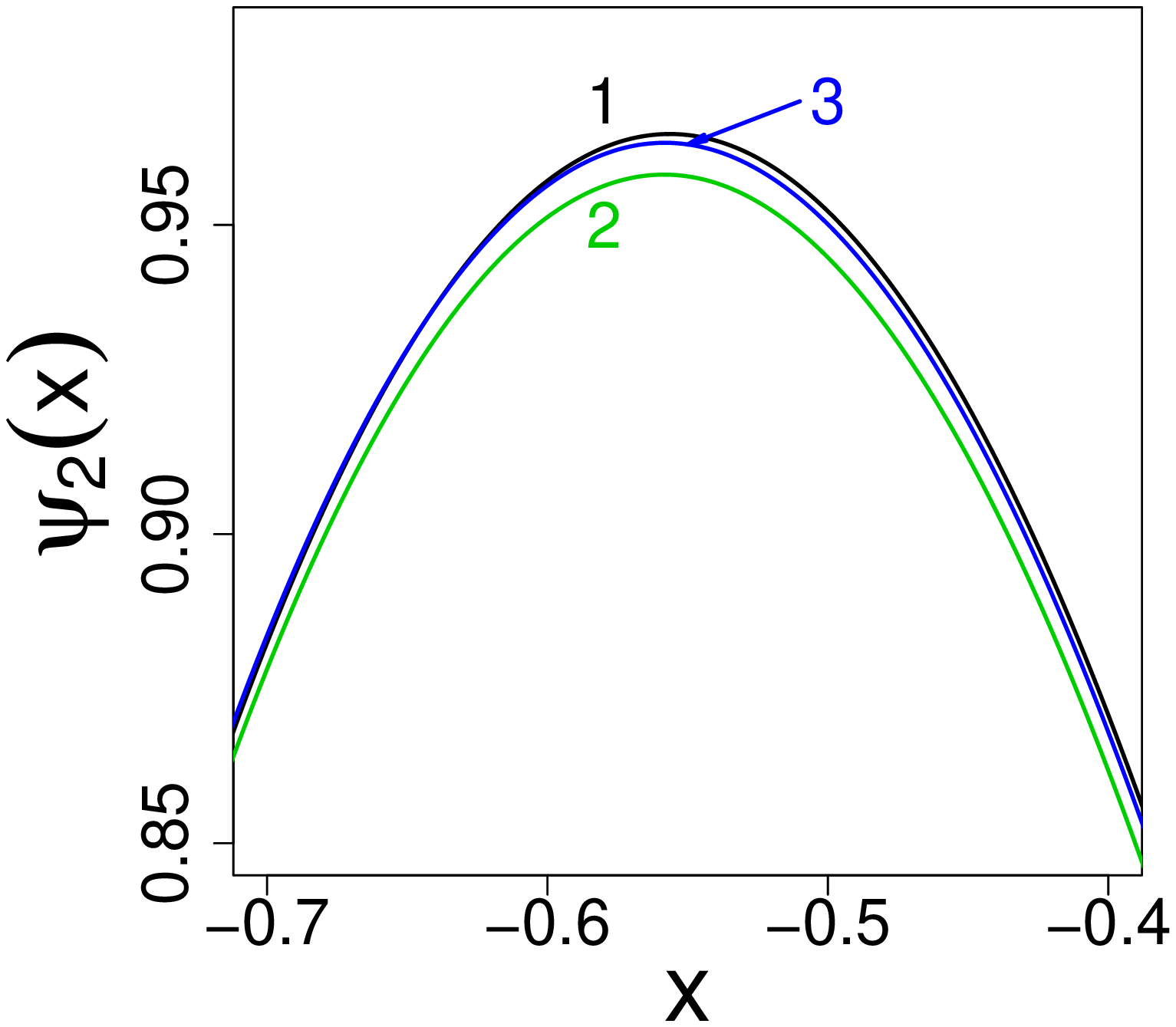}
\includegraphics[width=55mm,height=55mm]{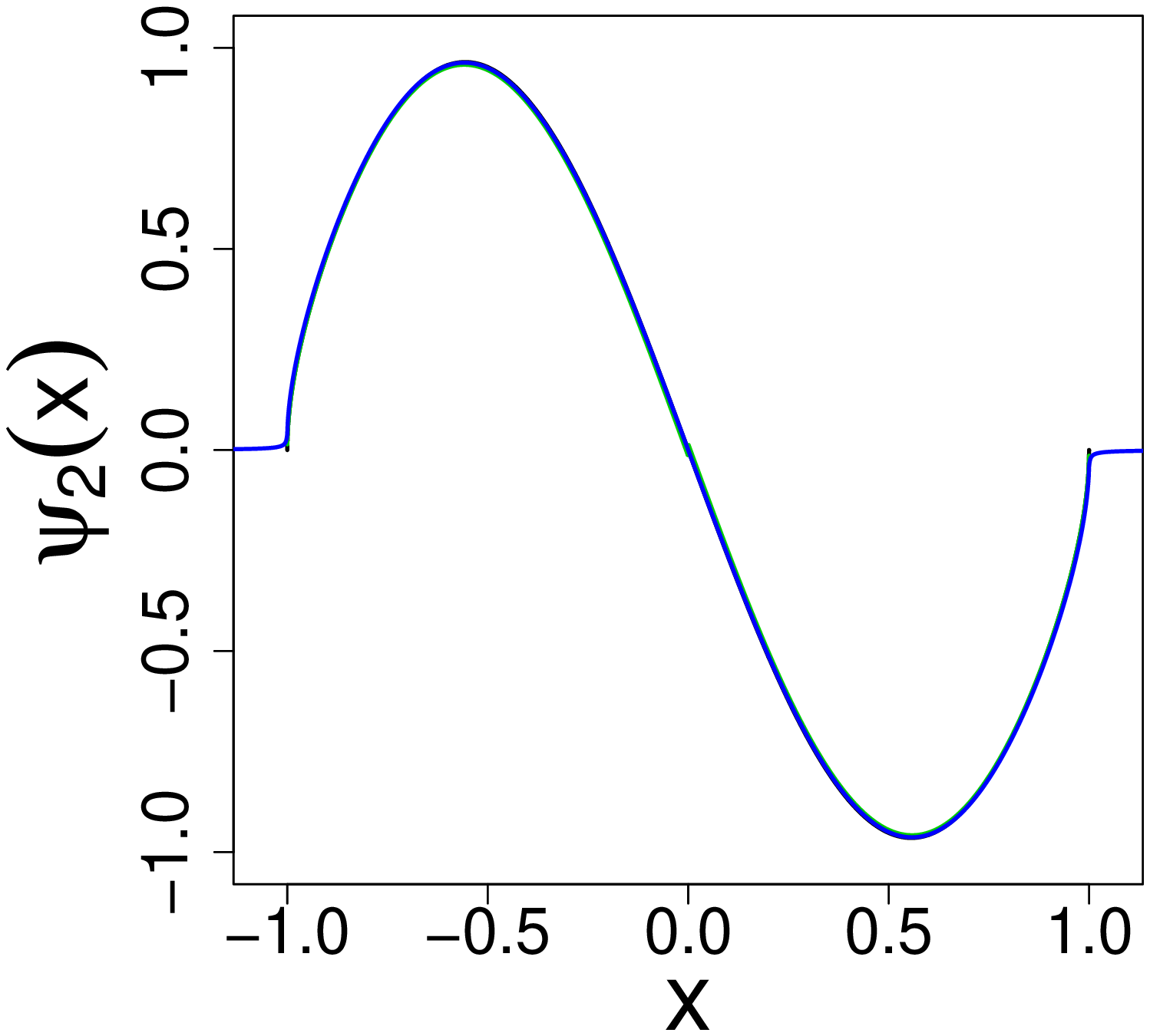}
\includegraphics[width=55mm,height=55mm]{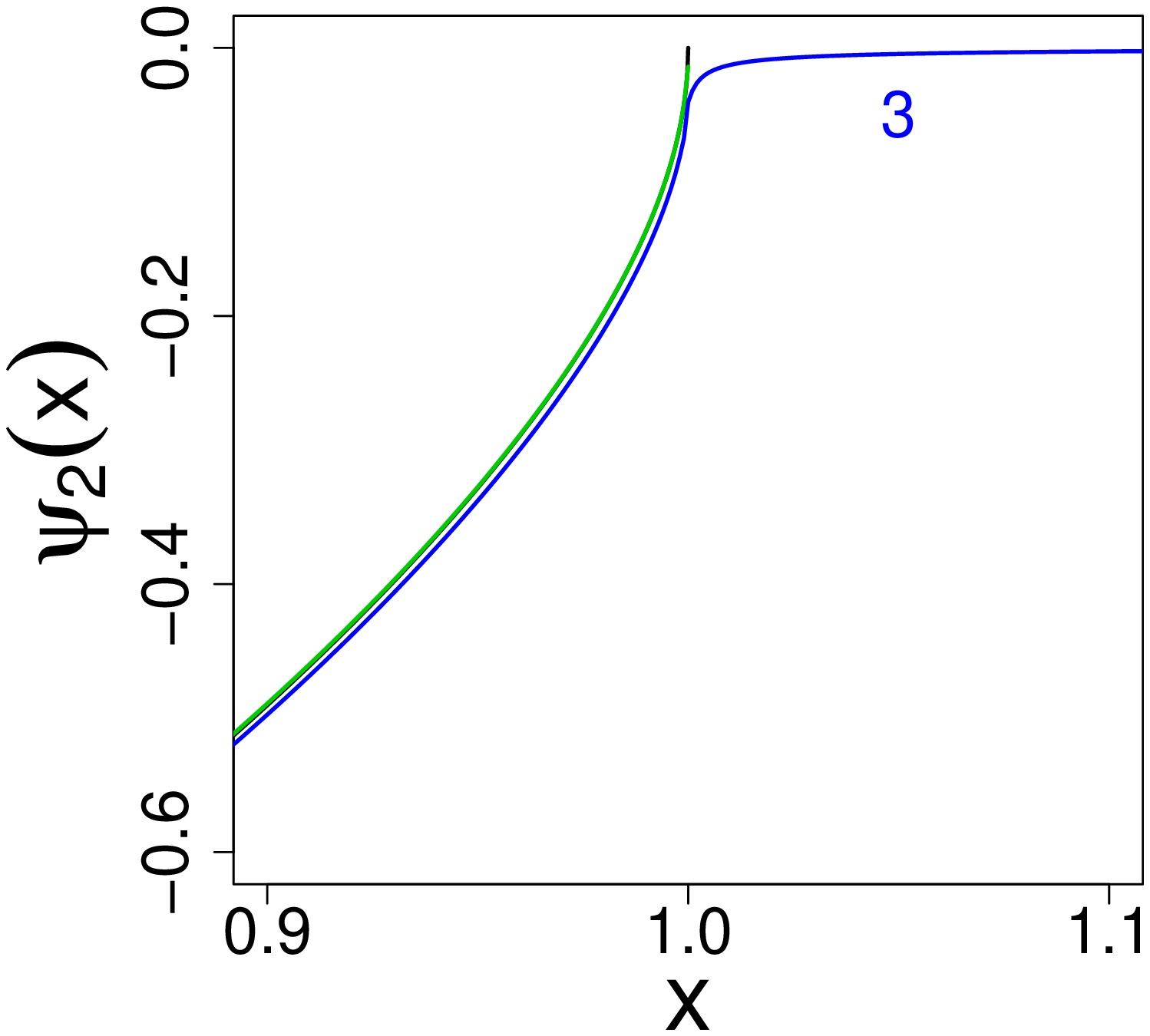}
\caption{$\psi _2(x)$, numbers refer to:  $1$ -  polynomial of degree  $501$, $2$ - according to   \cite{K}, $3$ - finite Cauchy well   $V_0=500$,
\cite{ZG}. Left  panel - enlargement of the minimum. Right panel - enlargement of the $x=1$ boundary.}
\end{center}
\end{figure}
\begin{figure}[h]
\begin{center}
\centering
\includegraphics[width=70mm,height=70mm]{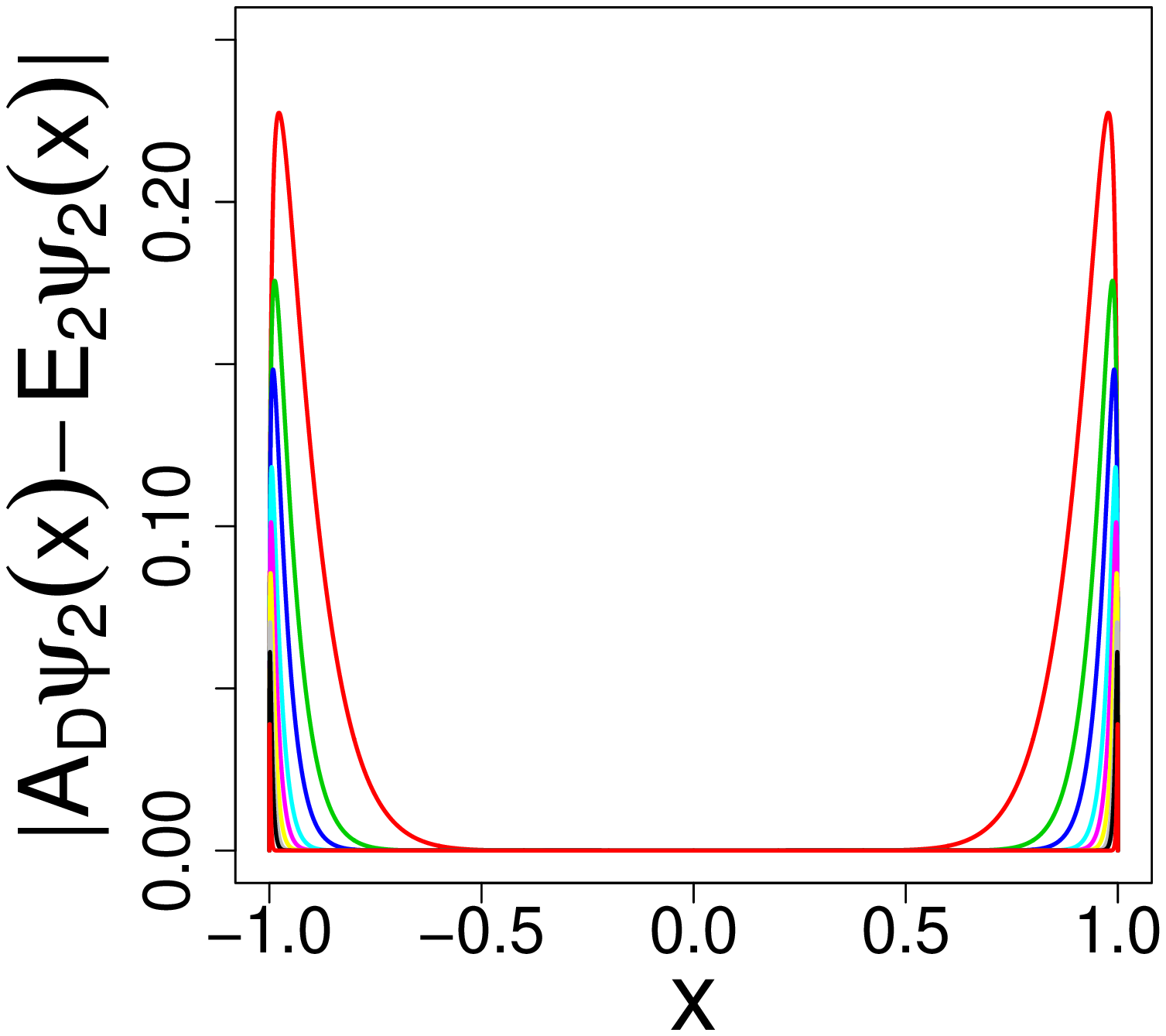}
\includegraphics[width=70mm,height=70mm]{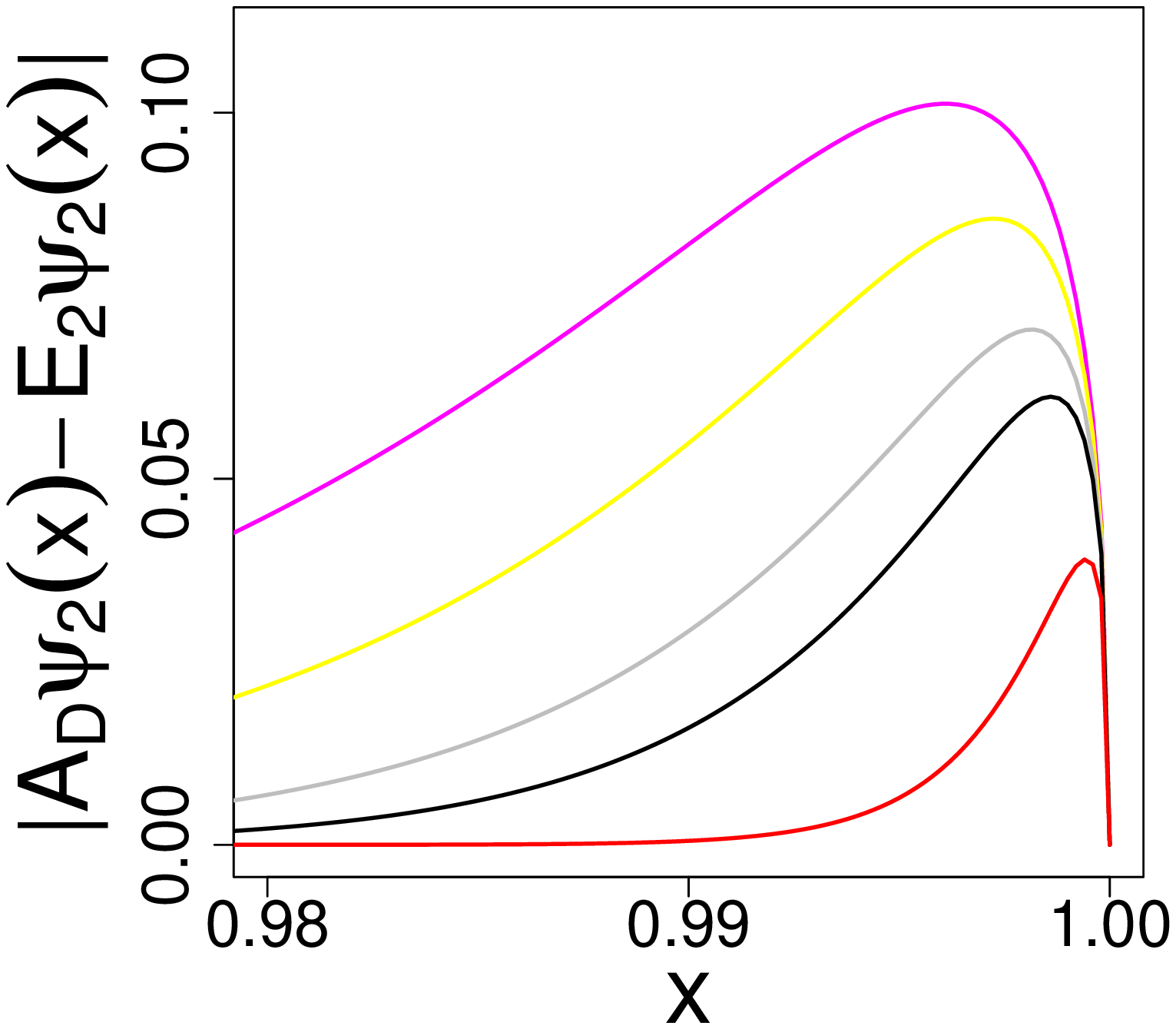}
\caption{$|A_D\psi_2(x)-E_2\psi_2(x)|$,   for  polynomial degrees $2n+1= 11,21,31,51,71,101,151,201,501$.
Right panel  - $2n+1= 71,101,151,201,501$.}
\end{center}
\end{figure}

In the notation of Section III.A, the odd eigenfunctions  case i can be  handled by  invoking:
\be
B_Dx\sqrt{1-x^2}=-\frac{2}{\pi}\frac{x\sqrt{1-x^2}}{1-x^2}+2x,
\ee
\be
B_Dx^3\sqrt{1-x^2}=-\frac{2}{\pi}\frac{x^3\sqrt{1-x^2}}{1-x^2}+2x\left(-\frac{1}{2}+2x^2\right),
\ee
\be
B_Dx^5\sqrt{1-x^2}=-\frac{2}{\pi}\frac{x^5\sqrt{1-x^2}}{1-x^2}+2x\left(-\frac{1}{8}-2\cdot\frac{1}{2}x^2+3x^4\right).
\ee
\be
B_Dx^7\sqrt{1-x^2}=-\frac{2}{\pi}\frac{x^7\sqrt{1-x^2}}{1-x^2}+2x\left(-\frac{1}{16}-2\cdot\frac{1}{8}x^2-3\cdot\frac{1}{2}x^4+4x^6\right).
\ee
i.e.
\be
B_Dx^{2n+1}\sqrt{1-x^2}=-\frac{2}{\pi}\frac{x^{2n+1}\sqrt{1-x^2}}{1-x^2}+2x(c_{2n}+2c_{2n-2}x^2+3c_{2n-4}+\ldots+(n+1)c_0x^{2n}),
\ee
where  $c_{2n}$  are Taylor series  coefficients for  $\sqrt{1-x^2}$.  We are interested in odd eigenfunctions and
seek them in the form:
\be
\psi(x)=C\sqrt{1-x^2}\sum_{n=0}^\infty\beta_{2n+1}x^{2n+1},\qquad \beta_1=1.
\ee

As in  the  case of even functions, we look for solutions of the eigenvalue problem  $A_D\psi (x)=  E\, \psi (x)$, so arriving at
\be
\sum_{n=0}^\infty\beta_{2n+1}\sum_{k=0}^n\frac{(2k)!(2n+2-2k)}{(1-2k)(k!)^24^k}x^{2n-2k+1}=E\sum_{k=0}^\infty\frac{(2k)!}{(1-2k)(k!)^24^k}x^{2k}\sum_{n=0}^\infty\beta_{2n+1}x^{2n+1},
\ee
and next at
\be
\sum_{k=0}^\infty\sum_{n=k}^\infty\beta_{2n+1}\frac{(2k)!(2n+2-2k)}{(1-2k)(k!)^24^k}x^{2n-2k+1}=\sum_{k=0}^\infty\sum_{n=0}^\infty E\,\beta_{2n+1}\frac{(2k)!}{(1-2k)(k!)^24^k}x^{2k+2n+1}.
\ee
Additionally, we  impose our  boundary  condition
\be
\lim\limits_{x\to\pm 1}A_D\psi(x)=0.
\ee

\begin{figure}[h]
\begin{center}
\centering
\includegraphics[width=55mm,height=55mm]{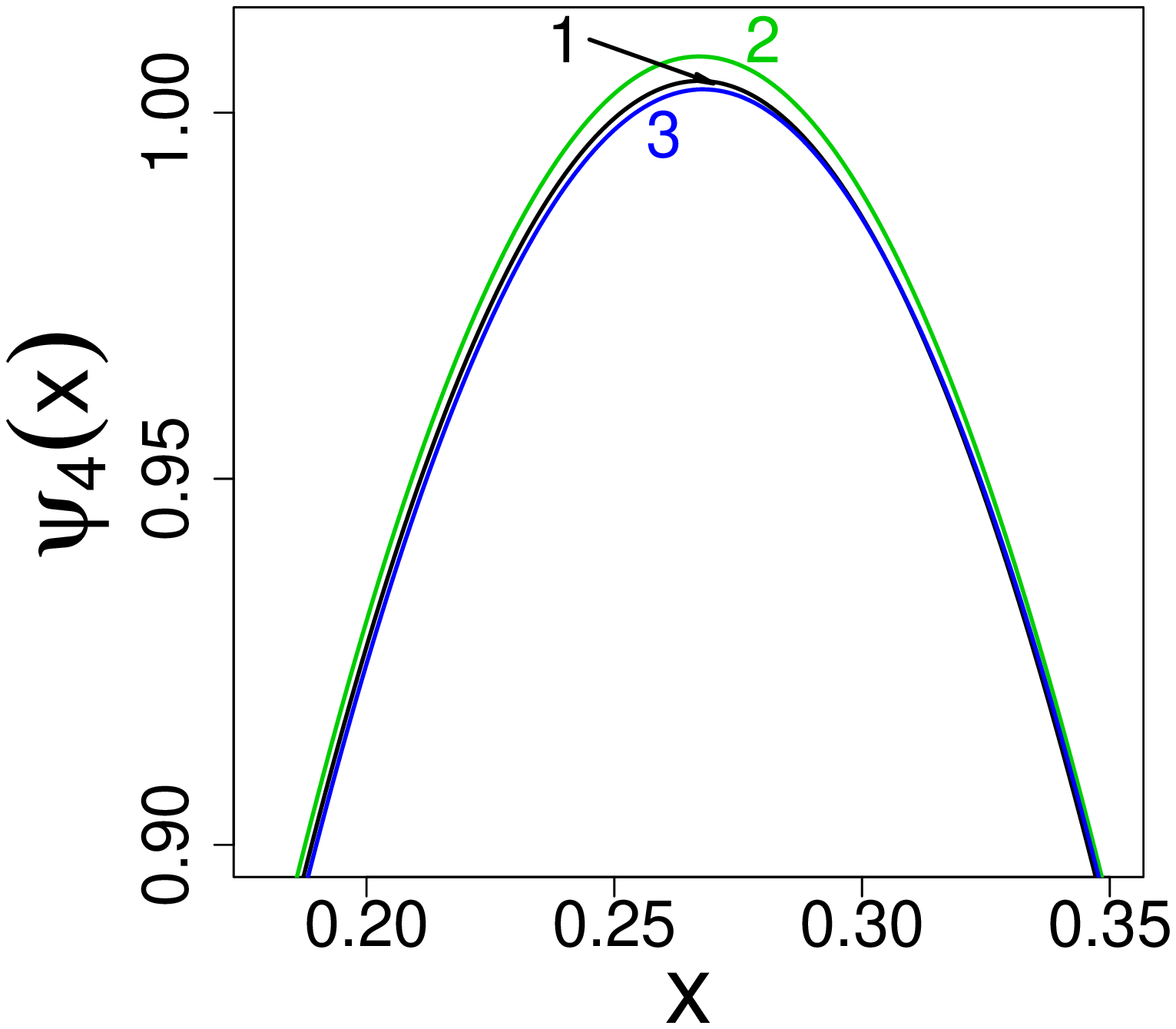}
\includegraphics[width=55mm,height=55mm]{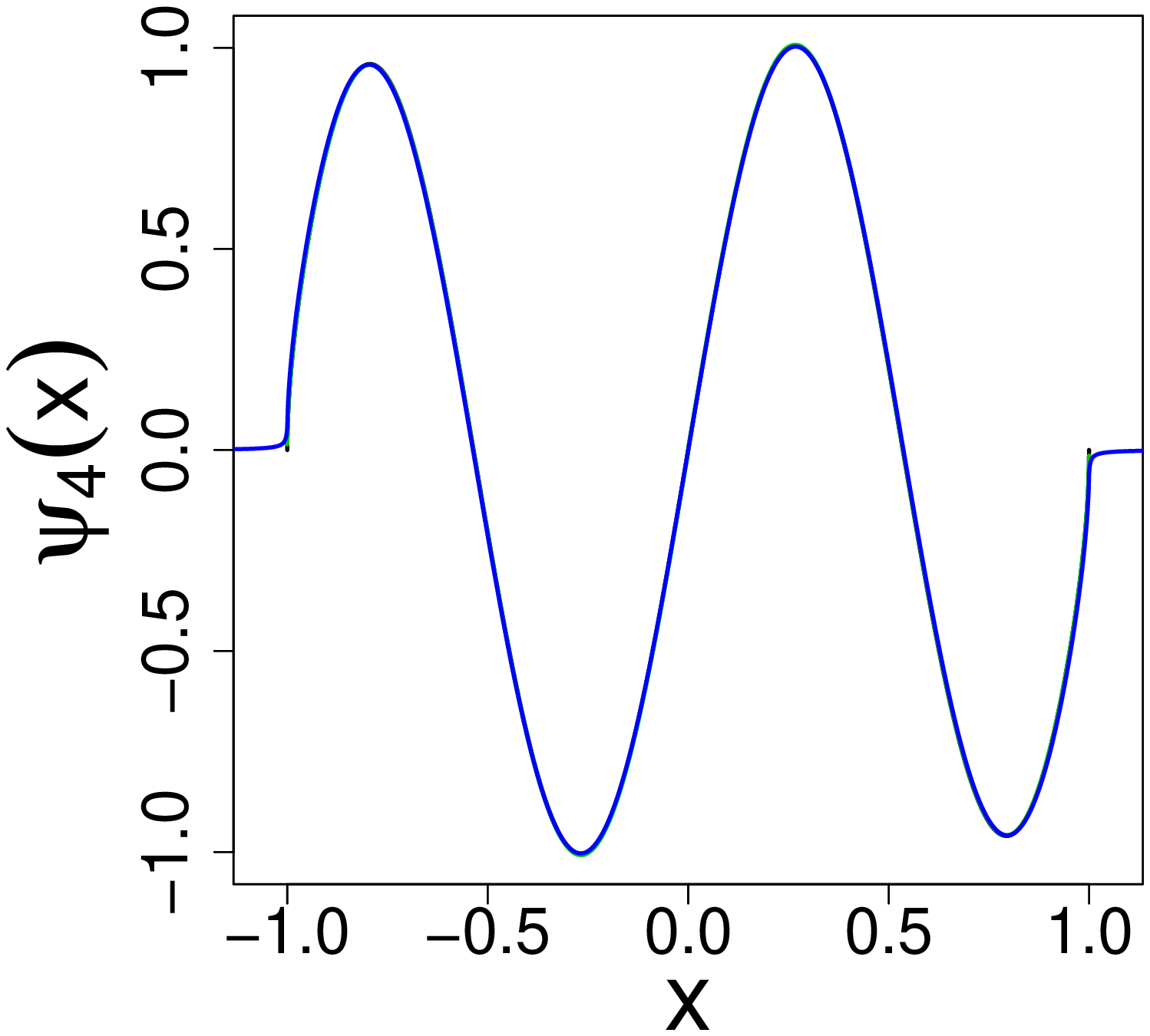}
\includegraphics[width=55mm,height=55mm]{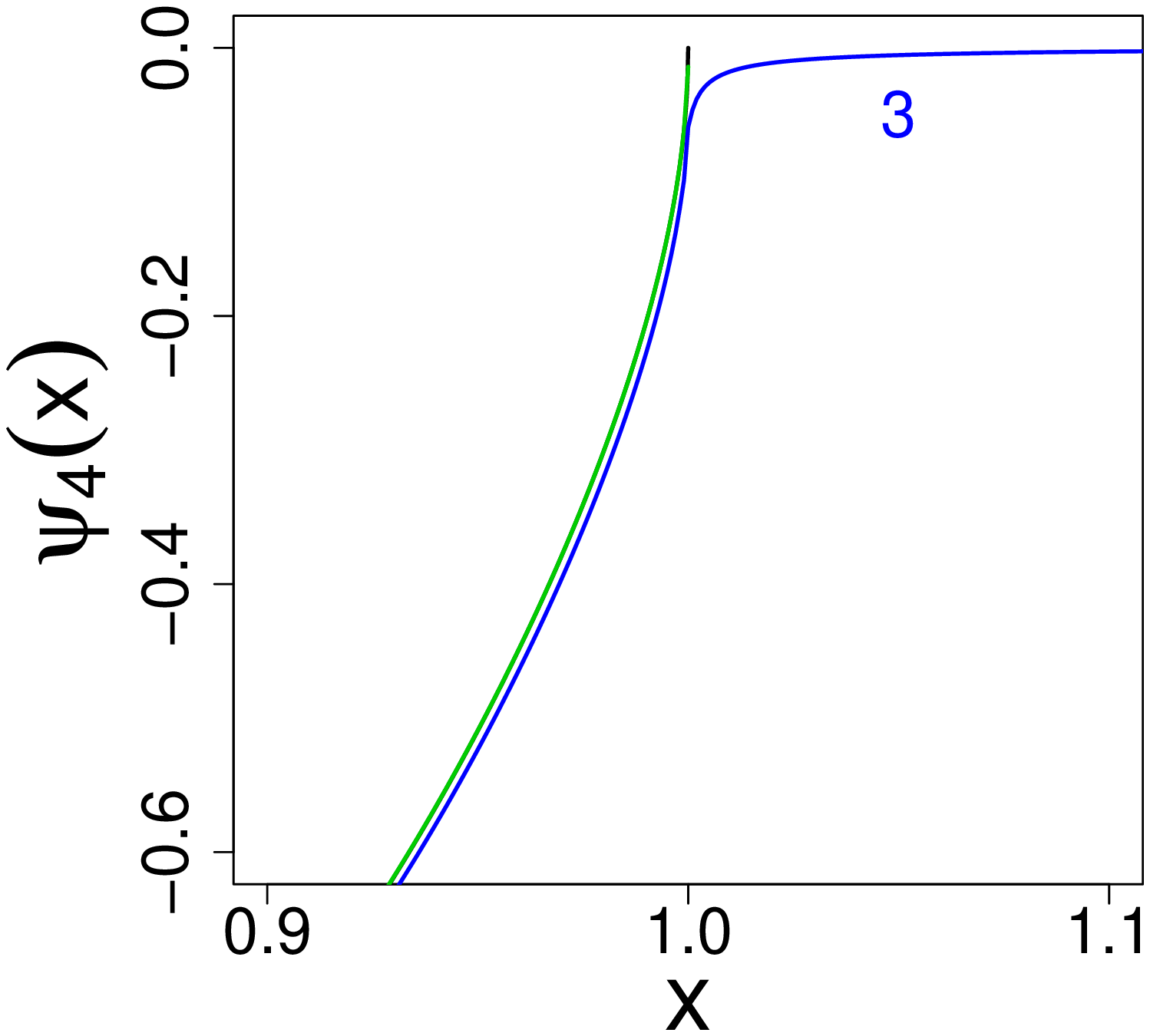}
\caption{$\psi _4(x)$, numbers refer to: $1$ - polynomial of degree  $501$, $2$  - $\psi _4(x)$ according to  \cite{K}, $3$ -
finite Cauchy well $V_0=500$ outcome, \cite{ZG}. Left  panel - minimum enlargememnt. Right panel -  right boundary enlargement.}
\end{center}
\end{figure}
\begin{figure}[h]
\begin{center}
\centering
\includegraphics[width=70mm,height=70mm]{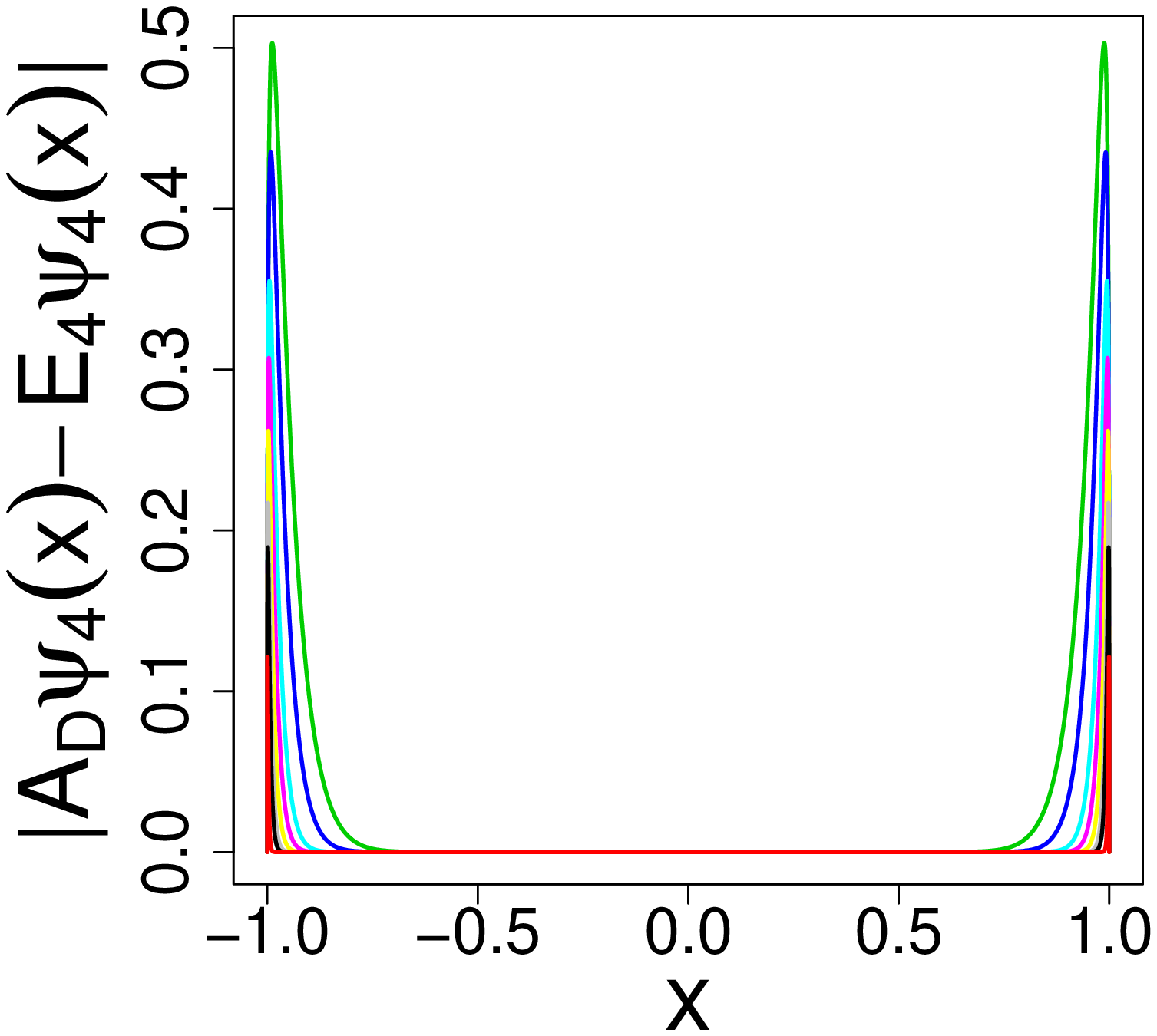}
\includegraphics[width=70mm,height=70mm]{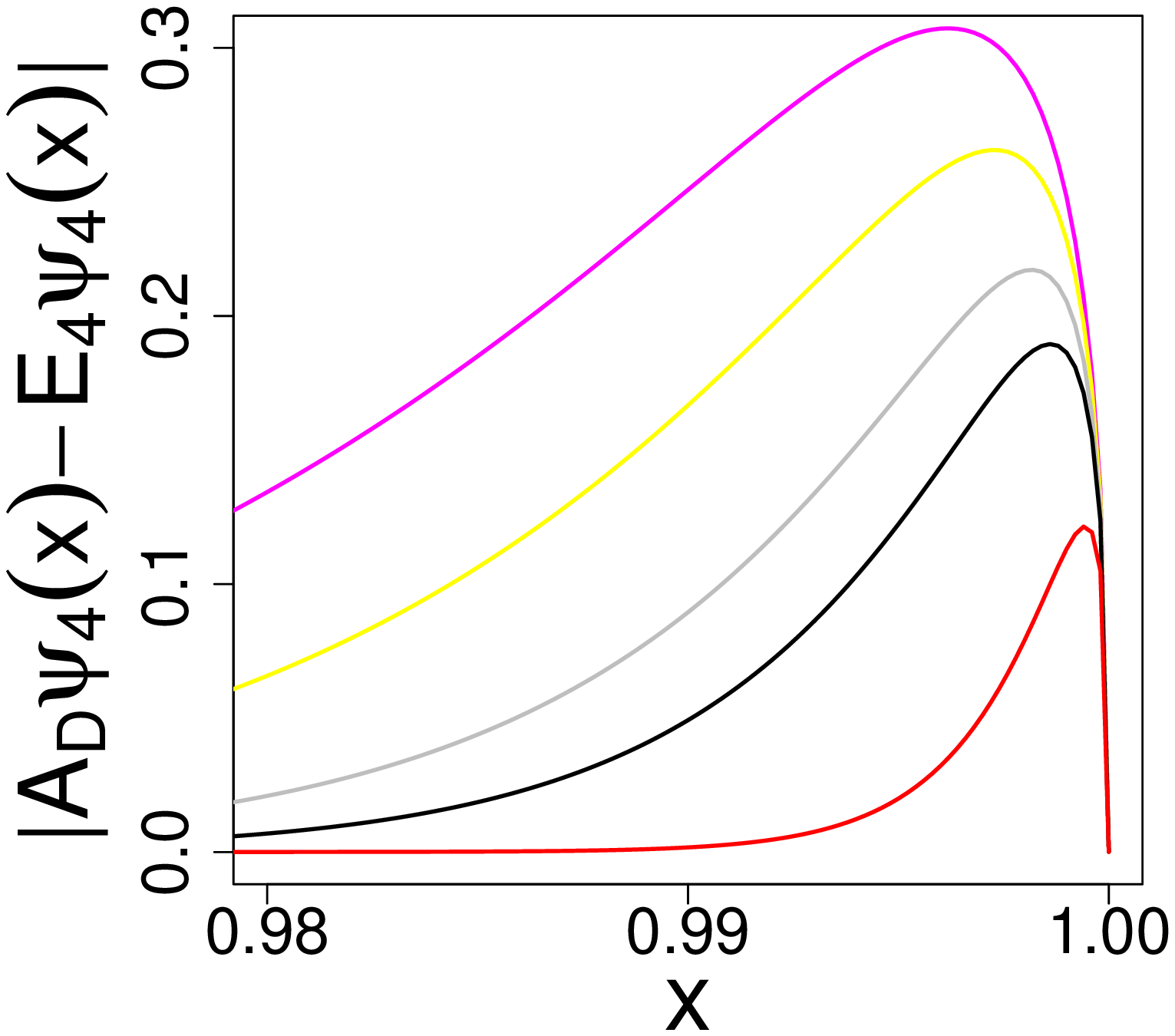}
\caption{$|A_D\psi_4(x)-E_4\psi_4(x)|$  where polynomial degrees are odd, $2n+1= 21,31,51,71,101,151,201,501$.
Right panel: $2n+1= 71,101,151,201,501$.}
\end{center}
\end{figure}

The procedure adopted to find polynomial approximations of eigenfunctions and eigenvalues in the even case,  can be extended to the
 odd case as well.
Let
\be
b_{k,n}=(2n+2-2k)c_k=\frac{(2k)!(2n+2-2k)}{(1-2k)(k!)^2 4^k},\qquad n\geqslant k,
\ee
where
\be
c_k=\frac{(2k)!}{(1-2k)(k!)^2 4^k}.
\ee
The linear system of equations from which all $\beta _{2n+1}$ and $E$ are to be inferred has the form   (we set $\beta_1=1$)
\begin{eqnarray}
\sum\limits_{k=i}^n \beta_{2k+1}b_{k-i,k}=E\sum\limits_{k=0}^i \beta_{2k+1}c_{i-k},\qquad i=0,1,\ldots,n-1,\nonumber\\
\sum\limits_{m=0}^n \left(\beta_{2m+1}\sum\limits_{k=0}^m b_{k,m}\right)=0.
\end{eqnarray}
Like in case of (50), the present system of equations has infinitely many solutions, real and complex-valued for each fixed $n$.
As before,  in the set of real solutions, an ordering relation is set by referring to an increasing order
of computed eigenvalues $E_{2k}\quad k\geq 1$.

\begin{table}[h]
\begin{center}
\begin{tabular}{|c||c|c|c|c|c|}
  \hline
  n & $E_1$ & $E_2$ & $E_3$ & $E_4$ & $E_5$ \\
  \hline
  $7$ & 1.160614 & 2.768252 & 4.351150 & 5.946117 & 7.337136 \\
  $8$ & 1.159993 & 2.765561 & 4.344362 & 5.934918 & 7.584192 \\
  $9$ & 1.159555 & 2.763594 & 4.339381 & 5.928041 & 7.512343 \\
  $10$ & 1.159234 & 2.762114 & 4.335613 & 5.922546 & 7.509991 \\
  $15$ & 1.158447 & 2.758299 & 4.325845 & 5.907535 & 7.485347 \\
  $20$ & 1.158159 & 2.756826 & 4.322066 & 5.901342 & 7.475242 \\
  $25$ & 1.158022 & 2.756110 & 4.320233 & 5.898233 & 7.470144 \\
  $30$ & 1.157948 & 2.755709 & 4.319211 & 5.896463 & 7.467238 \\
  $35$ & 1.157902 & 2.755463 & 4.318584 & 5.895363 & 7.465432 \\
  $40$ & 1.157872 & 2.755301 & 4.318173 & 5.894634 & 7.464235 \\
  $45$ & 1.157852 & 2.755188 & 4.317889 & 5.894127 & 7.463403 \\
  $50$& 1.157837 & 2.755107 & 4.317685 & 5.893759 & 7.462802 \\
  $75$& 1.157802 & 2.754913 & 4.317196 & 5.892875 & 7.461356 \\
  $100$& 1.157789 & 2.754844 & 4.317024 & 5.892559 & 7.460842 \\
  $150$& 1.157781 & 2.754795 & 4.316900 & 5.892331 & 7.460473 \\
  $200$& 1.157778 & 2.754777 & 4.316857 & 5.892251 & 7.460343 \\
  $250$& 1.157776 & 2.754769 & 4.316837 & 5.892214 & 7.460282 \\
  $ K $& 1.157773 &  2.754754 & 4.316801 & 5.892147 & 7.460175 \\
  \hline
\end{tabular}
\end{center}
\caption{We display the computed eigenvalues  $E_k, \quad k\geq 1$ under an assumption that polynomials of degree  $2n$ were employed in
the definition of even eigenfunctions and
 $2n+1$ for odd eigenfunctions. The capital K in the last line indicates data collected  from Ref. \cite{KKMS}.}
\end{table}

Skipping unnecessary repetitions of previous  arguments, we display our findings concerning $\psi _2(x)$ and $\psi _4(x)$,
 together with estimates for $|A_D\psi (x)- E\, \psi (x)|$.  A comparison is made with previously reported results on the shape
of corresponding approximate eigenfunctions. It appears that
our method provides most accurate to date data for both approximate eigenfunctions and eigenvalues and provides the
sharpest to date  point-wise estimates for $|A_D\psi (x)- E \psi (x)|$.

For completeness, in  Table III we report first five  computed eigenvalues ordered against the approximating polynomial
 degree and compare them with those obtained by other arguments in Refs. \cite{K,KKMS}.
It is clear that by increasing the polynomial degree $n$  we can achieve  still higher finesse level of  a computational
 accuracy with which both eigenfunctions and eigenvalues  can be retrieved. It is seen that with the growth $n$,
 there are definite stabilization symptoms in the  numerical outcomes for
the  eigenvalues.  We would like to point out that to increase  the reproduction accuracy
of higher "true" eigenfunctions, one should  pass to higher than $n=500$ polynomial degrees. The same pertains to the $0.01$
upper  bound for $A_D\psi (x)- E\psi (x)|$ if $\psi (x)$ is an approximation of the ground state. To push that bound closer to $0$, higher polynomiola
  degrees  are necessary.

\section{Outlook}

In the present paper we were largely motivated by: (i) on the one hand -successes in the approximate evaluation of eigenfunctions
 and eigenvalues for  the infinite Cauchy well problem  \cite{K,KKMS,ZG}, (ii) various drawbacks in the physics-motivated procedures to solve
that eigenvalue problem, summarized in \cite{GS,ZG}.   It has been often mentioned that the "true"
eigenfunctions show a striking similarity to  trigonometric  sine or cosine functions (identified as eigenfunctions of the
standard  Laplacian in the interval) when away from the boundaries of $D$, while  their fall-off towards zero
at the boundaries should be similar to $\sqrt{1-x^2}$.  Our trial function considerations of Section II proved that the
  trigonometric connection is somewhat deceiving, since  the  square root of a trigonometric  function has been involved.
In Section III we have resolved the away-from-the-boundary behavior by means of truncated polynomial expansions, that bear no obvious
similarity, neither to trigonometric functions nor to their square roots. Things became more complicated and subtle, since the polynomial shapes
actually dictate minute details of the eigenfunctions fall-off to $0$ at the boundaries. In this connection, we point out  the peculiar
fall-off of approximating polynomials around $\pm 1$, as depicted in Fig. 6.


\begin{thebibliography}{50}
\bibitem{GS} P. Garbaczewski and  V. Stephanovich, \textit{L\'{e}vy flights and nonlocal quantum dynamics}, J. Math. Phys. \textbf{54}, (2013) 072103.
\bibitem{ZRK}  A. Zoia, A. Rosso and M. Kardar, \textit{Fractional Laplacian in a bounded domain},
 Phys. Rev. E \textbf{76}, 021116,(2007).
\bibitem{jeng}  M. Jeng et al., \textit{On the nonlocality of the fractional
Schr\"{o}dinger equation}, J. Math. Phys. {\bf 51}, 062102 (2010).
\bibitem{luchko} Y. Luchko, \textit{Fractional Schr\"{o}dinger equation for a particle
moving in a potential well}, J. Math. Phys. {\bf 54}, 012111, (2013).
\bibitem{getoor} R. K. Getoor, \textit{First pasaage times for symmetric stable processes in space}, Trans. Amer. Math. Soc. {\bf 101}, 75, (1961).
 \bibitem{gar} P. Garbaczewski and V.  Stephanovich, \textit{L\'{e}vy flights in inhomogeneous environments}, Physica A {389}, 4419, (2010).
\bibitem{lorinczi} J. L\"{o}rinczi and J. Ma{\l}ecki, \textit{Spectral properties   of the massless relativistic harmonic oscillator},
 J. Diff. Equations, {\bf 251}, 2846, (2012).
\bibitem{K} M. Kwa\'{s}nicki, \textit{Eigenvalues of the fractional Laplace operator in the interval}, J. Funct. Anal. \textbf{262}, 2379, (2012).
\bibitem{BK} R. Ba\~{n}uelos and  T. Kulczycki, \textit{The Cauchy process and the Steklov problem}, J. Funct. Anal. \textbf{211}, 355-423, (2004).
\bibitem{BKM} R. Ba\~{n}uelos, T. Kulczycki and  P. J. M\'{e}ndez-Hern\'{a}ndez, \textit{On the shape of the Ground State Eigenfunction for
Stable Processes}, Potential Analysis \textbf{24}, 205-221, (2006).
\bibitem{KKMS} T. Kulczycki, M. Kwa\'{s}nicki, J. Ma{\l}ecki, A. St\'{o}s \textit{Spectral properties of the Cauchy process on half-line and interval},
 Proc. London. Math. Soc. \textbf{101}, 589-622, (2010).
\bibitem{dyda} B. Dyda, \textit{Fractional calculus form power functions and eigenvalues of the fractional Laplacian}, Fractional Calculus
\&  Applied Analysis, {\bf 15}(4), 536, (2012).
\bibitem{dyda1} B. Dyda, \textit{Fractional Hardy inequality with a remainder term}, Colloquium Math. {\bf 122}, (1), 59, (2011).
\bibitem{guan} Q.-Y. Guan and Z. -M. Ma, \textit{Reflected symmetric $\alpha $-stable processes and regional fractional Laplacian}, Probab. Theory
Related Fields, {\bf 134}, 649, (2006).
\bibitem{ZG} M. \.{Z}aba and  P. Garbaczewski, \textit{Solving fractional Schr\"{o}dinger-type spectral problems: Cauchy oscillator and Cauchy well},
 J. Math. Phys. \textbf{55}, (2014) 092103.
\bibitem{garolk} P. Garbaczewski and R. Olkiewicz, \textit{Cauchy noise and affiliated stochastic processes}, J. Math. Phys. {\bf 40}, 1057, (1999).
\bibitem{stein}E. M. Stein, \textit{Singular Integrals and Differentiability Properties of Functions}, (Princeton University Press, Princeton, 1970).
\bibitem{GK} P. Garbaczewski and W. Karwowski, Am. J. Phys., \textit{Impenetrable barriers and canonical quantization}, {\bf 72}, 924, (2004).
\bibitem{cohen} C. Cohen-Tannoudji, B. Diu and F. Lalo\"{e}, \textit{Quantum Mechanics}, (Wiley, Nw York, 1977), vol. 1.
\bibitem{robinett}  M. Belloni and R. W. Robinett, \textit{ The infnite well and Dirac delta function potentials as pedagogical, mathematical and
 physical models in  quantum  mechanics}, Physics Reports, {\bf 540 }, 25, (2014).
 \bibitem{Kbis} M. Kwa\'{s}nicki, private communication.
 \bibitem{KKM}  K. Kaleta, M. Kwa\'{s}nicki and J. Ma{\l}ecki, \textit{One-dimensional quasi-relativistic particle in the box}, Rev. Math. Phys.
 {\bf 25} (8), 1350014, (2013).
\bibitem{GR}  I.S. Gradshteyn and  I.M. Ryzhik, \textit{Table of Integrals, Series, and Products}, Eighth Edition by Daniel Zwillinger and
 Victor Moll  (2014).
\end{thebibliography}
\end{document}